\documentclass[12pt]{article}
\usepackage[latin9]{inputenc}
\usepackage[letterpaper]{geometry}
\usepackage{verbatim}
\usepackage{textcomp}
\usepackage{amsmath}
\usepackage{amssymb}
\usepackage{graphicx}
\usepackage{setspace}
\usepackage{subscript}
\makeatletter

\DeclareTextSymbolDefault{\textquotedbl}{T1}
\providecommand{\tabularnewline}{\\}

\usepackage{latexsym}
\usepackage{amsthm}
\usepackage{color}
\usepackage{amsfonts}

\newtheorem{proposition}{Proposition}\allowdisplaybreaks
\def\@biblabel#1{\hspace*{-\labelsep}}

\input{tcilatex}

\makeatother

\begin{document}
\begin{singlespace}

\title{{\Large{}\vspace{-1.2cm}
}}
\end{singlespace}

\title{Multi-Dimensional Pass-Through and Welfare Measures under Imperfect
Competition\thanks{{\footnotesize{}We are grateful to Yong Chao, Germain Gaudin, Makoto
Hanazono, Hiroaki Ino, Konstantin Kucheryavyy, Laurent Linnemer, Carol
McAusland, Babu Nataha, and Glen Weyl as well as conference and seminar
participants for helpful discussions. Adachi and Fabinger acknowledge
a Grant-in-Aid for Scientific Research (C) (15K03425) and a Grant-in-Aid
for Young Scientists (A) (26705003) from the Japan Society for the
Promotion of Science, respectively. Adachi also acknowledges financial
support from the Japan Economic Research Foundation. Any remaining
errors are solely ours. }}}

\author{Takanori Adachi\thanks{{\footnotesize{}School of Economics, Nagoya University, 1 Furo-cho,
Chikusa, Nagoya 464-8601, Japan; adachi.t@soec.nagoya-u.ac.jp.} } \and Michal Fabinger\thanks{{\footnotesize{}Graduate School of Economics, University of Tokyo,
7-3-1 Hongo, Bunkyo, Tokyo 113-0033, Japan. Fabinger is also a research
associate at CERGE-EI, Prague, Czechia; fabinger@gmail.com. }}}
\maketitle
\begin{abstract}
\begin{onehalfspace}
\vspace{0.12cm}
This paper provides a comprehensive\textbf{ }analysis of welfare
measures when oligopolistic firms face multiple policy interventions
and external changes under general forms of market demands, production
costs, and imperfect competition. We present our results in terms
of two welfare measures, namely,\textbf{ }marginal cost of public
funds and incidence, in relation to\textit{ multi-dimensional pass-through}.
Our arguments are best understood with two-dimensional taxation where
homogeneous firms face \textit{unit} and \textit{ad valorem} taxes.
The first part of the paper studies this leading case. We show, e.g.,
that there exists a simple and empirically relevant set of sufficient
statistics for the marginal cost of public funds, namely unit tax
and ad valorem pass-through and industry demand elasticity. We then
specialize our general setting to the case of price or quantity competition
and show how the marginal cost of public funds and the pass-through
are expressed using elasticities and curvatures of regular and inverse
demands. Based on the results of the leading case, the second part
of the paper presents a generalization with the tax revenue function
specified as a general function parameterized by a vector of multi-dimensional
tax parameters. We then argue that our results are carried over to
the case of heterogeneous firms and other extensions.%
\vspace{-0.05cm}
\end{onehalfspace}

\end{abstract}
\thispagestyle{empty}\pagebreak{}

\setcounter{page}{1}

\section{Introduction}

In many economic situations, it is important to understand the impact
of changes in government policies, market conditions, or production
technology. Both the impact on the economic equilibrium and the associated
change in the welfare of individual economic agents are of interest
to the policymaker. A convenient framework for studying these questions
is laid out in Weyl and Fabinger (2013) for the case of constant
changes in marginal cost such as a unit%
{} tax. In this paper, we provide a substantial generalization of this
framework under general forms of market demands, production costs,
and imperfect competition. We allow not only for changes in costs
proportional to the value of the goods such as an ad-valorem tax but
also for much more general interventions such as different kinds of
market regulation. In addition, we provide a new perspective on the
case of heterogeneous firms.%
{} 

The range of possibilities for governments' intervention is much richer
than just specific and ad-valorem taxes, on which the existing literature
has focused (see Section \ref{SectionTaxationAndWelfareInSymmetricallyDifferentiatedOligopoly}),
especially if we consider regulations of various kinds. Governments
often intervene in the marketplace by restricting sales. Examples
include restrictions of sales on Sundays and holidays in many European
countries, or restrictions of sales of alcohol, both in terms of time
of sales and in terms of locations where sales are allowed. Governments
also often regulate the labor market. Examples include restrictions
on the number of working hours or stipulation of a minimum wage. Besides
these, governments also impose reporting requirements that influence
the degree to which non-compliant firms misreport their marketplace
data to minimize their tax bill. 

Given that there are many possible interventions, it is convenient
to summarize the interventions of interest in a (multi-dimensional)
vector. The impact of infinitesimal changes in these interventions
on prices then corresponds to a vector (or more generally a matrix),
which may be termed \emph{multi-dimensional pass-through}. The relative
size of its components then provides insight into the relationship
between the impact of individual interventions, for example, the impact
of specific vs.\,\,ad-valorem taxes. We show that the multi-dimensional
pass-through is an important determinant of the welfare effects of
various kinds of interventions.\footnote{The usefulness of pass-through in welfare analysis has been verified
by related studies such as Cowan (2012); Miller, Remer, and Sheu (2013);
Weyl and Fabinger (2013); Gaudin and White (2014); MacKay, Miller,
Remer, and Sheu (2014); Adachi and Ebina (2014a,b); Chen and Schwartz
(2015); Gaudin (2016); Cowan (2016); Alexandrov and Bedre-Defolie
(2017); Mrázová and Neary (2017); and Chen, Li and Schwartz (2018).
See also Ritz (2018) for an excellent survey of theoretical studies
on pass-through and pricing under imperfect competition.\textbf{ }In
the context of antitrust analysis, Froeb, Tschantz, and Werden (2005)
theoretically compare the price effects when no synergies in cost
reduction realize when they are passed through as a form of price
reduction. See also Alexandrov and Koulayev (2015) for discussions
on the role of pass-through in antitrust analysis.}

Our arguments are best understood with two-dimensional taxation where
homogeneous firms face unit and ad valorem taxes. Therefore, in the
first part of the paper (Sections \ref{SectionTaxationAndWelfareInSymmetricallyDifferentiatedOligopoly}
and \ref{SectionExpressionsForPassThrough}), we derive succinct formulas
that relate the marginal cost of public funds to pass-through of these
taxes%
.\textbf{ }We also establish a relationship that connects pass-through
of unit taxes and pass-through of ad-valorem taxes in the same market.
Furthermore, we derive convenient expressions for values of unit and
ad valorem pass-through that are valid under a ``general'' type
competition and have not appeared in the previous literature. In
addition, we show how the marginal cost of public funds and pass-through
are expressed in terms of \textit{elasticities} and \textit{curvatures}
of demand and inverse demand when the setting is specialized to price
(differentiated Bertrand) or quantity (pure or differentiated Cournot)
competition. We also provide numerical examples. Our results also
apply without a change to symmetric oligopoly with multi-product firms.
Throughout the analysis, we allow for non-zero levels of unit and
ad valorem taxes. However, we also discuss some additional simplifications
that appear when instead the initial level of taxes is zero. 

Inspired by the results of this setting of two-dimensional taxation,
the second part of the paper generalizes our model in two directions
(Sections \ref{SectionMultiDimensionalPassThroughFramework}, \ref{SectionAsymmetricFirms},
and \ref{SectionPassThroughAndWelfareUnderProductionCostAndTaxationChanges}).
First, we allow for interventions that are more general than just
specific and ad valorem taxes. Second, we introduce firm heterogeneity.
We find that these more general relationships have a form that still
relatively simple and succinct. This substantially expands the applicability
of our results. 

From both theoretical and empirical standpoints, it is desirable to
be able to understand the welfare properties of oligopolistic markets
with a fairly general type of competition. In real-world situations,
firms' behavior may not simply be categorized into either the idealized
price competition or the idealized quantity competition. Price competition
does not allow for any friction in scaling production levels up or
down, yet in reality there tend to be substantial frictions, such
as those related to financial constraints or the labor market. Quantity
competition implies that the firm will not be able to increase production
levels when its competitor suddenly decides to increase prices. In
reality, such adjustment is probably feasible, since capacity utilization
is typically less than complete, and even if the firm is operating
at full capacity, boosting production levels is possible by overtime
work or by hiring temporary workers. Moreover, firms may behave, to
some extent, in a collusive way. Although the realities of competition
by firms may be complicated, it is possible to capture their essence
by working with a ``general'' type of competition, using the \textit{conduct
index}.\footnote{Our conduct index is a generalization of what is known as ``conduct
parameter'' in the empirical industrial organization literature,
where it is supposed to be constant for any level of industry output
as a target of estimation (see, e.g., Bresnahan 1989; Genesove and
Mullin 1998; Nevo 1998; Corts 1999; Delipalla and O'Donnell 2001;
and Shcherbakov and Wakamori 2017). It has also been successfully
applied to more general situations, such as selection markets (Mahoney
and Weyl 2017) or supply chains (Gaudin 2018). In this paper, we
generalize this concept by allowing it to vary across different levels
of output: we thus opt for the term ``conduct index'' to make it
explicit that it is a variable. In a less general setting, such ``conduct
index'' was used by Weyl and Fabinger (2013), where it was still
referred to as ``conduct parameter''. \label{FootnoteConduct}} 

Besides working with a more general type of competition, it is also
useful to relax the assumption of constant marginal costs that often
appears in the literature. Production technologies often have a non-trivial
structure, and so does the internal organization of the firm. For
example, if a firm decides to operate at a larger scale, it may take
advantage of technological and logistical economies of scale, but
at the same time, it may face more severe principal-agent problems
as top managers have to delegate responsibilities to lower-level managers.
The interplay between these forces can lead to a non-trivial dependence
of the marginal cost of production on the scale of the operation.
A notable benefit from our general framework is that one does not
necessarily have to assume constant marginal cost in conducting welfare
assessment in a precise manner.

In this spirit, the aim of Weyl and Fabinger's (2013) study is to
analyze imperfect competition in a way that does not rely on limiting
functional form assumptions.%
{} It is%
{} followed by Fabinger and Weyl (2018), who propose using%
{} further flexible functional forms to make economic models or their
parts solvable in closed form. The present paper goes in a different
direction: we focus on more general market interventions and exogenous
market changes without requiring that the models or their parts be
solved in closed form. As mentioned above, ad valorem taxes are not
studied in Weyl and Fabinger (2013). In this paper, we provide a more
general framework that allows not only for ad valorem taxes but for
other interventions in a general manner.

The rest of the paper mainly consists of two parts: the first part
focuses on the canonical situation of two-dimensional taxation where
symmetric firms face unit and ad valorem taxes, and the second part
extends these results to a more general framework of multiple government
interventions and external changes, allowing firm heterogeneity. First,
in the next section, we set up a framework of two-dimensional taxation
under imperfect competition and provide general formulas for marginal
cost of public funds and incidence in relation to unit tax and ad
valorem tax pass-throughs and industry demand elasticity. In Section
\ref{SectionExpressionsForPassThrough}, we specialize this setting
to the case of price or quantity competition. Then, in the second
part of the paper, Section \ref{SectionMultiDimensionalPassThroughFramework}
generalizes the results from unit and ad valorem taxation to much
more flexible taxation parameterized by $d$ different tax parameters
and discusses the implications of these general results. We also discuss
other types of government/exogenous interventions that are suitable
for our framework. Then, in Section \ref{SectionAsymmetricFirms}
we generalize our formulas to include heterogeneous firms. Finally,
Section \ref{SectionPassThroughAndWelfareUnderProductionCostAndTaxationChanges}
generalizes our previous results to include changes in both production
costs and taxes. Section \ref{SectionConcludingRemarks} concludes
the paper.

\section{Taxation and Welfare in Symmetric Oligopoly}

\label{SectionTaxationAndWelfareInSymmetricallyDifferentiatedOligopoly}

We begin with a canonical setting where firms face two types of taxation:\textit{
unit} and \textit{ad valorem} taxes. This issue has been in the central
part of the existing literature on government interventions. The welfare
cost of taxation has been extensively studied at least since Pigou
(1928). The majority of the studies simply assume perfect competition
(with zero initial taxes).\footnote{See, e.g., Vickrey (1963), Buchanan and Tullock (1965), Johnson and
Pauly (1969), and Browing (1976) for early studies. A study of unit
and ad valorem taxation under imperfect competition with homogeneous
products dates back to Delipalla and Keen (1992). See Auerbach and
Hines (2002) and Fullerton and Metcalf (2002) for comprehensive surveys
for this field.} As is widely known, under perfect competition, unit tax and ad valorem
tax are equivalent, and whether consumers or producers bear more is
determined by the relative elasticities of demand and supply (Weyl
and Fabinger, 2013, p.$\,$534). The initial attempt to relax the
assumption of perfect competition started with an analysis of \textit{homogeneous}-product
oligopoly under quantity competition, i.e., Cournot oligopoly. Notably,
Delipalla and Keen (1992), Skeath and Trandel (1994), and Hamilton
(1999) compare ad valorem and unit taxes in such a setting. Then,
Anderson, de Palma, and Kreider (2001a) extend these results to the
case of\textit{ differentiated} oligopoly under price competition.
In particular, Anderson, de Palma, and Kreider (2001a) find that whether
the after-tax price for firms and their profits rise by a change in
ad valorem tax depends importantly on the ratio of the curvature of
the firm's own demand to the elasticity of the market demand.\footnote{This curvature is denoted $\varepsilon_{m}$ in their notation, whereas
we instead use $\alpha_{F}$ below, and this elasticity is denoted
$\varepsilon_{DD}$ in their notation, whereas we instead use $\epsilon$
below.}

In this section, we extend Anderson, de Palma, and Kreider's (2001a)
setting and results in a number of important directions. First, we
consider a ``general'' mode of competition, captured by the conduct
index, including both quantity and price competition. Second, we provide
a complete characterization of tax burdens that enables one to quantitatively
compare consumers' burden with producers' burden, whereas Anderson,
de Palma, and Kreider's (2001a) focus only on the effective prices
for consumers and producers' profits. Third, while Anderson, de Palma,
and Kreider's (2001a) assume constant marginal cost, we allow non-constant
marginal cost and show how this generalization makes a difference
in our general formulas. Fourth, we further generalize the initial
tax level. When they analyze the effects of a unit tax, Anderson,
de Palma, and Kreider (2001a) assume that ad valorem tax is zero,
and vice versa. In contrast, we allow non-zero initial taxes in both
dimensions. Finally, and importantly, we generalize these results
to the case of a very general type of taxation, as well as to production
cost changes. This opens up the possibility to study a wider range
of interventions/taxes and to derive convenient sufficient statistics
for characteristics, including welfare characteristics, of the markets
of interest.\footnote{Our framework is in line with the \textquotedblleft sufficient-statistics\textquotedblright{}
approach to connecting structural and reduced-form methods, as advocated
by Chetty (2009), which has been successful in empirical economics.
For example, in the study by Atkin and Donaldson (2016), the pass-through
rate provides a sufficient statistic for welfare implications of intra-national
trade costs in low-income countries, without the need for a full demand
estimation. Similarly, Ganapati, Shapiro, and Walker (2018) examine
the welfare effects of input taxation, where a unit tax is levied
on the input. These effects are related to the effects of unit taxes
on output, but not identical. See also Fabra and Reguant (2014); Shrestha
and Markowitz (2016); Stolper (2016); Hong and Li (2017); Duso and
Szücs (2017); Gulati, McAusland, and Sallee (2017); and Muehlegger
and Sweeney (2017) for studies with the same spirit. In contrast,
Kim and Cotterill (2008) is among the first studies that structurally
estimate cost pass-through in differentiated product markets, followed
by Bonnet, Dubois, Villas-Boas, and Klapper (2013); Bonnet and Réquillart
(2013); Campos-Vázquez and Medina-Cortina (2015); Miller, Remer, Ryan,
and Sheu (2016); Conlon and Rao (2016); Miller, Osborne, and Sheu
(2017); and Griffith, Nesheim, and O'Connell (2018).}

\subsection{Setup}

Below, we study oligopolistic markets with $n$ symmetric firms and
a \textit{general} (first-order) mode of competition and the resulting
symmetric equilibria.\footnote{Although for brevity we speak of a general mode of competition, we
consider only ``first-order'' competition, in the sense of the firms
making decisions based on marginal cost and marginal revenue. This
excludes, for example, the possibility of vertical industries, an
important issue left for future research. } Our discussion applies to single-product firms as well as to \textit{multi-product}
firms if intra-firm symmetry conditions are satisfied, as discussed
in Appendix \ref{AppendixOligopolyWithMultiProductFirms}. For simplicity
of exposition, we use terminology corresponding to single-product
firms here, and later we discuss how to interpret the results in the
case of multi-product firms. 

Formally, the demand for firm $j$'s product $q_{j}=q_{j}(p_{1},...,p_{n})\equiv q_{j}\left(\mathbf{p}\right)$
depends on the vector of prices $\mathbf{p}\equiv\left(p_{1},...,p_{n}\right)$
charged by the individual firms. The demand system is symmetric and
the cost function $c(q_{j})$ is the same for all firms. We assume
that $q_{j}(\cdot)$ and $c(\cdot)$ are twice differentiable and
conditions for the uniqueness of equilibrium and the associated second-order
conditions are satisfied. We denote by $q(p)$ the per-firm industry
demand corresponding to symmetric prices: $q(p)\equiv q_{j}(p,...,p)$.
The elasticity of this function, defined as $\epsilon(p)\equiv-pq^{\prime}(p)/q(p)>0$
and referred to as the \textit{price elasticity of industry demand,}
should not be confused with the elasticity of the residual demand
that any of the firms faces.\footnote{The elasticity $\epsilon$ here corresponds to $\epsilon_{D}$ in
Weyl and Fabinger (2013, p.$\,$542). Note that $q^{\prime}(p)=\partial q_{j}(\mathbf{p})/\partial p_{j}+(n-1)\partial q_{j}(\mathbf{p})/\partial p_{j^{\prime}}|_{\mathbf{p}=\left(p,...,p\right)}$
for any two distinct indices $j$ and $j'$. We will define the firm's
elasticity and other related concepts in Section 3.\label{FootnoteIndustryDemandElasticity}} We also use the notation $\eta(q)=1/\epsilon\left(p\right)|_{q\left(p\right)=q}$
for the reciprocal of this elasticity as a function of $q$. For the
corresponding functional values, when we do not need to specify explicitly
their dependence on either $q$ or $p$, we use $\eta$ interchangeably
with $1/\epsilon$.

As mentioned above, we introduce two types\footnote{This specification corresponds to a two-dimensional (tax) instrument
$\left(t,v\right)$, which is a special case of multi-dimensional
instruments. For example, if the cost function had an additional technology
parameter $z$, we could describe the situation using a three-dimensional
instrument $\left(t,v,z\right).$\textbf{ }In Section \ref{SectionMultiDimensionalPassThroughFramework},
we introduce a framework for multi-dimensional pass-through. For now,
we specialize to the two-dimensional case of specific and ad valorem
taxes, which are very common: for example, in the United States both
types of taxes are imposed on the sales of soda, alcohol and cigarettes. } of taxation: a unit tax $t$ and an ad valorem tax $v$, with firm
$j$'s profit being $\pi_{j}=(1-v)p_{j}(q_{j})q_{j}-tq_{j}-c(q_{j})$.
At symmetric quantities the government tax revenue per firm is $R\left(q\right)\equiv tq+vp\left(q\right)q$,
and we denote by $\tau\left(q\right)$ the fraction of firm's pre-tax
revenue that is collected by the government in the form of taxes:
$\tau\left(q\right)\equiv R\left(q\right)/pq=v+t/p\left(q\right)$.
We now introduce the \emph{conduct index $\theta(q)$}, which measures
the industry's competitiveness (a lower \emph{$\theta$} corresponds
to a fiercer level of competition). Here, it is determined\emph{ independently}
of the cost side but potentially can change for different values of
the industry's output, $q$. Then, the symmetric equilibrium condition
is written as
\begin{equation}
\frac{1}{\eta\left(q\right)p\left(q\right)}\left(p\left(q\right)-\dfrac{t+mc\left(q\right)}{1-v}\right)=\theta(q)\text{,}\label{theta_M}
\end{equation}
where $mc(q)\equiv c^{\prime}(q)$ is the marginal cost of production.\footnote{As already noted in Footnote \ref{FootnoteConduct} above, \emph{$\theta(q)$}
is a generalization of conduct parameter in the sense that it is a
function of $q$ rather than a constant for any $q$. Hence, Equation
(\ref{theta_M}) \emph{should not be interpreted as an equation that
defines $\theta(q)$}. For our analysis we introduce\emph{ $\theta(q)$}
in an implicit manner:\emph{ $\theta(q)$} is a function independent
of the cost side of the problem such that Equation (\ref{theta_M})
is the symmetric first-order condition of the equilibrium. For specific
types of (\textquotedbl first-order\textquotedbl ) competition,
such as those discussed in Section \ref{SectionExpressionsForPassThrough},
it is possible to derive explicit expressions for \emph{$\theta(q)$
}that can replace our implicit definition.  Presumably, it is natural
to assume $\theta'(\cdot)<0$: a smaller amount of production is associated
with a smaller degree of competitiveness in the industry due to other
reasons than non-cooperative and simultaneous pricing (modeled here),
such as (unmodeled) tacit collusion. However, this restriction is
not necessary for the following analysis.  Also, note that $\theta(q)>1$
is not necessarily excluded, although in most interesting cases it
lies in $[0,1]$.} We denote by $\theta$ the functional value of $\theta(q)$ at the
equilibrium quantity. This is also understood as the\textit{ elasticity-adjusted
Lerner index}: the markup rate $[p-(t+mc)/(1-v)]/p$ should be adjusted
by the industry-wide elasticity to reflect the competitiveness in
the industry, where $(t+mc)/(1-v)$ is interpreted as the effective
marginal cost.\footnote{Accordingly, one can write the modified Lerner rule under $(v,t)$
as $(p-\frac{t+mc}{1-v})/p=\eta\theta,$ which implies the restriction
on $\theta$: $\theta\leq\epsilon$.} We emphasize here that once the conduct index is introduced, one
is able to describe oligopoly in a synthetic manner, without specifying
whether it is price or quantity setting, or whether it exhibits strategic
substitutability or complementarity. 

\subsection{The marginal cost of public funds and pass-through}

\textit{The marginal cost of public funds}, i.e., the marginal social
welfare loss associated with raising additional tax revenue, is a
crucial characteristic that a policymaker needs to take into account
when designing an optimal system of taxes.\footnote{In the absence of other considerations, the marginal cost of public
funds should be equalized across markets in order to maximize social
welfare.} In the special case of linear demand and constant marginal cost,
Häckner and Herzing (2016, p.\,147) explain that as long as the initial
level of taxes is zero, the marginal cost of public funds for unit
taxation equals $MC_{t}=\theta\rho_{t}$, where $\rho_{t}$ is the
unit-tax\textit{ pass-through} rate (the marginal effect of unit taxes
on prices), and $\theta$ is the conduct index. For ad valorem taxes,
Häckner and Herzing (2016) provide a similar formula. They argue%
, however, that if we let the initial level of taxes be non-zero,
those formulas are no longer valid. For this reason, they are forced
to analyze the magnitude of the marginal cost of public funds on a
case-by-case basis using explicit solutions to specific models.

This situation represents a puzzle. If there are simple formulas for
the marginal cost of public funds that were valid at zero taxes, is
there no compact generalization of these expressions in the case of
non-zero taxes? If there is no such generalization, that would be
an obstacle to empirical work, since we would have to make additional
modeling assumptions before obtaining empirical estimates of the marginal
cost of public funds. Our paper provides a solution to this problem.
In particular, Proposition \ref{PropositionMCt} below presents formulas
for the marginal cost of public funds that are valid even when the
initial level of (both ad valorem and unit) taxes is non-zero. They
are a bit longer than $MC_{t}=\theta\rho_{t}$, but still very manageable.
They also represent a starting point for the topics discussed in the
rest of the paper. These results with a non-zero initial taxes being
allowed, which are differentiated from Weyl and Fabinger (2013) and
Häckner and Herzing (2016), should be useful if one needs to evaluate
the marginal cost of taxation \textit{when some tax has been already
implemented}.

The marginal welfare cost $MC_{t}$ or $MC_{v}$ of raising government
revenue by the unit tax $t$ or the ad valorem tax $v$, i.e. the
marginal cost of public funds associated with such a tax, is defined
as 
\[
MC_{t}\equiv-\left(\frac{\partial R}{\partial t}\right)^{-1}\frac{\partial W}{\partial t},\qquad MC_{v}\equiv-\left(\frac{\partial R}{\partial v}\right)^{-1}\frac{\partial W}{\partial v},
\]
where $W$ is the social welfare per firm, which includes consumer
surplus, producer surplus, and government tax revenue. We define
the\textit{ unit tax pass-through rate} $\rho_{t}$ and the\textit{
ad valorem tax pass-through semi-elasticity} $\rho_{v}$ as:\footnote{Note that Häckner and Herzing (2016) use the symbol $\rho_{v}$ for
the ad valorem tax pass-through rate $\partial p/\partial v$, which
corresponds to $p\rho_{v}$ in our notation.}
\[
\rho_{t}=\frac{\partial p}{\partial t},\qquad\rho_{v}=\frac{1}{p}\ \frac{\partial p}{\partial v}.
\]

Consider an infinitesimal change in the unit tax, with the initial
tax level $(t,v)$. As mentioned in the introduction, in the special
case of zero initial taxes, linear demand, and constant marginal cost,
Häckner and Herzing (2016, p.\,147) show that $MC_{t}=\theta\rho_{t}$
and $MC_{v}=\theta\rho_{v}$, noting that at non-zero initial taxes
the formula no longer applies. In the absence of such formula, they
were forced to study the marginal cost of public funds on a case-by-case
basis, for different specifications of demand and cost.

Intuitively, there are at least two reasons why $\theta\rho_{t}$
fails to be an accurate measure of the marginal cost of public funds
when a unit tax is raised.\footnote{An analogous argument applies for $\theta\rho_{v}$ and the marginal
cost of public funds of tax $v$.} First, the expression is simply proportional to $\theta$, but when
$v$ is large, the firms sell at prices that are too high from the
social perspective not mainly because of a lack of competitiveness,
but primarily because the tax effectively raises their perceived cost.
When $v$ is large, we would expect the marginal cost of public funds
to be less sensitive to $\theta$, for a given value of $\rho_{t}$.
Second, the expression $\theta\rho_{t}$ does not explicitly feature
the level of the unit tax $t$. However, a situation where $t$ is
large and $mc$ small is very different from a situation where $t$
is small and $mc$ large, even if the equilibrium prices and quantities
are the same. In the former case, raising additional tax revenue
is quite harmful, since firms' production cuts will not substantially
decrease the total technological (i.e., pre-tax) cost of production.
In the latter case, raising additional tax revenue is \textit{less}%
{} harmful since it leads to reduced total technological cost. Based
on this intuition, we would expect the marginal cost of public funds
to be an increasing function of $t$.\footnote{In the sense of making the change $t\rightarrow t+\Delta t$, and
simultaneously $c\left(q\right)\rightarrow c\left(q\right)-q\Delta t$
in order to keep $q$, $\theta$, and $\rho_{t}$ at some fixed values.}

Thus, we are led to find a generalization of the formula $MC_{t}=\theta\rho_{t}$
and $MC_{v}=\theta\rho_{v}$ that would be applicable even at non-zero
initial taxes. It turns out that it is possible to identify a formula
with precisely these properties, as the following proposition shows.

\begin{proposition}\label{PropositionMCt}\textbf{Marginal cost and
total of public funds for unit and ad valorem taxations.} Under symmetric
oligopoly with a possibly non-constant marginal cost, the marginal
cost of public funds associated with a unit tax may be expressed as
\[
MC_{t}=\frac{(1-v)\thinspace\theta+\epsilon\thinspace\tau}{\frac{1}{\rho_{t}}+v-\epsilon\thinspace\tau},
\]
and the marginal cost of public funds associated with an ad valorem
tax may be expressed as 
\[
MC_{v}=\frac{(1-v)\thinspace\theta+\epsilon\tau}{\frac{1}{\rho_{v}}+v-\epsilon\tau}.
\]
\end{proposition}

\begin{figure}
\begin{centering}
\includegraphics{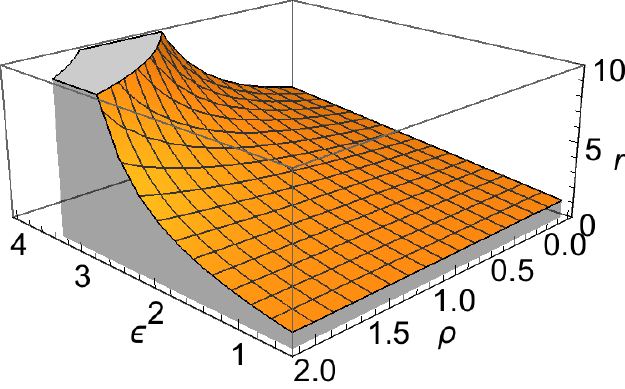}\qquad{}\includegraphics{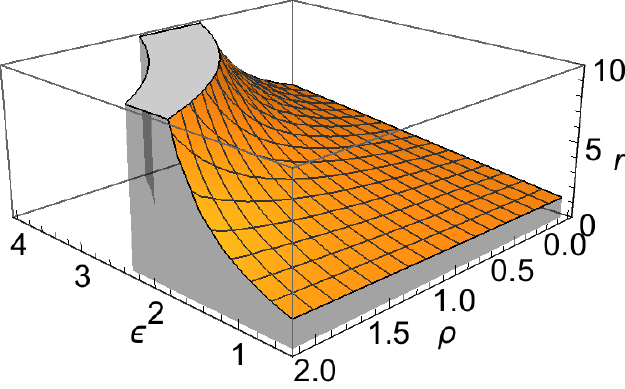}\bigskip{}
\bigskip{}
\par\end{centering}
\begin{centering}
\includegraphics{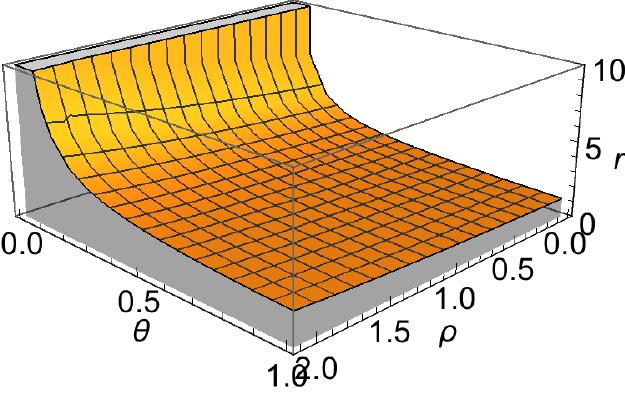}\qquad{}\includegraphics{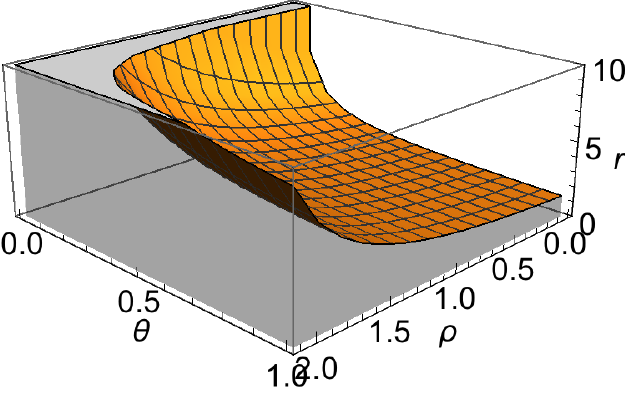}\bigskip{}
\bigskip{}
\par\end{centering}
\begin{centering}
\includegraphics{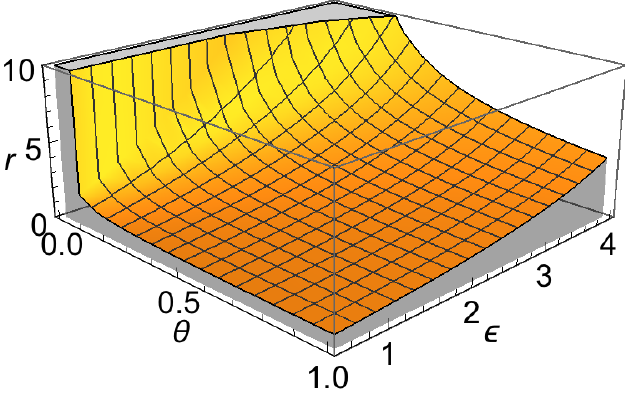}\qquad{}\includegraphics{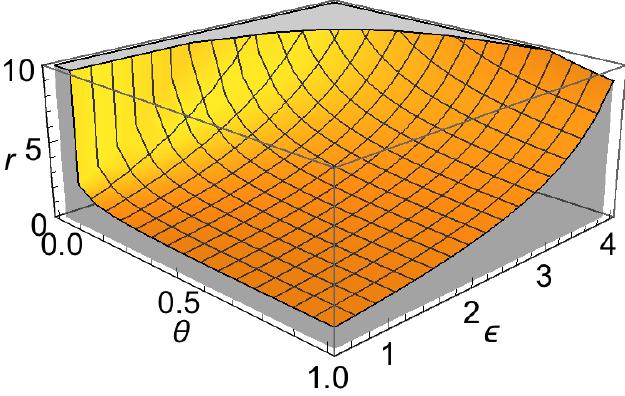}
\par\end{centering}
\centering{}\caption{\label{FigureMarginalCostOfPublicFunds}The ratio of the actual marginal
cost of public funds $MC$ and the naive expression $\theta\rho$
discussed just before Proposition \ref{PropositionMCt}, plotted as
a function of combinations of the conduct index $\theta$, the pass-through
$\rho$, and the industry demand elasticity $\epsilon$. The figures
on the left correspond to infinitesimal changes in unit taxation:
$\rho$ stands for $\rho_{t}$ and $MC$ stands for $MC_{t}$. The
numerical values were chosen to be $t=0,\ v=0.2,\ \tau=0.2$. The
figures on the right correspond to infinitesimal changes in ad valorem
taxation: $\rho$ stands for $\rho_{v}$ and $MC$ stands for $MC_{v}$.
The numerical values were chosen to be $t/p=0.2,\ v=0,\ \tau=0.2$.
The top figures correspond to $\theta=0.3$, the middle figures correspond
to $\epsilon=2$, and the bottom figures correspond to $\rho=1$.}
\end{figure}
The proposition is proven in Appendix \ref{AppendixProofOfPropositionForMCtAndMCv}
and the intuition behind it is discussed in detail in Appendix \ref{AppendixDiscussionofPropositionForMCtAndMCv}.
Here in the main text we just include Figure \ref{FigureMarginalCostOfPublicFunds},
which documents that these expressions for the marginal cost of public
funds $MC_{t}$ and $MC_{v}$ evaluated at realistic values of taxes
and other economic variables are very different from the values of
the expressions $\theta\rho_{t}$ and $\theta\rho_{v}$ (discussed
above) that would be equal to $MC_{t}$ and $MC_{v}$ if taxes were
zero. 

\subsection{Incidence and pass-through}

We define the incidence $I_{t}$ of unit taxation as the ratio of
changes $dCS$ in (per-firm) consumer surplus and changes $dPS$ in
(per-firm) producer surplus induced by an infinitesimal increase $dt$
in the unit tax $t$. The incidence $I_{v}$ of ad valorem taxation
is defined analogously. We obtain the following succinct results for
the incidence of taxation at non-zero unit and ad valorem taxes.

\begin{proposition}\label{PropositionConsumerAndProducerIncidence}\textbf{Incidence
of taxation.} Under symmetric oligopoly with a general type of competition
and with a possibly non-constant marginal cost, the incidence of unit
taxes $I_{t}$ and of ad valorem taxes $I_{v}$ is given by
\[
\frac{1}{I_{t}}=\frac{1}{\rho_{t}}-\left(1-v\right)\left(1-\theta\right),\qquad\frac{1}{I_{v}}=\frac{1}{\rho_{v}}-\left(1-v\right)\left(1-\theta\right).
\]
\end{proposition}

The proposition is proven in Appendix \ref{AppendixProofOfPropositionForItAndIv},
and we discuss it in detail in Appendix \ref{AppendixDiscussionOfPropositionOnItAndIv}.
Note that in the case of zero ad valorem tax, the expression for $I_{t}$
reduces to Weyl and Fabinger's (2013, p.$\:$548) Principle of Incidence
3. Next we show how $\rho_{t}$ and $\rho_{v}$ are related in the
following proposition.

\begin{proposition}\label{PropositionRelationshipBetweenPassThroughs}\textbf{Relationship
between pass-through of ad valorem and unit taxes.} Under symmetric
oligopoly with a possibly non-constant marginal cost, the pass-through
semi-elasticity $\rho_{v}$ of an ad valorem tax may be expressed
in terms of the unit tax pass-through rate $\rho_{t}$, the conduct
index $\theta$, and the industry demand elasticity $\epsilon$ as
\begin{equation}
\rho_{v}=\left(1-\frac{\theta}{\epsilon}\right)\ \rho_{t}.\label{EquationRelationshipBetweenPassThroughs}
\end{equation}
\end{proposition}

The proposition is proven Appendix \ref{AppendixProofOfPropositionRelatingRhotAndRhov},
and Appendix \ref{AppendixDiscussionOfPropositionRelatingPassThroughs}
provides a detailed discussion. Combined with Proposition \ref{PropositionMCt},
it is consistent with the well-known result that unit tax and ad valorem
tax are equivalent in the welfare effects under perfect competition:
if\textbf{ $\theta=0$}, then $\rho_{t}=\rho_{v}$. Under imperfect
competition, $\rho_{t}>\rho_{v}$, and $MC_{t}>MC_{v}$. This provides
another look of Anderson, de Palma, and Kreider's (2001b) result that
unit taxes are welfare-inferior to ad valorem taxes.\footnote{Under Cournot competition,  Equation (6.13) of Auerbach and Hines
(2002) coincides with Equation (\ref{EquationRelationshipBetweenPassThroughs})
above. Proposition \ref{PropositionRelationshipBetweenPassThroughs}
shows that this equation holds more generally. We thank Germain Gaudin
for pointing this out.}

By combining Propositions \ref{PropositionMCt} and \ref{PropositionRelationshipBetweenPassThroughs},
we find that $MC_{t}$ and $MC_{v}$ can be expressed without the
degree of competitiveness $\theta$.

\begin{proposition}\textbf{\label{Sufficient-statistics-for}Sufficient
statistics for marginal costs of public funds.} Under symmetric oligopoly
with a possibly non-constant marginal cost, the unit pass-through
rate $\rho_{t}$, the ad valorem pass-through semi-elasticity $\rho_{v}$,
and the elasticity $\epsilon$ of industry demand (together with the
tax rates and the fraction $\tau$ of the firm's pre-tax revenue collected
by the government in the form of taxes) serve as sufficient statistics
for the marginal cost of public funds both with respect to unit taxes
and ad valorem taxes. In particular: 
\[
MC_{t}=\frac{(1-v+\tau)\rho_{t}-(1-v)\rho_{v}}{1+(v-\epsilon\tau)\rho_{t}}\ \epsilon,\qquad MC_{v}=\frac{\left(1-v+\tau\right)\rho_{t}-(1-v)\rho_{v}}{1+\left(v-\epsilon\tau\right)\rho_{v}}\ \frac{\rho_{v}}{\rho_{t}}\epsilon.
\]

\end{proposition}

The proof is simple: Proposition \ref{PropositionRelationshipBetweenPassThroughs}
allows us to express the conduct index $\theta$ as $\theta=(1-\rho_{v}/\rho_{t})\epsilon$.
Substituting this into the relationships in Proposition \ref{PropositionMCt}
then gives the desired result. For a detailed discussion of related
intuition, see Appendix \ref{AppendixDiscussionOfPropositionOnSufficientStatistics}.

As the last result presented in this section, the following proposition
shows how the two forms of pass-through are characterized.

\begin{proposition}\label{PropositionGeneralSymmetricOligopolyPassThrough}\textbf{Pass-through
under symmetric oligopoly.} Under symmetric oligopoly with a general
(first-order) mode of competition and with a possibly non-constant
marginal cost: 
\[
\rho_{t}=\frac{1}{1-v}\ \frac{1}{\left[1+\frac{1-\tau}{1-v}\epsilon\chi\right]-\left(\eta+\chi\right)\theta+\epsilon q\left(\theta\eta\right)'},
\]
where the derivative is taken with respect to $q$ and $\chi\equiv mc^{\prime}q/mc$
is the elasticity of the marginal cost with respect to quantity. Further,
\[
\rho_{v}=\frac{\epsilon-\theta}{\left(1-v\right)\epsilon}\ \frac{1}{\left[1+\frac{1-\tau}{1-v}\epsilon\chi\right]-\left(\eta+\chi\right)\theta+\epsilon q\left(\theta\eta\right)'}.
\]
\end{proposition}

The proof and a related discussion are in Appendix \ref{AppendixProofOfPropositionGeneralOligopolyPassThrough}.
Further, we discuss a relationship with Weyl and Fabinger (2013) in
Appendix \ref{AppendixRelationshipToWeylAndFabinger}, provide a comparison
of perfect and oligopolistic competition in Appendix \ref{AppendixComparisonOfPerfectAndOligopolisticCompetition},
and show the applicability of these results to exchange rate changes
in Appendix \ref{AppendixApplicationToExchangeRateChanges}.

\subsection{Global changes in surplus measures}

So far, we have discussed local, i.e.$\,$infinitesimal, changes
in surplus measures ($CS$, $PS$, $R$, $W$). For larger changes
in some tax $T$, it is desirable to have an understanding of global
changes in surplus measures. Consider surplus measures $A$ and $B$.
Their finite changes $\Delta A=\int_{T_{1}}^{T_{2}}\frac{dA\left(T\right)}{dT}dT$
and $\Delta B=\int_{T_{1}}^{T_{2}}\frac{dB\left(T\right)}{dT}dT$
induced by a tax change from $T=T_{1}$ to $T=T_{2}$ are related
to their incidence ratios $\Theta{}_{AB}\equiv\frac{dA\left(T\right)}{dT}/\frac{dB\left(T\right)}{dT}$.
In particular, $\Delta A/\Delta B$ is a weighted average of $\Theta{}_{AB}$
over the interval $\left(T_{1},T_{2}\right)$:
\[
\frac{\Delta A}{\Delta B}=\int_{T\in\left(T_{1},T_{2}\right)}\!\!\!\Theta{}_{AB}\,dw_{B}^{\left(T_{1},T_{2}\right)}\!\left(T\right),
\]
where $dw_{B}^{\left(T_{1},T_{2}\right)}\!\left(T\right)\equiv\frac{dB\left(T\right)}{dT}dT/\int_{T_{1}}^{T_{2}}\frac{dB\left(T'\right)}{dT'}dT'$
is a weight normalized to one on the corresponding interval: $\int_{T_{1}}^{T_{2}}dw_{B}^{\left(T_{1},T_{2}\right)}\!\left(T\right)=1$.
The weight is positive as long as $\frac{dB\left(T\right)}{dT}$ has
the same sign as $\int_{T_{1}}^{T_{2}}\frac{dB\left(T\right)}{dT}dT$,
which is typically satisfied in useful applications. In many useful
cases $A$ and $B$ are zero at infinite $T$. Then
\[
\frac{A\left(T_{1}\right)}{B\left(T_{1}\right)}=\int_{T\in\left(T_{1},\infty\right)}\!\!\!\Theta{}_{AB}\,dw_{B}^{\left(T_{1},\infty\right)}\!\left(T\right).
\]
For example, 
\[
\frac{CS\left(T_{1}\right)}{PS\left(T_{1}\right)}=\int_{T\in\left(T_{1},\infty\right)}\!\!\!I_{T}\,dw_{PS}^{\left(T_{1},\infty\right)}\!\left(T\right),\qquad\frac{W\left(T_{1}\right)}{R\left(T_{1}\right)}=\int_{T\in\left(T_{1},\infty\right)}\!\!\!MC_{T}\,dw_{R}^{\left(T_{1},\infty\right)}\!\left(T\right).
\]
In the case of a per-unit tax, we obtain\footnote{Note that in this case $dw_{PS}^{\left(t,\infty\right)}=q\,dt/PS=q\,dt/\int_{t_{1}}^{\infty}q\,dt$}
\[
\frac{CS\left(t_{1}\right)}{PS\left(t_{1}\right)}=\frac{\int_{t_{1}}^{\infty}I{}_{T}\,q\,dt}{\int_{t_{1}}^{\infty}q\,dt}.
\]

\section{Taxation and Welfare under Specific Types of Competition }

\label{SectionExpressionsForPassThrough}In this section, we show
that for price competition and quantity competition in differentiated
oligopoly, our general expressions of the marginal cost of public
funds and pass-through lead to expressions in terms of demand primitives
such as the\textit{ elasticities} and the\textit{ curvatures}, and
the marginal cost elasticity $\chi$ defined above.\footnote{The question of whether quantity- or price-setting firms are more
appropriate depends on the nature of competition. As Riordan (2008,
p.\,176) argues, quantity competition is a more appropriate model
if one depicts a situation where firms determine the necessary capacity
for production. However, price-setting firms are more suitable if
firms in the industry of focus can quickly adjust to demand by changing
their prices. Although the real-world case of competition is not as
clear-cut as this, as we have emphasized in Introduction, we argue
below that it is possible to provide another useful characterization
for the marginal costs of public funds and the pass-through rates
by specifying the mode of competition.}  Throughout this section, we assume that firms' conduct is simply
described by one-shot Nash behavior, without any other further possibilities
such as tacit collusion. As seen below, this assumption enables one
to express the conduct index in terms of demand and inverse demand
elasticities, using Equation (\ref{theta_M}) directly (see Subsection
\ref{SubsectionExpressionConduct}). We also provide parametric examples
for these results.

\subsection{Elasticities and curvatures of the demand system}

\textbf{\label{SubsectionElasticitiesAndCurvatures}Direct demand.}
Following Holmes (1989, p.\,245), we define the \textit{own price
elasticity} $\epsilon_{F}(p)$ and \textit{the cross price elasticity
$\epsilon_{C}(p)$} \textit{of the firm's direct demand}\emph{ }by
\[
\epsilon_{F}(p)\equiv-\frac{p}{q\left(p\right)}\ \frac{\partial q_{j}(\mathbf{p})}{\partial p_{j}}|_{\mathbf{p}=\left(p,...,p\right)},\qquad\epsilon_{C}(p)\equiv\frac{(n-1)p}{q\left(p\right)}\ \frac{\partial q_{j^{\prime}}(\mathbf{p})}{\partial p_{j}}|_{\mathbf{p}=\left(p,...,p\right)},
\]
where $j$\textit{ }and\textit{ $j'$ }is an arbitrary pair of distinct
indices. These are related to the industry demand elasticity $\epsilon(p)$
by $\epsilon_{F}=\epsilon+\epsilon_{C}$.\footnote{Holmes (1989) shows this for two symmetric firms, but it is straightforward
to verify this relation more generally. See the equation in Footnote
\ref{FootnoteIndustryDemandElasticity} above. Note that the equation
$\epsilon_{F}=\epsilon+\epsilon_{C}$ simply means that the percentage
of consumers who cease to purchase firm\textit{ $j$}'s product in
response to its price increase is decomposed into (i) those who no
longer purchase from any of the firms ($\epsilon$) and (ii) those
who switch to (any of) the other firms' products ($\epsilon_{C}$).
Thus, $\epsilon_{F}$\textbf{ }measures the\textit{ firm's own competitiveness}:
it is decomposed into the industry elasticity and the degree of rivalry.
In this sense, these three price elasticities characterize \textquotedblleft first-order
competitiveness,\textquotedblright{} which determines whether the
equilibrium price is high or low, but one of them is not independently
determined from the other two elasticities.} Next, we define the \textit{curvature of the industry's direct demand}
$\alpha(p)$, the \textit{own} \textit{curvature $\alpha_{F}(p)$
of the firm's direct demand} and the \textit{cross curvature $\alpha_{C}(p)$
of the firm's direct demand}:\footnote{The curvature $\alpha_{F}(p)$ here corresponds to $\alpha(p)$ of
Aguirre, Cowan, and Vickers (2010, p.\,1603).} 
\[
\alpha(p)\!\equiv\!\frac{-pq^{\prime\prime}(p)}{q^{\prime}(p)}\!,\alpha_{F}(p)\!\equiv\!-p\!\left(\!\frac{\partial q_{j}(\mathbf{p})}{\partial p_{j}}\!\right)^{\!\!\!-1}\!\!\frac{\partial^{2}\!q_{j}(\mathbf{p})}{\partial p_{j}^{2}}\!,\alpha_{C}(p)\!\equiv\!-\!\left(n\!-\!1\right)\!p\!\left(\!\frac{\partial q_{j}(\mathbf{p})}{\partial p_{j}}\!\right)^{\!\!\!-1}\!\!\frac{\partial^{2}\!q_{j}(\mathbf{p})}{\partial p_{j}\,\!\partial p_{j'}},
\]
where again the derivatives are evaluated at $\mathbf{p}=\left(p,...,p\right)$,
and $j$\textit{ }and\textit{ $j'$ }is an arbitrary pair of distinct
indices. These curvatures satisfy $\alpha=(\alpha_{F}+\alpha_{C})\epsilon_{F}/\epsilon$
and are related to the elasticity of $\epsilon_{F}(p)$ by $p\,\epsilon_{F}'(p)/\epsilon_{F}(p)=1+\epsilon\left(p\right)-\alpha_{F}\left(p\right)-\alpha_{C}\left(p\right)$
(see Appendix \ref{subsec: DirectElasticitiesCurvature} for the derivation
and a related discussion). 
\begin{flushleft}
\textbf{Inverse demand.} We introduce analogous definitions for inverse
demand. We define the \textit{own quantity elasticity $\eta_{F}(q)$}
\textit{\emph{and the }}\textit{cross quantity elasticity} \textit{$\eta_{C}(q)$
of the firm's inverse demand }\textit{\emph{as}}
\[
\eta_{F}(q)\equiv-\frac{q}{p(q)}\ \frac{\partial p_{j}(\mathbf{q})}{\partial q_{j}}|_{\mathbf{q}=\left(q,...,q\right)},\qquad\eta_{C}(q)\equiv(n-1)\frac{q}{p(q)}\ \frac{\partial p_{j^{\prime}}(\mathbf{q})}{\partial q_{j}}|_{\mathbf{q}=\left(q,...,q\right)},
\]
for arbitrary distinct $j$ and $j'$. These satisfy $\eta_{F}=\eta+\eta_{C}$.\footnote{The identity $\eta_{F}=\eta+\eta_{C}$ means that as a response to
firm $j$'s increase in its output, the industry as a whole reacts
by lowering firm\textbf{ }$j$'s price (\textbf{$\eta$}). However,
each individual firm (other than $j$) reacts to this firm $j$'s
output increase by reducing its own output. This counteracts the initial
change in the price $(\eta_{C}<0$), and thus a percentage reduction
in the price for firm $j$ $(\eta_{F}$) is smaller than\textbf{ $\eta$},
which does not take into account strategic reactions. Note here that
$1/\eta_{F}$, not $\eta_{F}$, measures the industry's competitiveness.\textbf{
}Thus, these three quantity elasticities characterize \textquotedblleft first-order
competitiveness,\textquotedblright{} which determines whether the
equilibrium quantity is high or low.} We define the\textit{ curvature of the industry's inverse demand}
$\sigma(q)$, the \textit{own} \textit{curvature $\sigma_{F}(q)$
of the firm's inverse demand} and the \textit{cross curvature $\sigma_{C}(q)$
of the firm's inverse demand} by: 
\[
\sigma(q)\!\equiv\!\frac{-qp^{\prime\prime}(q)}{p^{\prime}(q)}\!,\,\sigma_{F}(q)\!\equiv\!-q\!\left(\!\frac{\partial p_{j}(\mathbf{q})}{\partial q_{j}}\!\right)^{\!\!\!-1}\!\!\frac{\partial^{2}p_{j}(\mathbf{q})}{\partial q_{j}^{2}}\!,\,\sigma_{C}(q)\!\equiv\!-\!\left(n\!-\!1\right)\!q\!\left(\!\frac{\partial p_{j}(\mathbf{q})}{\partial q_{j}}\!\right)^{\!\!\!-1}\!\!\frac{\partial^{2}\!p_{j}(\mathbf{q})}{\partial q_{j}\!\,\partial q_{j'}},
\]
where again the derivatives are evaluated at $\mathbf{q}=\left(q,...,q\right)$
and the indices $j$ and $j'$ are distinct. These curvatures represent
an oligopoly counterpart of monopoly $\sigma(q)$ of Aguirre, Cowan,
and Vickers (2010, p.\,1603). They satisfy the relationship $\sigma=(\sigma_{F}+\sigma_{C})(\eta_{F}/\eta)$
and are related to the elasticity of $\eta_{F}(q)$ by $q\,\eta_{F}'(q)/\eta_{F}(q)=1+\eta\left(q\right)-\sigma_{F}\left(q\right)-\sigma_{C}\left(q\right)$
(see Appendix \ref{subsec: IndirectElasticitiesCurvatures} for the
derivation and a related discussion).
\par\end{flushleft}

\subsection{Expressions for conduct index and pass-through}

\label{SubsectionExpressionConduct}In the case of price competition,
the conduct index $\theta$ is $\theta=\epsilon/\epsilon_{F}=1/(\eta\epsilon_{F})$,
which is verified by comparing the firm's first-order condition with
Equation (\ref{theta_M}). The marginal cost of public funds and the
incidence are obtained by substituting these expressions into those
of Propositions \ref{PropositionMCt} and \ref{PropositionConsumerAndProducerIncidence}.

\begin{proposition}\textbf{\label{PropositionPassThroughAndMCUnderPriceCompetition}Pass-through
under price competition.} Under symmetric oligopoly with price competition
and with a possibly non-constant marginal cost: 
\[
\rho_{t}=\frac{1}{1-v}\ \frac{1}{1+\frac{(1-\alpha/\epsilon_{F})\epsilon}{\epsilon_{F}}+\left(\frac{1-\tau}{1-v}-\frac{1}{\epsilon_{F}}\right)\epsilon\chi},
\]
\[
\rho_{v}=\frac{1}{1-v}\ \frac{1}{\frac{1}{1-1/\epsilon_{F}}+\frac{(1-\alpha/\epsilon_{F})\epsilon}{\epsilon_{F}-1}+\left(\frac{1-\tau}{1-v}\,\frac{\epsilon_{F}}{\epsilon_{F}-1}-\frac{1}{\epsilon_{F}-1}\right)\epsilon\,\chi}.
\]

\end{proposition}

The proof is in Appendix \ref{AppendixProofOfPassThroughFormulaUnderPriceCompetition}
and a detailed discussed in Appendix \ref{AppendixDiscussionOfPassThroughFormulaUnderPriceCompetition}.

Next, in the case of quantity competition, the conduct index $\theta$
is given by $\theta=\eta_{F}/\eta$, which is, again, verified by
comparing the firm's first-order condition with Equation (\ref{theta_M}).
Again, the marginal cost of public funds and the incidence are obtained
by substituting these expressions into those of Propositions \ref{PropositionMCt}
and \ref{PropositionConsumerAndProducerIncidence}. For the proof
and a related discussion, see Appendix \ref{AppendixPropositionProofOfPassThroughFormulaUnderQuantityCompetition}
and \ref{AppendixDiscussionOfPassThroughFormulaUnderQuantityCompetition}.

\begin{proposition}\textbf{\label{PropositionPassThroughAndMCUnderQuantityCompetition}Pass-through
under quantity competition.} Under symmetric oligopoly with quantity
competition and with a possibly non-constant marginal cost: 
\[
\text{\ensuremath{\rho_{t}\!=\!\frac{1}{1-v}\ \frac{1}{1+\frac{\eta_{F}}{\eta}-\sigma+\!\left(\frac{1-\tau}{1-v}-\eta_{F}\right)\frac{\chi}{\eta}},\qquad\rho_{v}\!=\!\frac{1}{1-v}\ \frac{\left(1-\eta_{F}\right)}{1+\frac{\eta_{F}}{\eta}-\sigma+\!\left(\frac{1-\tau}{1-v}-\eta_{F}\right)\frac{\chi}{\eta}}.}}
\]
\end{proposition}

\subsection{Simple parametric examples}

Although our formulas are general, it is illustrative to work through
a few simple examples. Below we provide two parametric examples with
$n$ symmetric firms and constant marginal cost: $\chi=0$. In this
case, the pass-through expressions are simplified to 
\[
\rho_{t}=\frac{1}{(1-v)\left[1+\left(1-\frac{\alpha}{\epsilon_{F}}\right)\theta\right]},\qquad\rho_{v}=\frac{\epsilon_{F}-1}{\epsilon_{F}\left\{ (1-v)\left[1+\left(1-\frac{\alpha}{\epsilon_{F}}\right)\theta\right]\right\} }
\]
under price competition, where $\theta=\epsilon/\epsilon_{F}$, and
\[
\rho_{t}=\frac{1}{(1-v)\left[1+\left(1-\frac{\sigma}{\theta}\right)\theta\right]},\qquad\rho_{v}=\frac{1-\eta_{F}}{(1-v)\left[1+\left(1-\frac{\sigma}{\theta}\right)\theta\right]}
\]
under quantity competition, where $\theta=\eta_{F}/\eta$. 

\subsubsection{Linear demand}

One is the case wherein each firm faces the following\textit{ linear
demand}, $q_{j}(p_{1},...,p_{n})=b-\lambda p_{j}+\mu\sum\nolimits _{j^{\prime}\neq j}p_{j^{\prime}}$,
where $b>mc$ and $\lambda>(n-1)\mu\geq0$, implying that all firms
produce substitutes and $\mu$ measures the degree of\textit{ substitutability}
(firms are effectively monopolists when $\mu=0$).\footnote{These linear demands are derived by maximizing the representative
consumer's net utility, $U(q\text{\textsubscript{1}},...,q_{n})-\sum_{j=1}^{n}pq_{j}$,
with respect to $q\text{\textsubscript{1}},...,$ and $q_{n}$. See
Vives (1999, pp.\,145-6) for details.}$^{,}$\footnote{In our notation below, the demand in symmetric equilibrium is given
by $q_{j}(p_{j},p_{-j})=b-\lambda p_{j}+\mu(n-1)p_{-j}$, whereas
it is written as
\[
q_{j}(p_{j},p_{-j})=\frac{\alpha}{1+\gamma(n-1)}-\frac{1+\gamma(n-2)}{(1-\gamma)[1+\gamma(n-1)]}p_{j}+\frac{\gamma(n-1)}{(1-\gamma)[1+\gamma(n-1)]}p_{-j}
\]
in Häckner and Herzing's (2016) notation, where $\gamma\text{\ensuremath{\in}}[0,1]$
is the parameter that measures substitutability between (symmetric)
products. Thus, if our $(b,\lambda,\mu)$ is determined by $b=\alpha/[1+\gamma(n-1)]$,
$\lambda=[1+\gamma(n-2)]/\left\{ (1-\gamma)[1+\gamma(n-1)]\right\} $,
and $\mu=\gamma/\left\{ (1-\gamma)[1+\gamma(n-1)]\right\} $, given
Häckner and Herzing's (2016) $(\alpha,\gamma)$, then our results
below can be expressed by Häckner and Herzing's (2016) notation as
well. Note here that our formulation is more flexible in the sense
that the number of the parameters is three. This is because the coefficient
for the own price is normalized to one: $p_{j}(q_{j},q_{-j})=\alpha-q_{j}-\gamma(n-1)q_{-j}$,
which is analytically innocuous, and Häckner and Herzing's (2016)
$\gamma$ is the normalized parameter.} Under symmetric pricing, the industry's demand is thus given by $q(p)=b-[\lambda-(n-1)\mu]p$.
The inverse demand system is given by
\[
p_{j}(q_{j},\mathbf{q}_{-j})=\frac{\lambda-(n-2)\mu}{\left(\lambda+\mu\right)[\lambda-(n-1)\mu]}(b-q_{j})+\frac{\mu}{\left(\lambda+\mu\right)[\lambda-(n-1)\mu]}\left[\sum_{j'\neq j}(b-q_{j'})\right],
\]
implying that $p(q)=(b-q)/[\lambda-(n-1)\mu]$ under symmetric production.
Obviously, both the direct and the indirect demand curvatures are
zero: $\alpha=0,\sigma=0$. Thus, the pass-through rates are simply
given by 
\[
\rho_{t}=\frac{1}{(1-v)\left(1+\theta\right)},\qquad\rho_{v}=\frac{\epsilon_{F}-1}{\epsilon_{F}(1-v)\left(1+\theta\right)}
\]
under price competition, where $\theta=[\lambda-(n-1)\mu]/\lambda$,
and $\epsilon_{F}=\lambda(p/q)$ (where $p$ and $q$ are the equilibrium
price and output under price setting), and 
\[
\rho_{t}=\frac{1}{(1-v)\left(1+\theta\right)},\qquad\rho_{v}=\frac{1-\eta_{F}}{(1-v)\left(1+\theta\right)}
\]
under quantity competition, where $\theta=[\lambda-(n-2)\mu]/(\lambda+\mu)$
and $\eta_{F}=\{[\lambda-(n-2)\mu](q/p)\}/\{(\lambda+\mu)[\lambda-(n-1)\mu]\}$
(where $q$ and $p$ are the equilibrium output and price under quantity
setting). 

Now, from Propositions \ref{PropositionMCt} and \ref{PropositionConsumerAndProducerIncidence},
the marginal cost of public funds and the incidence are given by
\[
MC_{t}=\frac{(1-v)\thinspace\theta+\epsilon\thinspace\tau}{1+(1-v)\theta-\epsilon\thinspace\tau},\qquad MC_{v}=\frac{(1-v)\theta+\epsilon\tau}{\frac{(1-v)\left(1+\theta\right)}{\epsilon_{F}-1}+v-\epsilon\tau}
\]
\[
I_{t}=\frac{1}{2(1-v)[1-(n-1)(\mu/\lambda)]},\qquad I_{v}=\frac{\epsilon_{F}-1}{(1-v)[2-\epsilon_{F}(1-\theta)]}
\]
under price competition, with $\epsilon=[\lambda-(n-1)\mu](p/q)$
is additionally provided, where $p$ and $q$ are the equilibrium
price and output under price setting, and
\[
MC_{t}=\frac{(1-v)\thinspace\theta+\frac{1}{\eta}\thinspace\tau}{1+(1-v)\theta-\frac{1}{\eta}\thinspace\tau},\qquad MC_{v}=\frac{(1-v)\theta+\frac{1}{\eta}\thinspace\tau}{\frac{(1-v)\left(1+\theta\right)}{1-\eta_{F}}+v-\frac{1}{\eta}\thinspace\tau}
\]
\[
I_{t}=\frac{\lambda+\mu}{2(1-v)[\lambda-(n-2)\mu]},\qquad I_{v}=\frac{1-\eta_{F}}{(1-v)[\eta_{F}+(2-\eta_{F})\theta]}
\]
under quantity competition, with $1/\eta=[\lambda-(n-1)\mu](p/q)$
is additionally provided, where $p$ and $q$ are the equilibrium
price and output under quantity setting. Thus, it suffices to solve
for the equilibrium price and output under both settings to compute
the pass-through rate and the marginal cost of public funds for all
four cases. 

\begin{table}
\centering{}\caption{Elasticities, Conduct Indices, and Curvatures }
\medskip{}

\centering{}%
\begin{tabular}{r@{\extracolsep{0pt}.}l|r@{\extracolsep{0pt}.}l}
\hline 
\multicolumn{4}{c}{(a) Linear Demand}\tabularnewline
\hline 
\multicolumn{2}{c|}{Price setting } & \multicolumn{2}{c}{Quantity setting}\tabularnewline
\hline 
\multicolumn{2}{c|}{$\text{\ensuremath{\epsilon}}=[\lambda-(n-1)\mu]\left(\frac{p}{q}\right)$ } & \multicolumn{2}{c}{$\text{\ensuremath{\eta}}=\frac{1}{\lambda-(n-1)\mu}\left(\frac{q}{p}\right)$}\tabularnewline
\multicolumn{2}{c|}{$\text{\ensuremath{\epsilon_{F}}}=\lambda\left(\frac{p}{q}\right)$} & \multicolumn{2}{c}{$\text{\ensuremath{\eta_{F}}}=\frac{\lambda-(n-2)\mu}{(\lambda+\mu)[\lambda-(n-1)\mu]}\left(\frac{q}{p}\right)$}\tabularnewline
\multicolumn{2}{c|}{$\theta=\text{\ensuremath{\epsilon}}/\text{\ensuremath{\epsilon_{F}}}=1-(n-1)\left(\frac{\mu}{\lambda}\right)$ } & \multicolumn{2}{c}{$\theta=\text{\ensuremath{\eta{}_{F}}}/\text{\ensuremath{\eta}}=\frac{\lambda-(n-2)\mu}{\lambda+\mu}$}\tabularnewline
\multicolumn{2}{c|}{$\text{\ensuremath{\alpha}}=0$ } & \multicolumn{2}{c}{$\text{\ensuremath{\sigma}}=0$}\tabularnewline
\hline 
\multicolumn{2}{c}{} & \multicolumn{2}{c}{}\tabularnewline
\hline 
\multicolumn{4}{c}{(b) Logit Demand}\tabularnewline
\hline 
\multicolumn{2}{c|}{Price setting } & \multicolumn{2}{c}{Quantity setting}\tabularnewline
\hline 
\multicolumn{2}{c|}{$\text{\ensuremath{\epsilon}}=\beta(1-ns)p$ } & \multicolumn{2}{c}{$\text{\ensuremath{\eta}}=\frac{1}{\beta(1-ns)p}$}\tabularnewline
\multicolumn{2}{c|}{$\text{\ensuremath{\epsilon_{F}}}=\beta(1-s)p$ } & \multicolumn{2}{c}{$\text{\ensuremath{\eta_{F}}}=\frac{1-\left(n-1\right)s}{\beta(1-ns)p}$}\tabularnewline
\multicolumn{2}{c|}{$\theta=\text{\ensuremath{\epsilon}}/\text{\ensuremath{\epsilon_{F}}}=\frac{1-ns}{1-s}$ } & \multicolumn{2}{c}{$\theta=\text{\ensuremath{\eta{}_{F}}}/\text{\ensuremath{\eta}}=1-\left(n-1\right)s$}\tabularnewline
\multicolumn{2}{c|}{$\text{\ensuremath{\alpha}}=\frac{(2ns-3)ns}{1-ns}p$ } & \multicolumn{2}{c}{$\text{\ensuremath{\sigma}}=\frac{1-2ns}{1-ns}$}\tabularnewline
\hline 
\end{tabular}\label{TableLinearDemandSummary} 
\end{table}

Table \ref{TableLinearDemandSummary} (a) summarizes the key variables
that determine the pass-through rates and the marginal costs of public
funds. It is verified that under both price and quantity competition,
$\partial\theta/\partial n<0$ and $\partial\theta/\partial\mu<0$.
To focus on the roles of these two parameters, $n$ and $\mu$, which
directly affect the degree of competition, we employ the following
simplification to compute the ratio $p/q$ in equilibrium: $b=1$,
$mc=0$, and $\lambda=1$ (see Appendix \ref{subsec: Eq P and Q under P and Q Competition with L Demand}
for the actual expressions of the equilibrium prices and outputs under
price and quantity competition).

\begin{figure}
\includegraphics[scale=0.65]{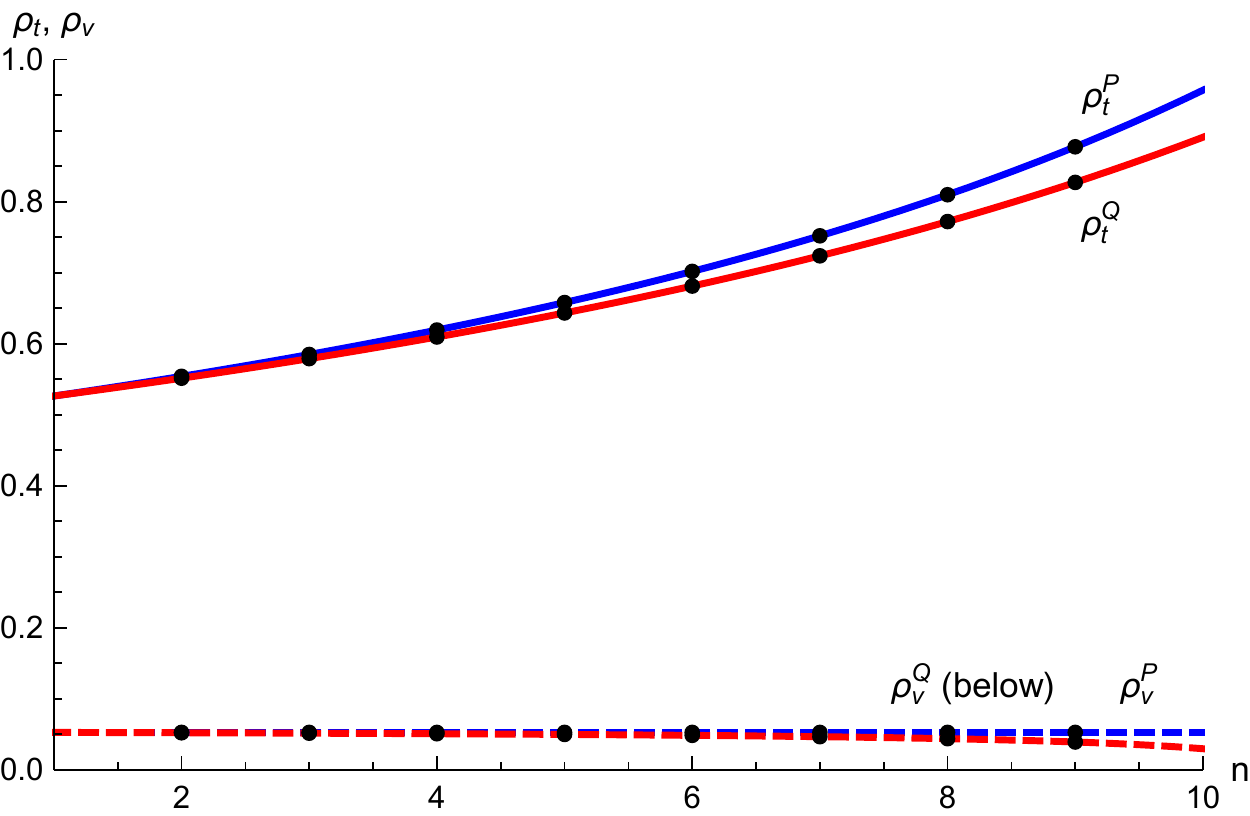}\quad{}\includegraphics[scale=0.65]{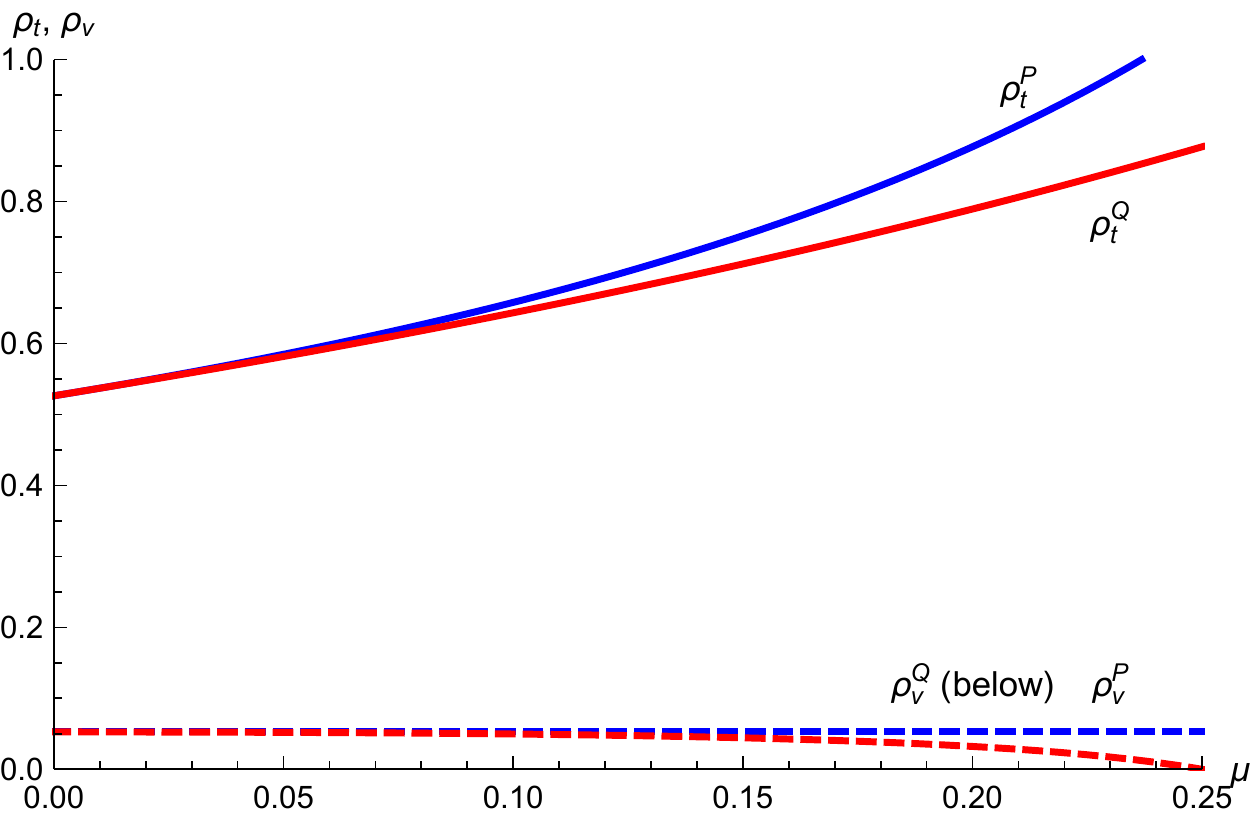}

\bigskip{}

\includegraphics[scale=0.65]{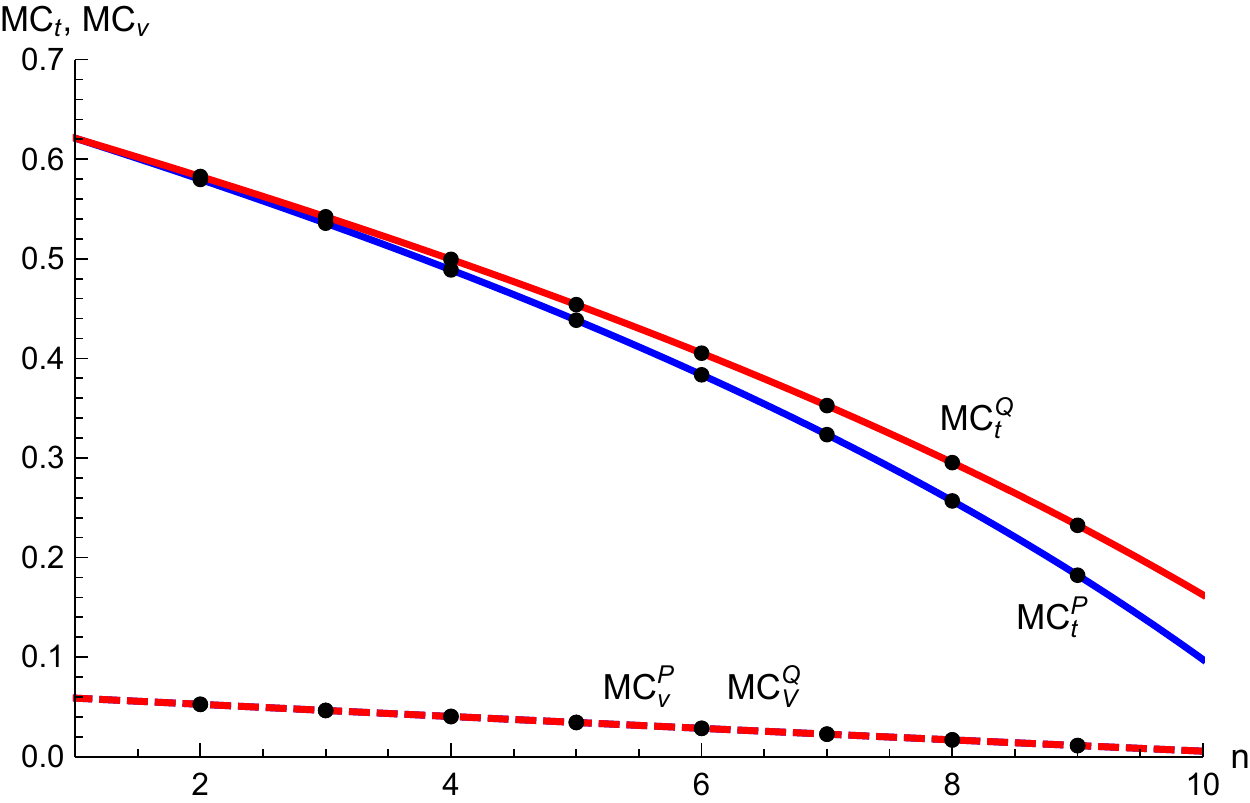}\quad{}\includegraphics[scale=0.65]{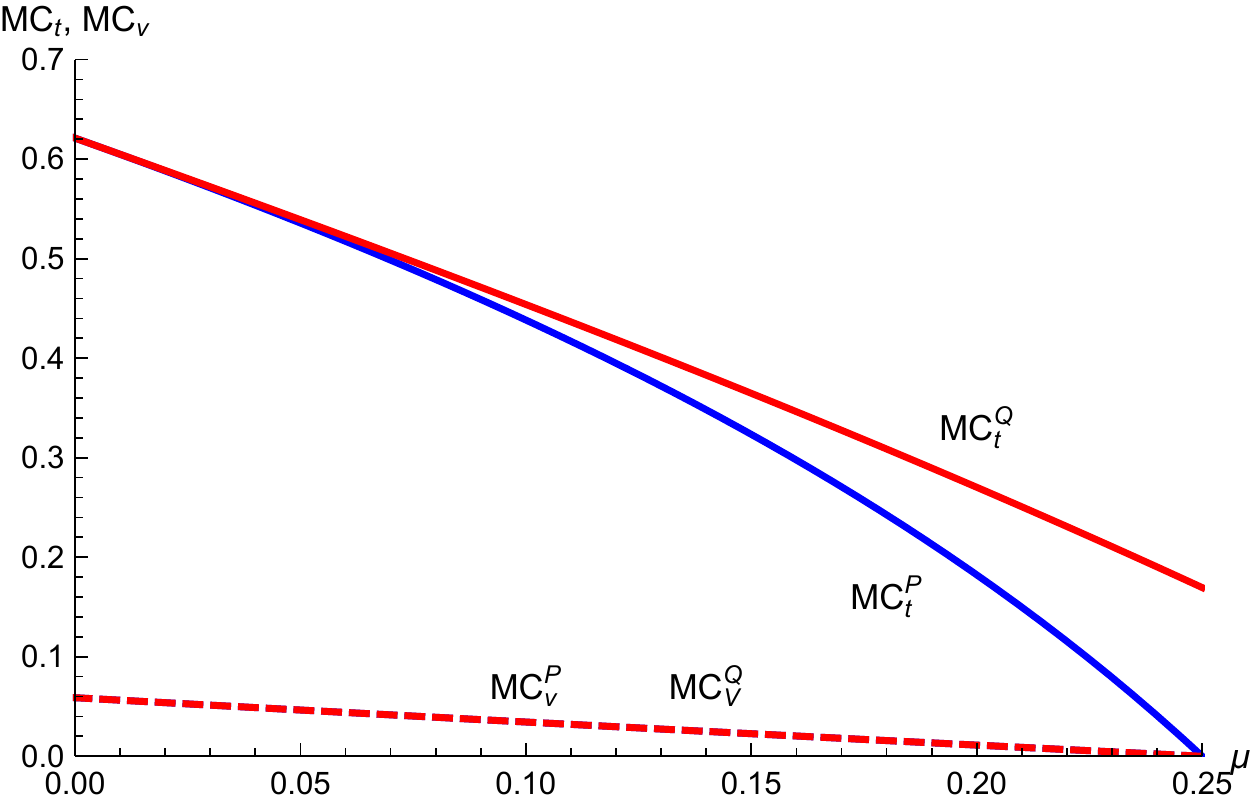}

\bigskip{}

\includegraphics[scale=0.65]{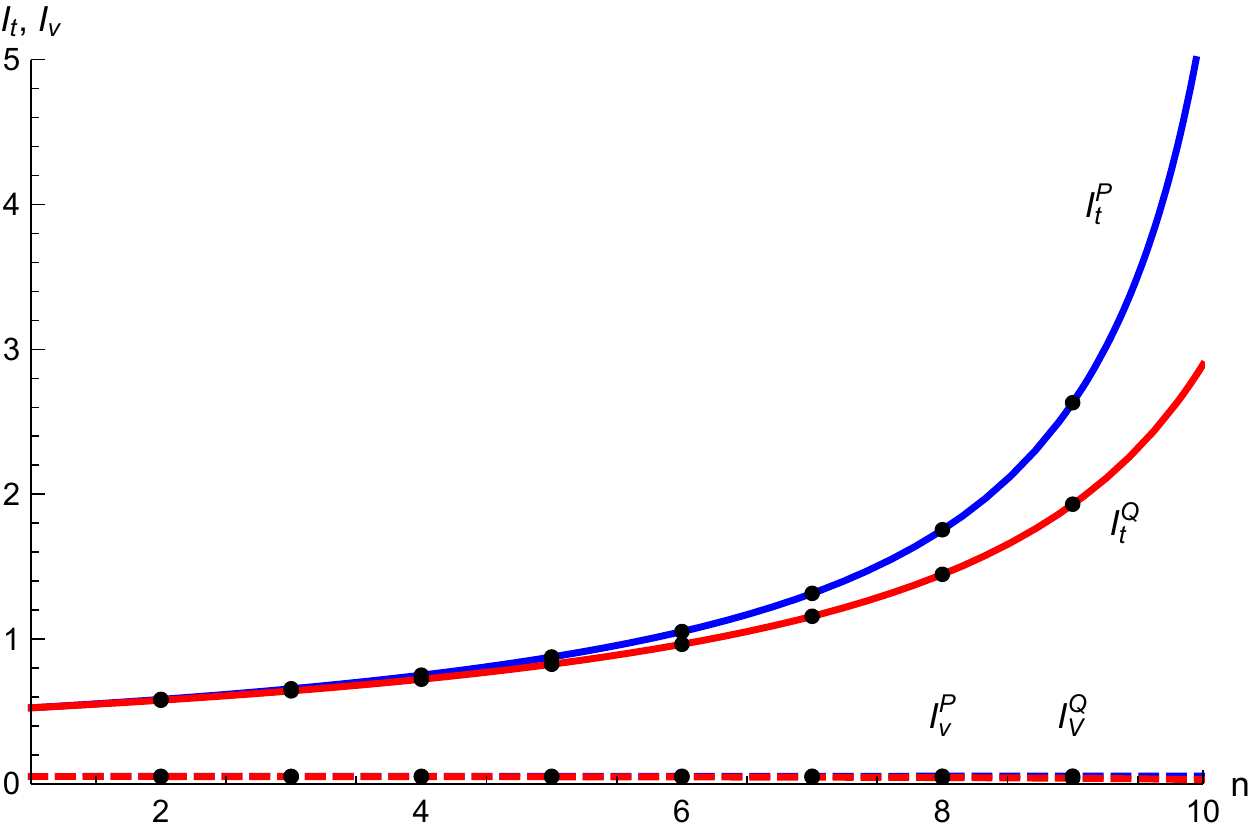}\quad{}\includegraphics[scale=0.65]{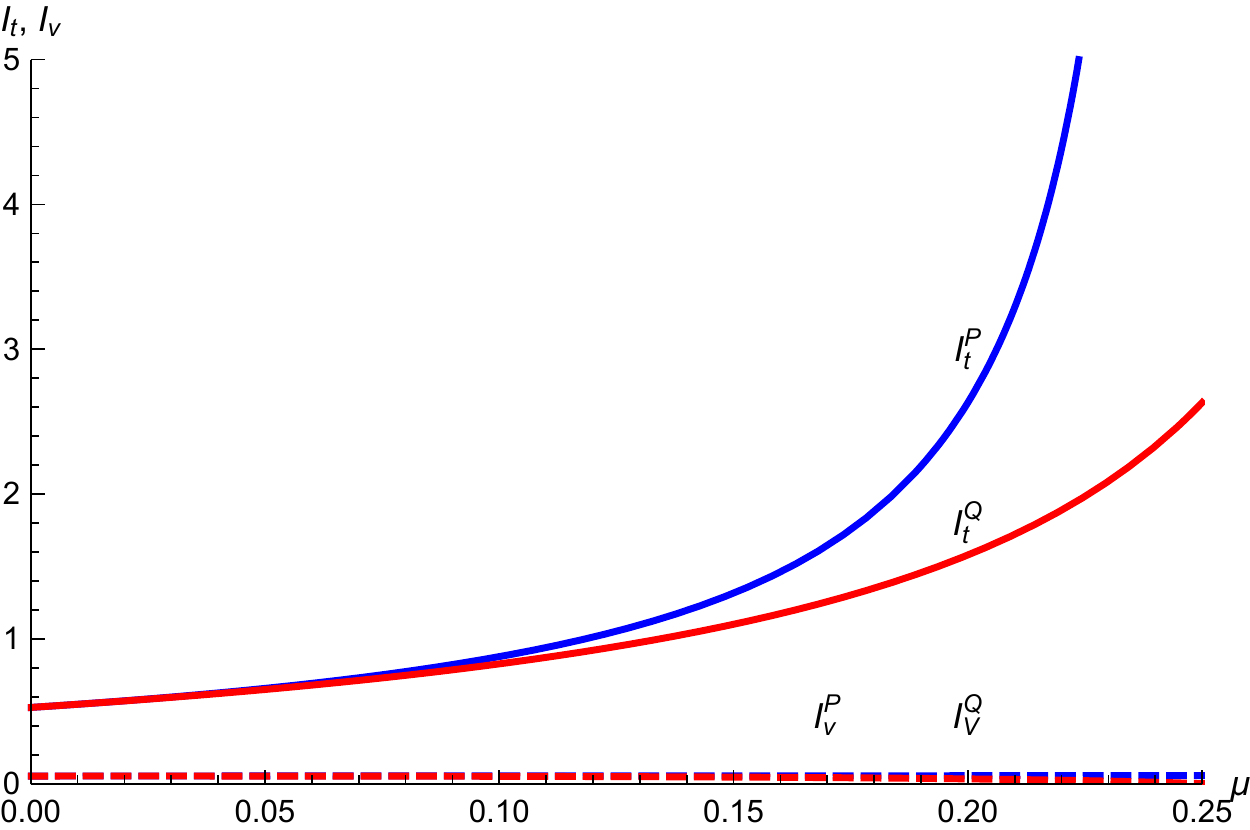}\bigskip{}

\caption{\label{fig:linear}Pass-through rate (top), marginal cost of public
funds (middle), and incidence (bottom) with linear demand. The horizontal
axes on the left and the right panels correspond to the number of
firms ($n$) and the substitutability parameter ($\mu$), respectively.}
\end{figure}

The top two panels in Figure \ref{fig:linear} illustrate how $\rho_{t}$
and $\rho_{v}$ behave as the number of firms ($n$; the left) or
the sustainability parameter ($\mu$; the left) increases, with the
superscript denoting price ($P$) or quantity ($Q$) setting. Similarly,
the middle and the bottom panels draw $MC_{t}$ and $MC_{v}$, and
$I_{t}$ and $I_{v}$, respectively. It is observed that the ad valorem
tax pass-through rates are close to zero because in this case both
$\epsilon_{F}$ and $\eta_{F}$ are close to 1. As competition becomes
fiercer, both $\rho_{t}^{P}$ and $\rho_{t}^{Q}$ become larger, although
the discrepancy also becomes larger. In the case of linear demand,
the difference in the mode of competition does not yield a significant
difference in each of the three measures. As is verified by Anderson,
de Palma, and Kreider (2001b), the ad valorem tax is more efficient
than the unit tax: the dashed lines in the two middle panels lie below
the solid lines. This ranking is related inversely to the pass-through
and the incidence: as the pass-through or the incidence becomes larger,
the marginal cost of public funds becomes smaller. 

\subsubsection{Logit demand}

\begin{figure}
\includegraphics[scale=0.68]{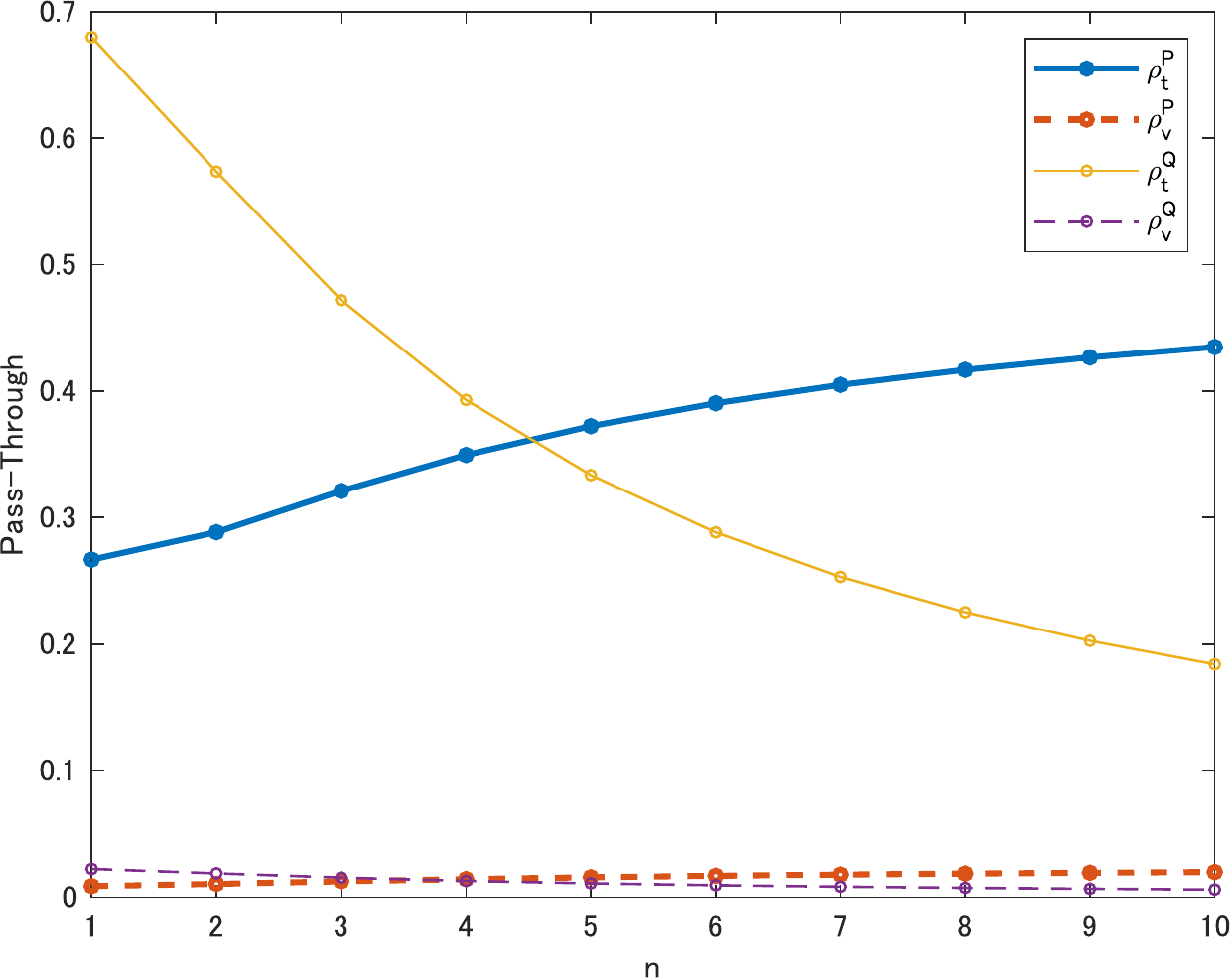}\quad{}\includegraphics[scale=0.68]{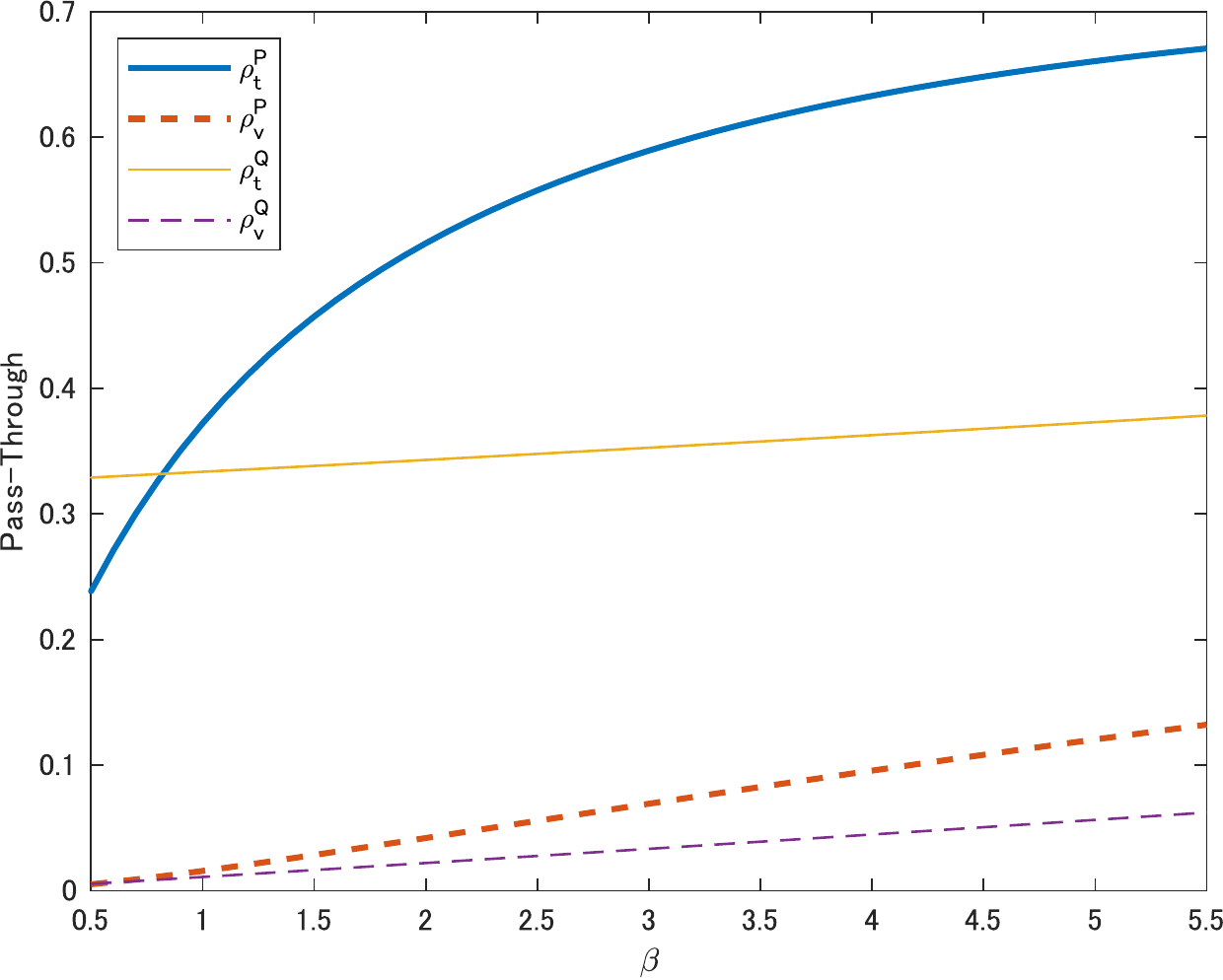}

\smallskip{}

\includegraphics[scale=0.68]{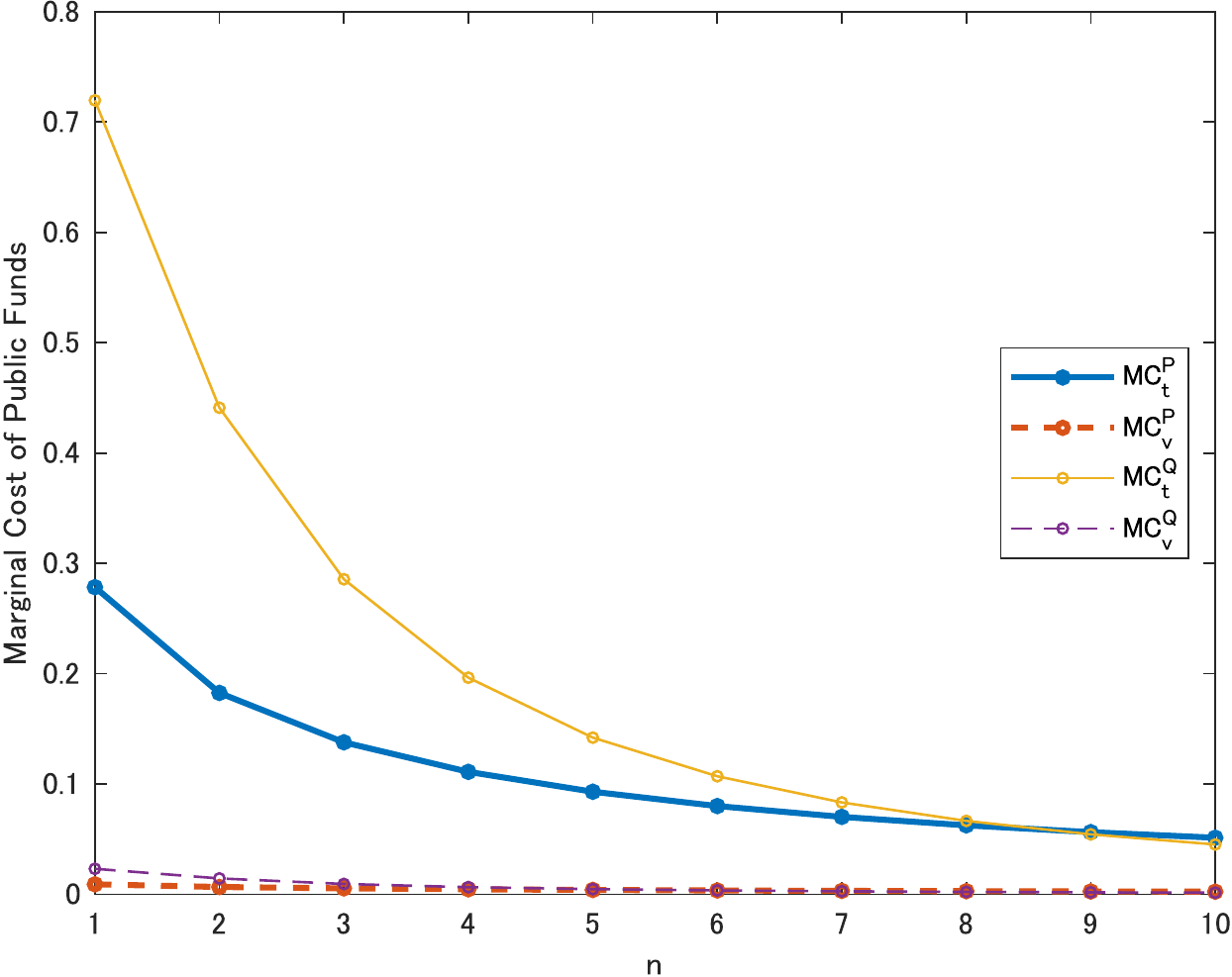}\quad{}\includegraphics[scale=0.67]{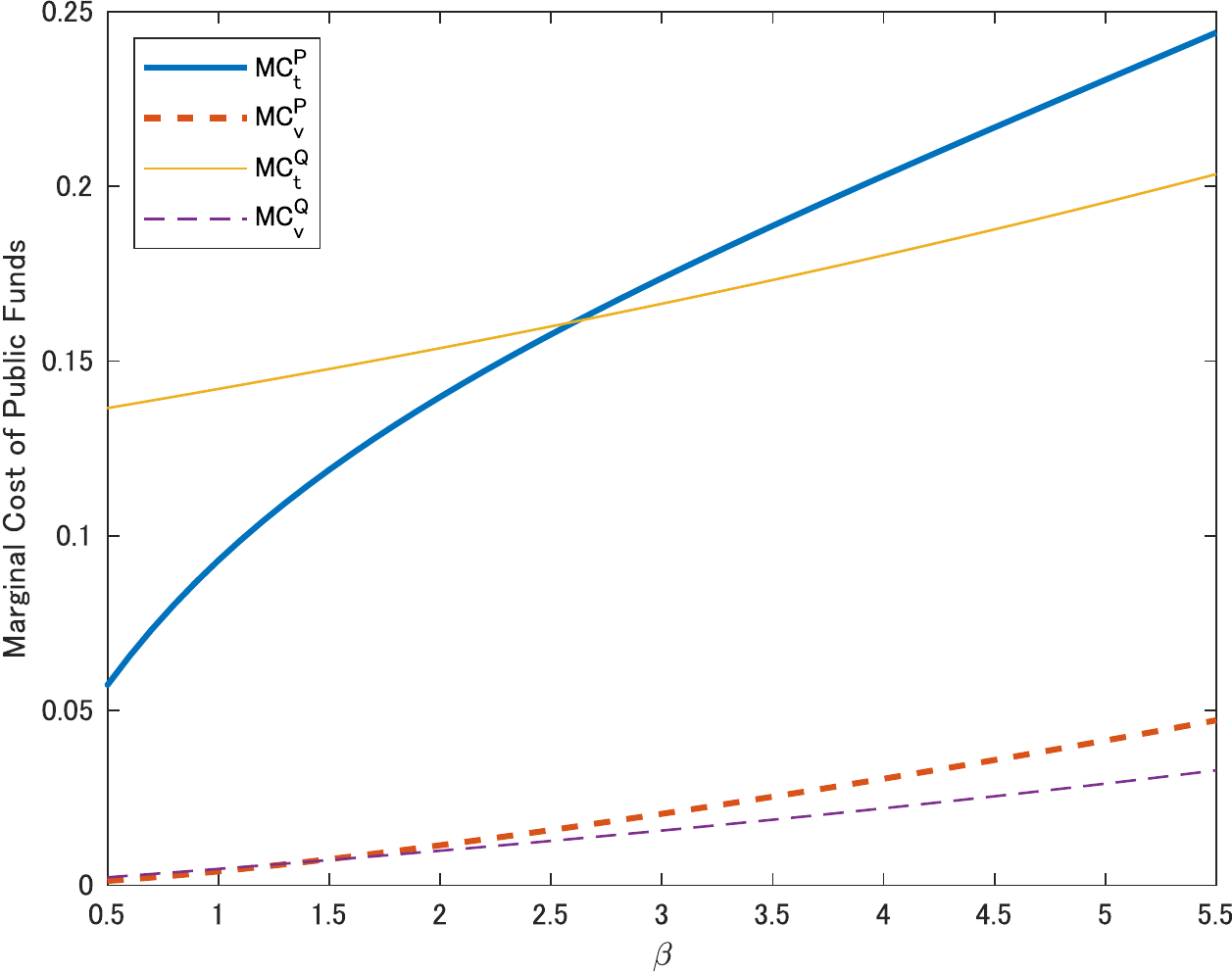}

\smallskip{}

\includegraphics[scale=0.68]{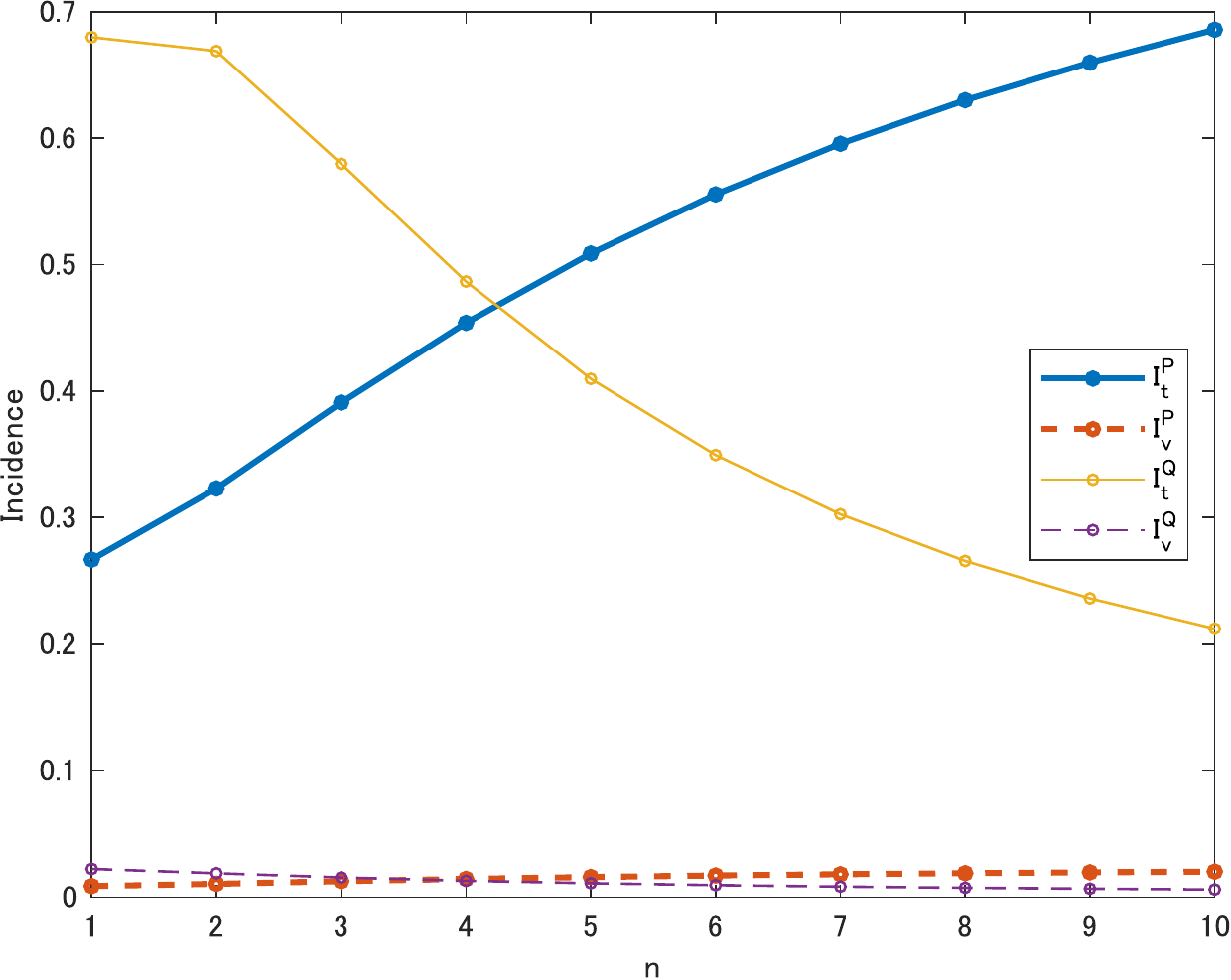}\quad{}\includegraphics[scale=0.68]{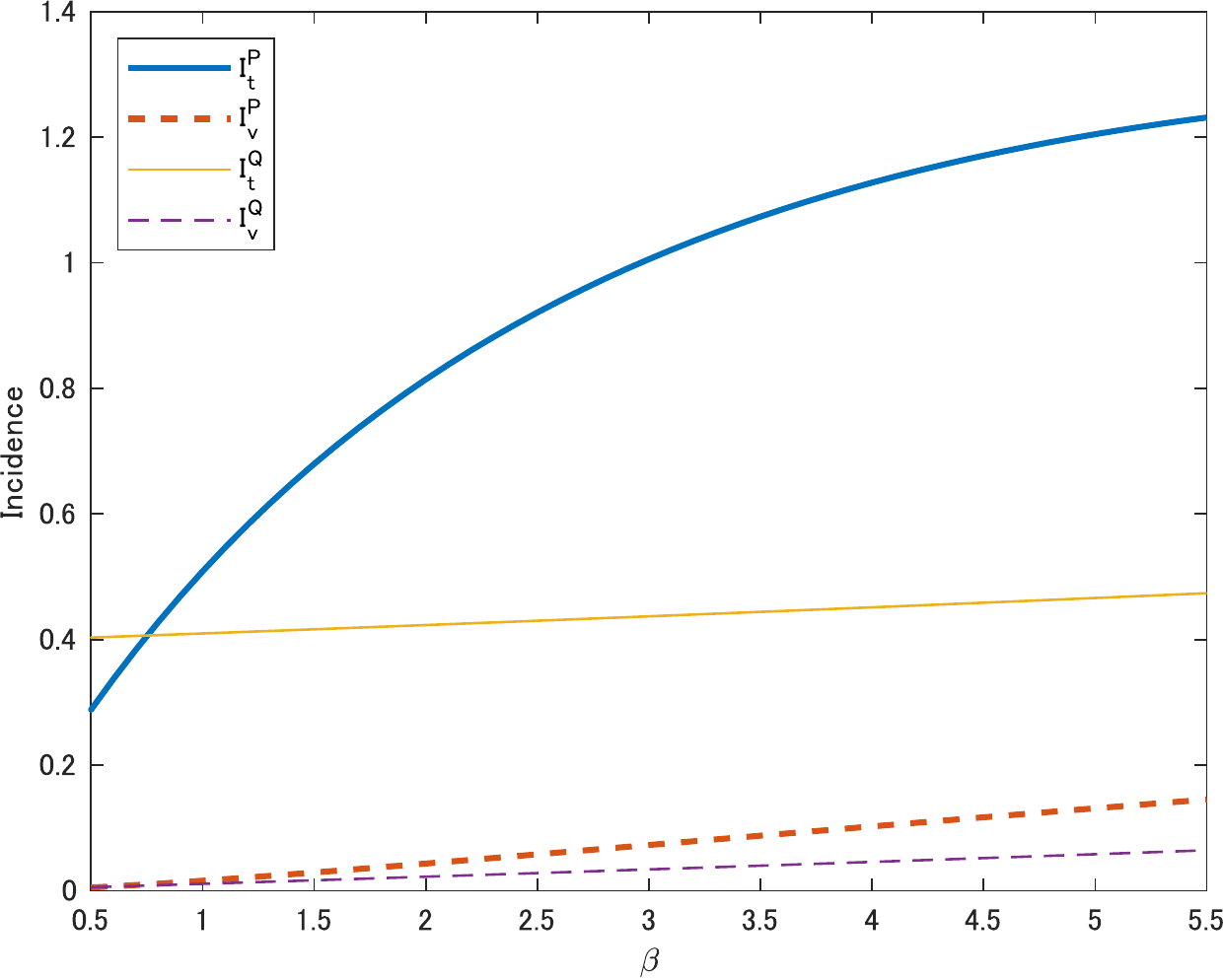}

\caption{\label{fig:logit}Pass-through rates (top), marginal costs of public
funds (middle), and incidence (bottom) with logit demands. The horizontal
axes on the left and the right panels are the number of firms ($n$)
and the price coefficient ($\beta$), respectively.}
\end{figure}

The next parametric example is the\textit{ logit demand}. Each firm
$j=1,...,n$ faces the following demand: $s_{j}(\mathbf{p})=\exp(\delta-\beta p_{j})/[1+\text{\ensuremath{\sum}}_{j\text{\textasciiacute}=1,...,n}\exp(\delta-\beta p_{j\text{\textasciiacute}})]\text{\ensuremath{\in}}(0,1)$,
where $\delta$ is the (symmetric) product-specific utility and $\beta>0$
is the\textit{ responsiveness} to the price.\footnote{Here, $q_{j}(p\text{\textsubscript{1}},...,p_{n})$ is derived by
aggregating over individuals who choose product $j$ (the total number
of individuals is normalized to one): individual $i$'s net utility
from consuming $j$ is given by $u_{ij}=\delta-\beta p_{j}+\tilde{\varepsilon}_{ij}$,
whereas $u_{i0}=\tilde{\varepsilon}_{i0}$ is the net utility from
consuming nothing, and $\tilde{\varepsilon}_{i0},\tilde{\varepsilon}_{i1},...,\tilde{\varepsilon}_{in}$
are independently and identically distributed according to the Type
I extreme value distribution for all individuals. See Anderson, de
Palma, and Thisse (1992, pp.\,39-45) for details. We work in terms
of market share variables $s_{j}$ and $s$, instead of $q_{j}$ and
$q$, which is consistent with the usual notation in the industrial
organization literature.} We define $s\text{\textsubscript{0}}=1-\text{\ensuremath{\sum}}_{j=1,...,n}s_{j}<1$
as the share of all outside goods. Table \ref{TableLinearDemandSummary}
(b) summarizes the key variables that determine the pass-through rates
and the marginal costs of public funds. We need to numerically solve
for the equilibrium price and market share under both settings to
compute the pass-through rate, the marginal cost of public funds,
and incidence for all four cases. To focus on the two parameters,
$\beta$ and $n$, we assume that $\delta=1$ and $mc=0$. Because
$\partial s_{j}(\mathbf{p})/\partial p_{j}|_{\mathbf{p}=\left(p,...,p\right)}=-\beta s(1-s)$,
the first-order conditions for the symmetric equilibrium price and
the market share satisfy $p-t/(1-v)=1/[\beta(1-s)]$ and $s=\mathrm{\exp}(1-\beta p)/[1+n\exp(1-\beta p)]$.
If $p$ and $s$ are solved numerically, then $\epsilon$, $\epsilon_{F}$,
$\theta$ and $\alpha$ can also be numerically computed.\footnote{It can be verified that $s_{j}(\cdot;\mathbf{p}_{-j})$ is convex
as long as $s_{j}<1/2$ because $\partial^{2}s_{j}/\partial p_{j}^{2}=-\beta(\partial s_{j}/\partial p_{j})(1-2s_{j})>0$.
However, the second-order condition is always satisfied because $\partial^{2}\pi_{j}/\partial p_{j}^{2}=-\beta s_{j}<0$.
In symmetric equilibrium with $\delta=1$ and $mc=0$, the largest
market share is attained as $1/(n+1)$ when the equilibrium price
is zero, which implies that the market share of the outside goods
$s\text{\textsubscript{0}}$ is no less than each firm's market share:
$s\text{\textsubscript{0}}>s$.} Next, we consider the inverse demands under quantity competition.
Then, as in Berry (1994), firm $j$'s inverse demand is given by $p_{j}(\mathbf{s})=[\delta-\log(s_{j}/s\text{\textsubscript{0}})]/\beta$,
which implies that $\partial p_{j}(\mathbf{s})/\partial s_{j}|_{\mathbf{s}=\left(s,...,s\right)}=-[1-(n-1)s]/[\beta s(1-ns)]$.
Thus, the first-order conditions for the symmetric equilibrium price
and the market share satisfy $p-t/(1-v)=[1-(n-1)s]/[\beta(1-ns)]$
and $p=[1-\mathrm{\log}(s/[1-ns])]/\beta$. Then, as above, $\eta$,
$\eta_{F}$, $\theta$ and $\sigma$ are computed by numerically solving
the first-order conditions for $p$ and $s$. Interestingly, it is
verified that in symmetric equilibrium under quantity setting, $\partial p/\partial n=0$:
the equilibrium price is the same irrespective of the number of firms,
whereas the individual market share is decreasing in the number of
firms: $\partial s/\partial n<0$. On the other hand, both the equilibrium
price and market share are decreasing in the price coefficient, $\beta$.

Figure \ref{fig:logit} illustrates the pass-through rate, the marginal
cost of public funds, and the incidence, in analogy with Figure \ref{fig:linear}.
The right panels now show the variables' dependence on the price coefficient
$\beta$. Overall, as in the case of linear demand, an increase in
the ad valorem tax has a small impact on these measures for each of
$n$ and $\beta$, whereas an increase in the unit tax has a large
effect. However, there are important differences between the cases
of linear and logit demand. First, the unit tax pass-through under
quantity competition $\rho_{t}^{Q}$ is\textit{ decreasing} in the
number of firms. To understand this, compare the difference in the
denominators of $\rho_{t}^{P}=1/\{(1-v)\left[1+(1-\alpha/\epsilon_{F})\theta\right]\}$
and $\rho_{t}^{Q}=(1-v)\left[1+\theta-\sigma\right]$. As $\theta$
decreases (i.e., as competition becomes fiercer), the second term
in the denominator of $\rho_{t}^{P}$ decreases, and thereby $\rho_{t}^{P}$\textit{
increases} as $n$ increases. However, $\theta-\sigma$ increases
as $\theta$ decreases, and thus $\rho_{t}^{Q}$\textit{ decreases}.
This difference in the denominators is also reflected in the fact
that $I_{t}^{Q}$ is decreasing in $n$ as well. Naturally, $MC_{t}^{Q}$
is decreasing in $n$ as in the case of linear demand because $1/\rho_{t}^{Q}$
becomes larger (see the formulas in Proposition \ref{PropositionMCt}).
Second, while the pass-through rate and the incidence increase as
$\beta$ increases, the marginal cost of public funds is also increasing
in contrast to the case of linear demands. The reason is that the
effect on $MC$ of decreases in $\theta$ is weaker than the effect
of the increase in $\epsilon$: the industry's demand becomes elastic
quickly as consumers become more sensitive to a price increase. 

\section{Multi-Dimensional Pass-Through Framework}

\label{SectionMultiDimensionalPassThroughFramework} Now,%
{} we generalize our previous results to a more general specification
of taxation that involves multiple tax instruments. We define two
different types of pass-through vectors: (i) the\textit{ pass-through
rate vector} and (ii) \textit{pass-through quasi-elasticity}. We study
their properties and show that they play a central role in evaluating
welfare changes in response to changes in taxation.

\subsection{Pass-through, conduct index, and welfare: A general discussion }

\subsubsection{Generalized pass-through and tax sensitivities}

Consider a tax structure under which a firm's tax payment is expressed
as $\phi(p,q,\mathbf{T})$, where $\mathbf{T}\equiv(T_{1},...,T_{d})$
is a $d$-dimensional vector of tax instruments\footnote{To be precise, $\phi(p,q,\mathbf{T})$ represents a simplified notation
for a function $\phi(p,q,T_{1},...,T_{d})$ with $d+2$ arguments.} so that the firm's profit in symmetric equilibrium is written as
$\pi=pq-c(q)-\phi(p,q,\mathbf{T})$. Note that the argument so far
is a special case of two dimensional pass-through: $\phi(p,q,\mathbf{T})=tq+vpq$,
where $\mathbf{T}=(t,v)$. The components of the (per-firm) tax revenue
gradient vector $\nabla\phi(p,q,\mathbf{T})$ are 
\[
\phi_{T_{\ell}}(p,q,\mathbf{T})\equiv\frac{\partial\phi(p,q,\mathbf{T})}{\partial T_{\ell}}.
\]
Here, as in other parts of the paper, we use the symbol $\nabla$
for the $d$-dimensional gradient with respect to $\mathbf{T}$. The
arguments $p$ and $q$ in $\phi(p,q,\mathbf{T})$ are treated as
fixed for the purposes of taking this gradient. We also denote by
$\mathbf{f}$ a vector components $\phi_{T_{\ell}}(p,q,\mathbf{T})/q$.
We denote the equilibrium price function\footnote{Unlike the inverse demand function $p\left(q\right)$, the function
$p^{\star}\left(\mathbf{T}\right)$ takes the vector of taxes as arguments,
and its functional value is the price in the resulting equilibrium.} by $p^{\star}\left(\mathbf{T}\right)$ and its gradient, the \emph{pass-through
rate vector}, by $\tilde{\boldsymbol{\rho}}\equiv\nabla p^{\star}\left(\mathbf{T}\right).$
Further, we use the components of the $\mathbf{f}$ and $\boldsymbol{\tilde{\rho}}$
to define the \emph{pass-through quasi-elasticity vector} as 
\[
\boldsymbol{\rho}\equiv\left(\rho_{T_{1}},...,\rho_{T_{d}}\right),\qquad\rho_{T_{\ell}}\equiv\frac{\tilde{\rho}_{T_{\ell}}}{f_{T_{\ell}}}=\frac{q}{\phi_{T_{\ell}}(p,q,\mathbf{T})}\ \frac{\partial p^{\star}}{\partial T_{\ell}}.
\]
Note that the components of $\boldsymbol{\rho}$ are all dimensionless.
We define the (\textit{first-order})\textit{ price sensitivity} $\nu$
of the tax revenue and the (\textit{first-order})\textit{ quantity
sensitivity} $\tau$ of the (per-firm) tax revenue as follows: 
\[
\nu(p,q,\mathbf{T})\equiv\frac{1}{q}\phi_{p}(p,q,\mathbf{T}),\qquad\tau(p,q,\mathbf{T})\equiv\frac{1}{p}\phi_{q}(p,q,\mathbf{T}),
\]
and their derivatives are
\[
\nu_{T_{\ell}}(p,q,\mathbf{T})\equiv\frac{\partial\nu(p,q,\mathbf{T})}{\partial T_{\ell}},\qquad\tau_{T_{\ell}}(p,q,\mathbf{T})\equiv\frac{\partial\tau(p,q,\mathbf{T})}{\partial T_{\ell}}.
\]
The analogous definitions for the\textit{ second-order sensitivities}
are:
\[
\nu_{\text{(2)}}(p,q,\mathbf{T})\equiv\frac{p}{q}\ \frac{\partial^{2}\phi(p,q,\mathbf{T})}{\partial p^{2}},\,\tau_{\text{(2)}}(p,q,\mathbf{T})\equiv\frac{q}{p}\ \frac{\partial^{2}\phi(p,q,\mathbf{T})}{\partial q^{2}},\:\kappa(p,q,\mathbf{T})\equiv\frac{\partial^{2}\phi(p,q,\mathbf{T})}{\partial p\:\partial q}.
\]
The first-order and second-order sensitivities are dimensionless,
as are the components of $\boldsymbol{\rho}$. In this section, we
keep the same definition of the elasticities $\epsilon$ and $\eta$
as before.

\subsubsection{Generalized conduct index}

We introduce the conduct index $\theta$ as a function, independently
of the cost-side of the oligopoly game, so that in equilibrium the
following condition holds:
\begin{equation}
\left[1-\tau-\left(1-\nu\right)\eta\,\theta\right]p=mc.\label{EquationDefinitionOfGeneralizedConduct}
\end{equation}
In the case of unit and ad valorem taxation, this definition reduces
to the conduct index defined earlier (Equation \ref{theta_M}), where
$\tau=v+t/p$ and $\nu=v$: this is the reason why we can keep using
$\tau$ below. In principle, there are many possible definitions that
agree with the earlier definition in the case of unit and ad valorem
taxation. However, we find the specification of Equation \ref{EquationDefinitionOfGeneralizedConduct}
particularly convenient.

\subsubsection{Relationships for the pass-through vector components}

We now establish the following relationship for the relative size
of pass-through vector component.

\begin{proposition}\label{PropositionRelativeSizeOfPassThroughVectorComponents}The
pass-through rates and quasi-elasticities satisfy$\:$\footnote{If the denominators are zero, the fractions become ill-defined. In
that case, of course, the statement does not apply.} 
\[
\frac{\tilde{\rho}_{T_{\ell}}}{\tilde{\rho}_{T_{\ell'}}}=\frac{\tau_{T_{\ell'}}-\nu_{T_{\ell'}}\eta\:\theta}{\tau_{T_{\ell}}-\nu_{T_{\ell}}\eta\:\theta},\qquad\frac{\rho_{T_{\ell}}}{\rho_{T_{\ell'}}}=\frac{f_{T_{\ell'}}}{f_{T_{\ell}}}\ \frac{\tau_{T_{\ell}}-\eta\:\theta\:\nu_{T_{\ell}}}{\tau_{T_{\ell'}}-\eta\:\theta\:\nu_{T_{\ell'}}}.
\]

\end{proposition}The proposition is proven in Appendix \ref{AppendixProofOfPropositionRelativeSizeOfPassThroughVectorComponents}.
Since the components have known proportions, we can write them using
a common factor $p\rho_{\left(0\right)}$ as
\begin{equation}
\tilde{\rho}_{T_{\ell}}=\left(\tau_{T_{\ell}}-\nu_{T_{\ell}}\eta\:\theta\right)p\rho_{\left(0\right),}\qquad\rho_{T_{\ell}}=\frac{p}{f_{T_{\ell}}}\left(\tau_{T_{\ell}}-\nu_{T_{\ell}}\eta\:\theta\right)\rho_{\left(0\right),}\label{MultiDimentionalPassThroughInTermsOfRho0}
\end{equation}
with the factor $\rho_{\left(0\right)}$ determined in the following
proposition.\begin{proposition}\label{PropositionCommonFactorInPassThrough}The
value of the factor $\rho_{\left(0\right)}$ introduced in Equation
(\ref{MultiDimentionalPassThroughInTermsOfRho0}) is given by:
\begin{equation}
\frac{1}{\rho_{\left(0\right)}}=1-\kappa+\epsilon\tau_{\left(2\right)}+(1-\tau)\epsilon\chi+\left[\nu-\kappa+\eta\nu_{\left(2\right)}+\left(\omega-\eta-\chi\right)(1-\nu)\right]\theta,\label{MultiDimentionalPassThroughFactorRho0}
\end{equation}
where $\omega\equiv q\:(\eta\theta)'/\left(\eta\theta\right),$ with
the prime denoting a derivative with respect to the quantity $q.$\end{proposition}The
proof is in Appendix \ref{AppendixProofOfPropositionCommonFactorInPassThrough}.\footnote{If $\phi(p,q,\mathbf{T})=tq+vpq$, then $\tau=(t+vp)/p=t/p+v$ and
$\nu=vq/q=v.$ First, $\rho_{t}=\frac{q}{\partial\phi/\partial t}\tilde{\rho}_{t}=\frac{q}{q}\tilde{\rho}_{t}=\tilde{\rho}_{t},$
and $\rho_{v}=\frac{q}{\partial\phi/\partial v}\tilde{\rho}_{t}=\frac{q}{pq}\tilde{\rho}_{t}=\frac{1}{p}\tilde{\rho}_{t}.$
Next, $\rho_{t}=\frac{pq}{\partial\phi/\partial t}\left[\tau_{t}-\nu_{t}\left(\frac{\theta}{\epsilon^{D}}\right)\right]\rho_{\left(0\right)}=\rho_{\left(0\right)}$
because $\tau_{t}=1/p$ and $\nu_{t}=0$. Then,
\[
\frac{1}{\rho_{\left(0\right)}}=[\underset{=1-v}{(\underbrace{1-\kappa})}+\underset{=0}{\underbrace{\epsilon^{D}\tau_{\left(2\right)}}}+(1-\tau)\left(\frac{\epsilon^{D}}{\epsilon^{S}}\right)]+\left[\underset{=0}{\underbrace{\nu-\kappa+\eta\nu_{\left(2\right)}}}+\left(\omega-\frac{1}{\epsilon^{D}}-\frac{1}{\epsilon^{S}}\right)\underset{=1-v}{(\underbrace{1-\nu})}\right]\theta
\]
\[
=(1-v)\left\{ \left[1+\frac{1-\tau}{1-v}\left(\frac{\epsilon^{D}}{\epsilon^{S}}\right)\right]-\left(\frac{1}{\epsilon^{D}}+\frac{1}{\epsilon^{S}}\right)\theta+\epsilon^{D}q\frac{\partial(\theta/\epsilon^{D})}{\partial q}\right\} \text{\quad\quad}
\]
since $\kappa(p,q,\mathbf{T})\equiv\frac{\partial^{2}\phi(p,q,\mathbf{T})}{\partial p\:\partial q}=v,$
$\tau_{\text{(2)}}(p,q,\mathbf{T})\equiv\frac{q}{p}\cdot\frac{\partial^{2}\phi(p,q,\mathbf{T})}{\partial q^{2}}=0,$
and $\nu_{\text{(2)}}(p,q,\mathbf{T})\equiv\frac{p}{q}\cdot\frac{\partial^{2}\phi(p,q,\mathbf{T})}{\partial p^{2}}=0.$}

\subsubsection{Welfare changes and their relationship to pass-through vectors}

Now, we establish the general formulas for the marginal cost of public
fund and incidence in the multi-dimensional pass-through framework.
Welfare component changes in response to an infinitesimal change in
taxes can be found as follows. The (per-firm) consumer surplus change
in response to an infinitesimal change $dT_{\ell}$ in the tax $T_{\ell}$
is
\[
dCS=-qdp=-q\tilde{\rho}_{T_{\ell}}dT_{\ell},
\]
which means that in vector notation, $\frac{1}{q}\,\nabla CS=-\tilde{\boldsymbol{\rho}}$.
The change in (per-firm) producer surplus is 
\[
dPS=d\left(pq-c\left(q\right)-\phi(p,q,\mathbf{T})\right)=\left[\phi_{T_{\ell}}(p,q,\mathbf{T})-\left(1-\nu\right)\left(1-\theta\right)\tilde{\rho}_{T_{\ell}}\right]dT_{\ell},
\]
where we utilize Equation (\ref{EquationDefinitionOfGeneralizedConduct})
to eliminate the marginal cost. In vector notation, this is $\frac{1}{q}\,\nabla PS=\left(1-\nu\right)\left(1-\theta\right)\tilde{\boldsymbol{\rho}}-\mathbf{f}$,
since $\mathbf{f}=\frac{1}{q}\nabla\phi(p,q,\mathbf{T})$. The change
in tax revenue is 
\[
dR=\phi_{p}(p,q,\mathbf{T})dp+\phi_{q}(p,q,\mathbf{T})dq+\phi_{T_{\ell}}(p,q,\mathbf{T})dT_{\ell}=\left[\phi_{T_{\ell}}(p,q,\mathbf{T})-\left(\epsilon\tau-\nu\right)\tilde{\rho}_{T_{\ell}}\right]dT_{\ell}.
\]
In vector notation, $\frac{1}{q}\,\nabla R=\mathbf{f}-\left(\epsilon\tau-\nu\right)\tilde{\boldsymbol{\rho}}$.
Finally, for the change in social welfare, we have
\[
dW=\left(p-mc\right)dq=\left[\epsilon\tau+\theta\left(1-\nu\right)\right]\tilde{\rho}_{T_{\ell}}dT_{\ell}.
\]
In vector notation, $\frac{1}{q}\,\nabla W=-\left[\epsilon\tau+\theta\left(1-\nu\right)\right]\tilde{\boldsymbol{\rho}}$.

Note that the welfare components $CS\left(\mathbf{T}\right),\ PS\left(\mathbf{T}\right),\:R\left(\mathbf{T}\right)$,
and $W\left(\mathbf{T}\right)=CS\left(\mathbf{T}\right)+PS\left(\mathbf{T}\right)+R\left(\mathbf{T}\right)$
are all treated as functions of taxes only and represent the equilibrium
outcomes. This is different from the tax revenue function $\phi\left(p,q,\mathbf{T}\right)$,
which has also $p$ and $q$ as arguments and which is specified by
the government irrespective of the equilibrium. We summarize these
findings in the following proposition.

\begin{proposition}\label{PropositionGeneralTaxWelfareComponentChanges}The
tax gradients of consumer surplus, producer surplus, tax revenue,
and social welfare with respect to the taxes all belong to a two-dimensional
vector space spanned by $\mathbf{f}$ and $\tilde{\boldsymbol{\rho}}$.
The precise linear combinations of $\mathbf{f}$ and $\tilde{\boldsymbol{\rho}}$
are
\[
\frac{1}{q}\,\nabla CS=-\tilde{\boldsymbol{\rho},}
\]
\[
\frac{1}{q}\,\nabla PS=\left(1-\nu\right)\left(1-\theta\right)\tilde{\boldsymbol{\rho}}-\mathbf{f},
\]
\[
\frac{1}{q}\,\nabla R=\mathbf{f}+\left(\nu-\epsilon\tau\right)\tilde{\boldsymbol{\rho}},
\]
\[
\frac{1}{q}\,\nabla W=-\left[\left(1-\nu\right)\theta+\epsilon\tau\right]\tilde{\boldsymbol{\rho}}.
\]
\end{proposition}

These relationships, considered component-wise, immediately imply
the following results for welfare change ratios and generalize Propositions
\ref{PropositionMCt} and \ref{PropositionConsumerAndProducerIncidence}.\footnote{Remember that the $T_{\ell}$ component of the vector $\mathbf{f}$
is $\phi_{T_{\ell}}(p,q,\mathbf{T})/q=$$\tilde{\rho}_{T_{\ell}}/\rho_{T_{\ell}}$.}

\begin{proposition}\label{PropositionGeneralTaxWelfareChangeRatios}The
marginal cost of public funds of a tax $T_{\ell}$, $MC_{T_{\ell}}=\left(\nabla W\right)_{T_{\ell}}/\left(\nabla R\right)_{T_{\ell}}$,
is
\[
MC_{T_{\ell}}=\frac{\left(1-\nu\right)\theta+\epsilon\tau}{\frac{1}{\rho_{T_{\ell}}}+\nu-\epsilon\tau}.
\]
The incidence of this tax, $I_{T_{\ell}}=\left(\nabla CS\right)_{T_{\ell}}/\left(\nabla PS\right)_{T_{\ell}}$,
equals:
\[
I_{T_{\ell}}=\frac{1}{\frac{1}{\rho_{T_{\ell}}}-\left(1-\nu\right)\left(1-\theta\right)}.
\]
Similarly, the social incidence, $SI_{T_{\ell}}=\left(\nabla W\right)_{T_{\ell}}/\left(\nabla PS\right)_{T_{\ell}}$,
equals:
\[
SI_{T_{\ell}}=\frac{\left(1-\nu\right)\theta+\epsilon\tau}{\frac{1}{\rho_{T_{\ell}}}-\left(1-\nu\right)\left(1-\theta\right)}.
\]
\end{proposition}

\subsection{Pass-through, conduct index, and welfare: special cases}

The results of the previous subsection contain our results for ad
valorem and unit taxes as special cases, but provide much greater
generality, since the taxes (government interventions) may be specified
in a very flexible way.

\subsubsection{Exogenous competition and depreciating licenses}

Weyl and Fabinger's (2013) results under symmetric oligopoly can be
interpreted as special cases of the present results. In particular,
Weyl and Fabinger's (2013) analysis considers either unit taxes or
exogenous competition (an exogenous quantity supplied to the market).
The case of unit taxes is clearly included in the present results,
which has motivated this paper. At the same time, it turns out that
the case of exogenous competition is included as well. The reasoning
is as follows.

Consider a tax $T_{1}=\tilde{q}$ of the form: $\phi\left(p,q,\tilde{q}\right)=\tilde{q\,}p+c(\text{\ensuremath{q-\tilde{q}}})-c(q)$.
Then, the firm's profit is given by:
\[
pq-c\left(q\right)-\phi\left(p,q,\tilde{q}\right)=p\left(q-\tilde{q}\right)+c(\text{\ensuremath{q-\tilde{q}}}).
\]
The firm, therefore, has the same profit function as in the case of
exogenous competition $\tilde{q}$ in Weyl and Fabinger (2013). Proposition
\ref{PropositionGeneralTaxWelfareChangeRatios} above (specialized
to constant marginal cost and zero initial $\tilde{q}$) then implies
the social incidence result in Principle of Incidence 3 in Weyl and
Fabinger (2013, p.\,548). 

Similarly, the relationships between pass-through of unit taxes and
of exogenous competition are implied by the general result of Proposition
\ref{PropositionRelativeSizeOfPassThroughVectorComponents} for the
tax specification $T_{1}=t$, $T_{2}=\tilde{q}$,
\[
\phi\left(p,q,t,\tilde{q}\right)=tq+\tilde{q\,}p+c(\text{\ensuremath{q-\tilde{q}}})-c(q).
\]
To obtain the absolute size of the two types of pass-through, one
can straightforwardly use Proposition \ref{PropositionCommonFactorInPassThrough}.
More generally, $\phi$ is extended as

\[
\phi=c\left(q-\tilde{q}\right)+v\left(q-\tilde{q}\right)p(q)+(1-v)\tilde{q}p(q)+t\left(q-\tilde{q}\right)-c(q),
\]
where an ad valorem tax is also considered. As an example, one can
think of a government which procures goods from abroad and supplies
them to the market in order to lower domestic prices.

In the special case of a monopolist with constant marginal cost, the
mathematics allow for another interesting interpretation: It is isomorphic
to the case of ``depreciating licenses'' in Weyl and Zhang (2017).
Depreciating licenses correspond to a tax scheme where the owner of
an asset announces a reservation price at which she is willing to
sell it and gets taxed a fixed fraction of that prices. Another agent
in the economy may buy the asset at the announced price. The owner
then faces a tradeoff between announcing a low price and paying low
taxes and announcing a high price in order to be able to keep the
asset and derive utility from it. The optimization problem then leads
to exactly the same mathematical form as the problem of a monopolist
with constant marginal cost facing exogenous competition. We include
a more detailed explanation in Appendix \ref{AppendixDepreciatingLicenses}.\footnote{We thank Glen Weyl for suggesting this relationship between Weyl and
Zhang (2017) and our analysis.}

\subsubsection{Sales restrictions}

Governments often regulate when, where, and to whom products may be
sold. For example, there are restrictions on weekend sales, store
locations, etc. A simple way of modeling this situation is to assume
that due to the restrictions a firm loses a fixed proportion of its
customers. If the absence of the regulation and taxation, the profit
function is $p\left(q\right)q-c\left(q\right)$. The new profit function
will be $\left(1-v\right)p([1+\kappa]q)q-tq-c\left(q\right)$, where
$1-1/\left(1+\kappa\right)$ is the fraction of customers lost. The
only change is in the argument of the inverse demand function: for
the firm to sell quantity $q$, each remaining customer needs to buy
$\left(1+\kappa\right)$ times more than in the absence of the regulation,
and the price would have to be correspondingly lower. This change
may be described using as:
\[
\phi\left(p,q,t,v,\kappa\right)=\left(1-\left(1-v\right)h\left(q,\kappa\right)\right)pq+tq,\qquad h\left(q,\kappa\right)\equiv\frac{p\left(q\right)-p\left(\left(1+\kappa\right)q\right)}{p\left(q\right)}.
\]
For demand with constant elasticity $\epsilon$, $h\left(q,\kappa\right)=1-(1+\kappa)^{-1/\epsilon}$,
independently of $q$.

\subsubsection{Tax evasion/tax avoidance and concealment costs}

Tax evasion, clearly, is a very important problem in many situations
since economic agents do not always strictly follow the law (Choi,
Furusawa, and Ishikawa 2017).

For simplicity, consider a firm that needs to pay an ad valorem tax
$v\tilde{p}q$, where $\tilde{p}$ is the price reported to the government
and may differ from the true price $p$. We capture the cost associated
with deceiving the government by introducing a concealment cost of
the form:

\[
\frac{1}{4\lambda}p^{-\zeta}(\tilde{p}-p)^{2}q^{1-\xi}.
\]
The firm then chooses the reported price $\tilde{p}$ to minimize
the sum of these two additional costs:
\[
v\tilde{p}q+\frac{1}{4\lambda}p^{-\zeta}q^{1-\xi}(\tilde{p}-p)^{2}.
\]
The corresponding first-order condition implies $\tilde{p}=p-2\lambda vp^{\zeta}q^{\xi}$,
which gives the effective additional cost
\[
pqv-\lambda v^{2}p^{\zeta}q^{1+\xi}.
\]
\[
\phi\left(p,q,t,v,\lambda\right)=tq+pqv-\lambda v^{2}p^{\zeta}q^{1+\xi}
\]
\[
\tilde{\phi}\left(p,q,t,v,\lambda\right)=tq+pqv-2\lambda v^{2}p^{\zeta}q^{1+\xi}
\]
The government needs to pay additional enforcement cost inversely
related to $\lambda$, which needs to be remembered in the welfare
analysis.

\section{Heterogeneous Firms}

\label{SectionAsymmetricFirms}In this section, we extend our results
to the case of $n$ heterogeneous firms (i.e. asymmetric firms), where
each firm $i$ controls a strategic variable $\sigma_{i}$, which
could be, for example, the price or quantity of its product. We allow
for the tax function $\phi_{i}\left(p_{i},q_{i},\mathbf{T}\right)$
to depend explicitly on the identity of the firm; we write $f_{i\:T_{\ell}}\left(p_{i},q_{i},\mathbf{T}\right)=\frac{1}{q_{i}}\,\frac{\partial}{\partial T_{\ell}}\phi_{i}\left(p_{i},q_{i},\mathbf{T}\right)$
for its derivative with respect to tax $T_{\ell}$. Similarly, the
sensitivities $\tau_{i}\left(p_{i},q_{i},\mathbf{T}\right)$, $\nu_{i}\left(p_{i},q_{i},\mathbf{T}\right)$,
etc., now also have the firm index $i$. The marginal cost $mc_{i}\left(q_{i}\right)$
of firm $i$ is also allowed to depend on the identity of the firm,
and we denote its elasticity $\chi_{i}\left(q_{i}\right)\equiv q_{i}\,mc_{i}'\left(q_{i}\right)/mc_{i}\left(q_{i}\right).$

\subsection{Pricing strength index and pass-through}

We define the \emph{pricing strength index} $\psi_{i}\left(\mathbf{q}\right)$
of firm $i$ to be a function independent of the cost side of the
economic problem such that the first-order condition for firm $i$
is:
\[
\left\{ 1-\tau_{i}\left(p_{i}\left(\mathbf{q}\right),q_{i},\mathbf{T}\right)-\psi_{i}\left(\mathbf{q}\right)\left[1-\nu_{i}\left(p_{i}\left(\mathbf{q}\right),q_{i},\mathbf{T}\right)\right]\right\} p_{i}\left(\mathbf{q}\right)=mc\left(q_{i}\right).
\]
In the special case of symmetric firms, this definition reduces to
$\psi_{i}=\eta\,\theta$.  

We express the pass-through rate in terms of these pricing strength
indices. Specifically, the pass-through rate is an $n\times d$ matrix
$\mathbf{\tilde{\rho}}$ with rows $\tilde{\mathbf{\boldsymbol{\rho}}}_{T_{\ell}}\equiv\partial\mathbf{p}/\partial T_{\ell}$
and elements $\tilde{\rho}_{i\,T_{\ell}}=\partial p_{i}/\partial T_{\ell}$.
It is shown that the pass-through rate equals
\begin{equation}
\tilde{\mathbf{\rho}}_{T_{\ell}}=\mathbf{b}^{-1}\,.\,\mathbf{\iota}_{T_{\ell}},\label{PassThroughRateForHeterogeneousFirms}
\end{equation}
where the factors on the right-hand side are defined as follows. The
matrix $\mathbf{b}$ is an $n\times n$ matrix, independent of the
choice of $T_{\ell}$, with elements 
\begin{eqnarray*}
b_{ij} & = & \left[1-\kappa_{i}-\left(1-\nu_{i}-\nu_{\text{(2)}i}\right)\!\psi_{i}\right]\delta_{ij}-\left(1-\nu_{i}\right)\!\psi_{i}\Psi_{ij}\\
 & + & \left\{ \tau_{\text{(2)}i}+\nu_{i}\psi_{i}-\kappa_{i}+\left[1-\tau_{i}-\left(1-\nu_{i}\right)\psi_{i}\right]\!\chi_{i}\right\} \epsilon_{ij},
\end{eqnarray*}
where   $\delta_{ij}$ is the Kronecker delta, and
\[
\epsilon_{ij}=-\,\frac{p_{i}}{q_{i}}\,\frac{\partial q_{i}\left(\mathbf{p}\right)}{\partial p_{j}},\qquad\Psi_{ij}=\frac{p_{i}}{\psi_{i}}\frac{\partial\psi_{i}\left(\mathbf{q}\left(\mathbf{p}\right)\right)}{\partial p_{j}}.
\]
For each tax $T_{\ell}$, $\mathbf{\iota}_{T_{\ell}}$ is an $n$-dimensional
vector with components
\[
\iota_{i\,T_{\ell}}=p_{i}\,\frac{\partial\tau_{i}\left(p_{i},q_{i},\mathbf{T}\right)}{\partial T_{\ell}}-p_{i}\,\psi_{i}\,\frac{\partial\nu_{i}\left(p_{i},q_{i},\mathbf{T}\right)}{\partial T_{\ell}}.
\]
In the case of symmetric firms and at symmetric prices, the pass-through
rate expression in Equation (\ref{PassThroughRateForHeterogeneousFirms})
agrees with the expression represented by Equations (\ref{MultiDimentionalPassThroughInTermsOfRho0})
and (\ref{MultiDimentionalPassThroughFactorRho0}) in Section \ref{SectionMultiDimensionalPassThroughFramework}.\footnote{To confirm this agreement, note that at symmetric prices, $\sum_{j=1}^{n}\Psi_{ij}=-\epsilon\omega$.
Note also that $\epsilon_{ii}\left(\mathbf{p}\right)|_{\mathbf{p}=\left(p,...,p\right)}=\epsilon_{F}(p),$
and for $j\neq i$, $\epsilon_{ij}\left(\mathbf{p}\right)|_{\mathbf{p}=\left(p,...,p\right)}=-\,\frac{1}{n-1}\,\epsilon_{C}(p)$.}

To generalize the notion of pass-through quasi-elasticity to the case
of heterogeneous firms, we define the pass-through quasi-elasticity
matrix $\mathbf{\rho}$ as an $n\times d$ matrix with elements 
\[
\rho_{i\,T_{\ell}}=\frac{1}{f_{i\,T_{\ell}}\left(p_{i},q_{i},\mathbf{T}\right)}\ \frac{\partial p_{i}}{\partial T_{\ell}},
\]
and with rows denoted $\mathbf{\rho}_{T_{\ell}}$.

\subsection{Welfare changes}

\label{SubsectionHeterogeneousFirmsWelfareChanges}In the following,
for each $i$, $\mathbf{\epsilon}_{i}$ is an $n$-dimensional vector
with its $j$-th component equal to $\epsilon_{ij}$. For the tax
gradients of welfare components corresponding to individual firms
we obtain:
\[
\frac{1}{q_{i}}\nabla CS_{i}=-\mathbf{e}_{i}.\tilde{\boldsymbol{\rho},}
\]
\[
\frac{1}{q_{i}}\nabla PS_{i}=\left(1-\nu_{i}\right)\left(\mathbf{e}_{i}-\psi_{i}\,\mathbf{\epsilon}_{i}\right).\tilde{\boldsymbol{\rho}}-\mathbf{f}_{i},
\]
\[
\frac{1}{q_{i}}\nabla R_{i}=\mathbf{f}_{i}+\left(\nu_{i}\,\mathbf{e}_{i}-\tau_{i}\,\mathbf{\epsilon}_{i}\right).\tilde{\boldsymbol{\rho}},
\]
\[
\frac{1}{q_{i}}\nabla W_{i}=-\left[\tau_{i}+\psi_{i}\:\left(1-\nu_{i}\right)\right]\mathbf{\epsilon}_{i}.\tilde{\boldsymbol{\rho}}.
\]
The corresponding gradients of total welfare components are then obtained
by adding up contributions from individual firms. For example, $\nabla CS=\sum_{i=1}^{n}\nabla CS_{i}$.
Denoting the total quantity as $Q\equiv\sum_{i=1}^{n}q_{i}$, this
means that $\frac{1}{Q}\nabla CS$ is a weighted average of $-\mathbf{e}_{i}.\tilde{\boldsymbol{\rho}}$,
with the weights proportional to $q_{i}$. The arguments for the other
welfare components are similar. This generalizes Proposition \ref{PropositionGeneralTaxWelfareComponentChanges}
above.

We can also consider ratios of welfare changes corresponding to some
tax $T_{\ell}$:
\[
MC_{i\,T_{\ell}}=\frac{\left[\tau_{i}+\left(1-\nu_{i}\right)\,\psi_{i}\right]\,\epsilon_{i\,T_{\ell}}^{\rho}}{\frac{1}{\rho_{i\,T_{\ell}}}+\nu_{i}-\tau_{i}\,\epsilon_{i\,T_{\ell}}^{\rho}},
\]
\[
I_{i\,T_{\ell}}=\frac{1}{\frac{1}{\rho_{i\,T_{\ell}}}-\left(1-\nu_{i}\right)(1-\psi_{i}\,\epsilon_{i\,T_{\ell}}^{\rho})},
\]
\[
SI_{i\,T_{\ell}}=\frac{\left[\tau_{i}+\left(1-\nu_{i}\right)\,\psi_{i}\right]\,\epsilon_{i\,T_{\ell}}^{\rho}}{\frac{1}{\rho_{i\,T_{\ell}}}-\left(1-\nu_{i}\right)(1-\psi_{i}\,\epsilon_{i\,T_{\ell}}^{\rho})},
\]
where $\epsilon_{i\,T_{\ell}}^{\rho}\equiv\mathbf{\epsilon}_{i}.\tilde{\boldsymbol{\rho}}_{T_{\ell}}/\tilde{\rho}_{i\,T_{\ell}}=\mathbf{\epsilon}_{i}.\boldsymbol{\rho}_{T_{\ell}}/\rho_{i\,T_{\ell}}.$
The ratios of the corresponding total welfare changes will be weighted
averages of these firm-specific ratios. The weights correspond to
the sizes of the denominators times $q_{i}$. For example, $MC_{T_{\ell}}$
will lie between $\min_{i}MC_{i\,T_{\ell}}$ and $\max_{i}MC_{i\,T_{\ell}}$.
The same reasoning also holds for the other ratios. This generalizes
Proposition \ref{PropositionGeneralTaxWelfareChangeRatios} above.

\subsection{Conduct index and welfare changes}

For heterogeneous firms, we introduce the conduct index of firm $i$
so that
\[
\theta_{i}=-\frac{\sum_{j=1}^{n}\left\{ p_{j}\left[1-\tau_{j}\left(p_{j},q_{j},\mathbf{T}\right)\right]-mc\left(q_{j}\right)\right\} \frac{dq_{j}}{d\sigma_{i}}}{\sum_{j=1}^{n}\left[1-\nu_{j}\left(p_{j},q_{j},\mathbf{T}\right)\right]q_{j}\,\frac{dp_{j}}{d\sigma_{i}}}
\]
holds. In the special case of only unit taxation, this definition
reduces to Weyl and Fabinger's (2013, p.\,552) Equation (4). In the
special case of symmetric firms the definition reduces to our Equation
(\ref{EquationDefinitionOfGeneralizedConduct}) with $\theta_{i}=\theta$. 

The conduct index $\theta_{i}$ is closely connected to the pricing
strength index $\psi_{i}$, but not as closely as it would be in the
case of symmetric oligopoly. Using the definitions of the indices,
it is shown that
\[
\theta_{i}=-\frac{\sum_{j=1}^{n}\left(1-\nu_{j}\right)\psi_{j}\,p_{j}\frac{dq_{j}}{d\sigma_{i}}}{\sum_{j=1}^{n}\left(1-\nu_{j}\right)q_{j}\frac{dp_{j}}{d\sigma_{i}}}.
\]
For symmetric oligopoly, this equation reduces simply to $\theta=\epsilon\psi.$

The conduct index is used to express welfare component changes in
response to infinitesimal changes in taxes. The relationships are
a bit more complicated than in the case of using the pricing strength
index: they can be expressed as follows. We define the price response
to an infinitesimal change in the strategic variable $\sigma_{k}$
of firm $j$ as
\[
\zeta_{ij}=\frac{dp_{i}}{d\sigma_{j}}.
\]
Since the vectors $\zeta_{i1}$, $\zeta_{i2}$, ... , $\zeta_{in}$
form a basis in the $n$-dimensional vector space to which $\tilde{\rho}_{i\,T_{\ell}}$
for a given $\ell$ belongs, we can write $\tilde{\rho}_{i\,T_{\ell}}$
as a linear combination of them for some coefficients $\lambda_{i\,T_{\ell}}$:
\[
\tilde{\rho}_{i\,T_{\ell}}=\sum_{j=1}^{n}\lambda_{j\,T_{\ell}}\zeta_{ij}.
\]
For changes in consumer and producer surplus, we obtain: 
\[
\frac{dCS}{dT_{\ell}}=-\sum_{i=1}^{n}q_{i}\tilde{\rho}_{i\,T_{\ell}}=-\sum_{j=1}^{n}\left(\sum_{i=1}^{n}q_{i}\zeta_{ij}\right)\lambda_{j\,T_{\ell}},
\]
\[
\frac{dPS}{dT_{\ell}}=-\sum_{i=1}^{n}f_{i\,T_{\ell}}\left(p_{i},q_{i},\mathbf{T}\right)-\sum_{j=1}^{n}\hat{\zeta}_{j}\,\left(1-\theta_{j}\right)\lambda_{j\,T_{\ell}},
\]
where we used the notation
\[
\hat{\zeta}_{j}\equiv\sum_{i=1}^{n}\left[1-\nu_{i}\left(p_{i},q_{i},\mathbf{T}\right)\right]q_{i}\,\zeta_{ij}.
\]
These surplus change expressions represent a generalization of the
surplus expressions in Weyl and Fabinger's (2013) Section 5.

\subsection{Aggregative games}

In the case of oligopoly in the form of aggregative games, where all
other firms' actions are summarized as an aggregator in each firm's
profit, we can further manipulate the above formulas for pricing strength
and conduct indices.\footnote{Here, we consider a setup in Anderson, Erkal, and Piccinin's (2016)
Section 2.} We identify the firm's strategic variable $\sigma_{i}$ with an action
$a_{i}\equiv\sigma_{i}$ the firm can take, which contributes to an
aggregator $A=\sum_{i=1}^{n}a_{i}$. The prices and quantities are
functions of just two arguments: $p_{i}\left(A,a_{i}\right)$ and
$q_{i}\left(A,a_{i}\right)$. Their derivatives that take into account
the dependence of $A$ on the action of firm $i$ are $\frac{dq_{j}}{d\sigma_{i}}=\frac{\partial q_{j}}{\partial a_{i}}+\frac{\partial q_{j}}{\partial A}$
and $\frac{dp_{j}}{d\sigma_{i}}=\frac{\partial p_{j}}{\partial a_{i}}+\frac{\partial p_{j}}{\partial A}$.
The firm's first-order condition is:
\[
0=\left(\frac{\partial p_{j}}{\partial a_{i}}\left(A,a_{i}\right)+\frac{\partial p_{j}}{\partial A}\left(A,a_{i}\right)\right)q_{i}\left(A,a_{i}\right)\left(\nu_{i}\left(p_{i}\left(A,a_{i}\right),q_{i}\left(A,a_{i}\right),\mathbf{T}\right)-1\right)+\qquad\qquad\qquad
\]
\[
\qquad\left(\frac{\partial q_{j}}{\partial a_{i}}\left(A,a_{i}\right)+\frac{\partial q_{j}}{\partial A}\left(A,a_{i}\right)\right)\left(mc\left(q_{i}\left(A,a_{i}\right)\right)+p_{i}\left(A,a_{i}\right)\left(\tau_{i}\left(p_{i}\left(A,a_{i}\right)\!,\!q_{i}\left(A,a_{i}\right)\!,\!\mathbf{T}\right)-1\right)\right),
\]
which gives us a relatively simple expression for the pricing strength
index:
\[
\psi_{i}\left(A,a_{i}\right)=-\frac{q_{i}\left(A,a_{i}\right)}{p_{i}\left(A,a_{i}\right)}\ \ \frac{\frac{\partial p_{j}}{\partial a_{i}}\left(A,a_{i}\right)+\frac{\partial p_{j}}{\partial A}\left(A,a_{i}\right)}{\frac{\partial q_{j}}{\partial a_{i}}\left(A,a_{i}\right)+\frac{\partial q_{j}}{\partial A}\left(A,a_{i}\right)}.
\]
The expression for the conduct index also simplifies:
\[
\theta_{i}=\sum_{j=1}^{n}w_{j}\ \frac{\gamma_{j}\left(A,a_{i}\right)}{\gamma_{j}\left(A,a_{j}\right)},
\]
where $w_{i}$ is a normalized version of unnormalized \textquotedblleft weights\textquotedblright{}
$\tilde{w}_{j}$,
\[
w_{i}\equiv\frac{\tilde{w}_{i}}{\sum_{j=1}^{n}\tilde{w}_{j}},\qquad\tilde{w}_{j}\equiv\left(1-\nu_{j}\right)q_{j}\left(A,a_{j}\right)\,\left(\frac{\partial p_{j}}{\partial a_{i}}\left(A,a_{i}\right)+\frac{\partial p_{j}}{\partial A}\left(A,a_{i}\right)\right),
\]
and
\[
\gamma_{j}\left(A,a_{i}\right)\equiv\frac{\partial q_{j}}{\partial a_{i}}\left(A,a_{i}\right)+\frac{\partial q_{j}}{\partial A}\left(A,a_{i}\right).
\]
These simplified formulas would be used for further analysis of pass-through
and welfare in aggregative oligopoly games.

\section{Pass-Through and Welfare under Production-Cost and Taxation Changes}

\label{SectionPassThroughAndWelfareUnderProductionCostAndTaxationChanges}In
the previous sections, we have studied changes in taxation, but not
changes in production costs. Here, we generalize our main results
to incorporate both taxation and production costs. As shown below,
the generalization to production cost changes turns out to be%
{} straightforward. These more general formulas may be applied to a
range of economic situations such as cost changes due to exchange
rate movements or movements in the world prices of commodities. 

The additional cost to the firm is denoted $\phi\left(p,q,\mathbf{T}\right)$
as before, but the tax bill of firm $i$, denoted $\tilde{\phi}\left(p,q,\mathbf{T}\right)$,
is different, in general. Here $\mathbf{T}$ is a vector of interventions
(by the government or by external circumstances), which may or may
not include traditional taxes. We recover the previous case of only
taxation by setting $\tilde{\phi}\left(p,q,\mathbf{T}\right)=\phi\left(p,q,\mathbf{T}\right)$.
If all of the additional cost to the firm comes from the production
side, we have $\tilde{\phi}\left(p,q,\mathbf{T}\right)=0$. In general,
$\phi\left(p,q,\mathbf{T}\right)-\tilde{\phi}\left(p,q,\mathbf{T}\right)$
is the production part of the additional cost $\phi\left(p,q,\mathbf{T}\right)$. 

\subsection{Symmetric firms}

In addition to the notation used in the previous section, we define
$\tilde{\mathbf{f}}=\frac{1}{q}\nabla\tilde{\phi}(p,q,\mathbf{T})$.
First, we obtain a generalization of the formulas for the tax gradients
of welfare components in Proposition \ref{PropositionGeneralTaxWelfareComponentChanges}.
The equilibrium outcome depends only on the additional cost $\phi\left(p,q,\mathbf{T}\right)$
and not on its split between taxes and production costs. For this
reason, the formulas for consumer and producer surplus will be unchanged.
The government revenue and therefore also total social welfare%
{} depends on $\tilde{\phi}\left(p,q,\mathbf{T}\right)$.%
{} In the formula for the gradient of government revenue, $\mathbf{f}$
will be replaced by $\tilde{\mathbf{f}}$, and the formula for social
welfare will be adjusted to reflect this difference. Hence, the generalization
of the results in Proposition \ref{PropositionGeneralTaxWelfareComponentChanges}
is:
\[
\frac{1}{q}\,\nabla CS=-\tilde{\boldsymbol{\rho},}
\]
\[
\frac{1}{q}\,\nabla PS=\left(1-\nu\right)\left(1-\theta\right)\tilde{\boldsymbol{\rho}}-\mathbf{f},
\]
\[
\frac{1}{q}\,\nabla R=\mathbf{\tilde{f}}+\left(\nu-\epsilon\tau\right)\tilde{\boldsymbol{\rho}},
\]
\[
\frac{1}{q}\,\nabla W=-[\left(1-\nu\right)\theta+\epsilon\tau]\tilde{\boldsymbol{\rho}}+\mathbf{\tilde{f}}-\mathbf{f}.
\]

We further define $g_{T_{\ell}}\equiv\tilde{f}_{T_{\ell}}/f_{T_{\ell}}$,
which represents the fraction of an increase in additional cost ($\phi$)
to the firm (due to a change in the tax parameter $T_{\ell}$) that
is collected by the government in the form of taxes ($\tilde{\phi}$).
In other words, $g_{T_{\ell}}$ is the government's share in increases
of the additional costs induced by marginal changes in $T_{\ell}$.
If $\phi$ is a pure tax, then $g_{T_{\ell}}=1$, and if $\phi$ is
a pure production cost with no tax component, then $g_{T_{\ell}}=0$.
By taking ratios of the components of the tax gradients above, we
obtain a generalization of Proposition \ref{PropositionGeneralTaxWelfareChangeRatios}:
The marginal cost of public funds associated with intervention $T_{\ell}$,
$MC_{T_{\ell}}=\left(\nabla W\right)_{T_{\ell}}/\left(\nabla R\right)_{T_{\ell}}$,
is
\[
MC_{T_{\ell}}=\frac{\frac{1-g_{T_{\ell}}}{\rho_{T_{\ell}}}+\left(1-\nu\right)\theta+\epsilon\tau}{\frac{g_{T_{\ell}}}{\rho_{T_{\ell}}}+\nu-\epsilon\tau}.
\]
The incidence of this intervention, $I_{T_{\ell}}=\left(\nabla CS\right)_{T_{\ell}}/\left(\nabla PS\right)_{T_{\ell}}$,
equals:
\[
I_{T_{\ell}}=\frac{1}{\frac{1}{\rho_{T_{\ell}}}-\left(1-\nu\right)\left(1-\theta\right)}.
\]
Similarly, the social incidence, $SI_{T_{\ell}}=\left(\nabla W\right)_{T_{\ell}}/\left(\nabla PS\right)_{T_{\ell}}$,
equals:
\[
SI_{T_{\ell}}=\frac{\frac{1-g_{T_{\ell}}}{\rho_{T_{\ell}}}+\left(1-\nu\right)\theta+\epsilon\tau}{\frac{1}{\rho_{T_{\ell}}}-\left(1-\nu\right)\left(1-\theta\right)}.
\]

\subsection{Heterogeneous firms}

The adjustments to our formulas needed to generalize the results of
Subsection \ref{SubsectionHeterogeneousFirmsWelfareChanges} are analogous
to the case of symmetric firms we just discussed. For each firm $i$,
we define $\tilde{\mathbf{f}}_{i}=\frac{1}{q}\nabla\tilde{\phi}_{i}(p,q,\mathbf{T}).$
For the welfare gradients, we obtain:
\[
\frac{1}{q_{i}}\nabla CS_{i}=-\mathbf{e}_{i}.\tilde{\boldsymbol{\rho},}
\]
\[
\frac{1}{q_{i}}\nabla PS_{i}=\left(1-\nu_{i}\right)\left(\mathbf{e}_{i}-\psi_{i}\,\mathbf{\epsilon}_{i}\right).\tilde{\boldsymbol{\rho}}-\mathbf{f}_{i},
\]
\[
\frac{1}{q_{i}}\nabla R_{i}=\left(\nu_{i}\,\mathbf{e}_{i}-\tau_{i}\,\mathbf{\epsilon}_{i}\right).\tilde{\boldsymbol{\rho}}+\mathbf{\tilde{f}}_{i},
\]
\[
\frac{1}{q_{i}}\nabla W_{i}=-\left[\tau_{i}+\psi_{i}\:\left(1-\nu_{i}\right)\right]\mathbf{\epsilon}_{i}.\tilde{\boldsymbol{\rho}}+\mathbf{\tilde{f}}_{i}-\mathbf{f}_{i}.
\]
 Similarly, for each firm $i$, we define $g_{i\,T_{\ell}}\equiv\tilde{f}_{i\,T_{\ell}}/f_{i\,T_{\ell}}$.
For the firm-specific welfare change ratios, we obtain:
\[
MC_{i\,T_{\ell}}=\frac{\frac{1-g_{i\,T_{\ell}}}{\rho_{i\,T_{\ell}}}+(\tau_{i}+\left(1-\nu_{i}\right)\,\psi_{i})\,\epsilon_{i\,T_{\ell}}^{\rho}}{\frac{g_{i\,T_{\ell}}}{\rho_{i\,T_{\ell}}}+\nu_{i}-\tau_{i}\,\epsilon_{i\,T_{\ell}}^{\rho}},
\]
\[
I_{i\,T_{\ell}}=\frac{1}{\frac{1}{\rho_{i\,T_{\ell}}}-\left(1-\nu_{i}\right)(1-\psi_{i}\,\epsilon_{i\,T_{\ell}}^{\rho})},
\]
\[
SI_{i\,T_{\ell}}=\frac{\frac{1-g_{T_{\ell}}}{\rho_{i\,T_{\ell}}}+(\tau_{i}+\left(1-\nu_{i}\right)\,\psi_{i})\,\epsilon_{i\,T_{\ell}}^{\rho}}{\frac{1}{\rho_{i\,T_{\ell}}}-\left(1-\nu_{i}\right)(1-\psi_{i}\,\epsilon_{i\,T_{\ell}}^{\rho})}.
\]

\section{Concluding Remarks}

\label{SectionConcludingRemarks}In this paper, we characterize the
welfare measures%
{} of taxation and other external changes%
{} in oligopoly with a general specification of competition, market
demand and production cost. For symmetric oligopoly, we first derive
formulas for marginal welfare losses from unit and ad valorem taxation,
$MC_{t}$ and $MC_{v}$, using the unit tax pass-through rate $\rho_{t}$
and the ad valorem tax pass-through semi-elasticity $\rho_{v}$ (Proposition
\ref{PropositionMCt}) as well as the formulas for tax incidence,
$I_{t}$ and $I_{v}$ (Proposition \ref{PropositionConsumerAndProducerIncidence}).
We then show that $\rho_{v}$ can be expressed in terms of $\rho_{t}$
(Proposition \ref{PropositionRelationshipBetweenPassThroughs}). These
relationships are used to derive sufficient statistics for $MC_{t}$
and $MC_{v}$ (Proposition \ref{Sufficient-statistics-for}). The
pass-through is also characterized, generalizing Weyl and Fabinger's
(2013) formula (Proposition \ref{PropositionGeneralSymmetricOligopolyPassThrough}).
In the case of price or quantity competition, we explain how $\rho_{t}$
and $\rho_{v}$ can be written only in terms of the demand elasticities,
the demand curvatures, and the marginal cost elasticity (Propositions
\ref{PropositionPassThroughAndMCUnderPriceCompetition} and \ref{PropositionPassThroughAndMCUnderQuantityCompetition}).
We have discussed the relationships to other quantities of interest,
as well as illustrative special cases. 

The second part of the paper extends the results beyond the two-dimensional
taxation problem. Specifically, we show that the previous results
have a very natural generalization to a general specification of the
tax revenue function as a function parameterized by a vector of tax
parameters (Propositions \ref{PropositionRelativeSizeOfPassThroughVectorComponents},
\ref{PropositionCommonFactorInPassThrough}, \ref{PropositionGeneralTaxWelfareComponentChanges},
and \ref{PropositionGeneralTaxWelfareChangeRatios}). We further discuss
an extension of our analysis to the case of asymmetric oligopoly,
where the firms face different costs and possibly also different taxes
(Section \ref{SectionAsymmetricFirms}).\footnote{By allowing (constant) asymmetric marginal costs, Anderson, de Palma,
and Kreider (2001b) show that under quantity competition with homogeneous
products (i.e., Cournot competition), ad valorem taxation is still
preferable to unit taxation, although they were not able to verify
if the same conclusion held under quantity competition with product
differentiation. However, Anderson, de Palma, and Kreider (2001b)
discuss a specific demand system (with perfectly inelastic individual
demand) under which unit taxation is preferable to ad valorem taxation
if the required tax revenue is sufficiently high. We conjecture that
one could obtain further generalization by allowing the conduct index
$\theta$ to be firm-specific. See also Zimmerman and Carlson (2010)
for a parametric analysis of asymmetric firms.}$^{,}$\footnote{Interestingly, Tremblay and Tremblay (2017) study tax incidence in
an asymmetric duopoly where one firm competes in price and the other
firm competes in quantity, focusing on unit taxation. The pass-through
rates can be different for the two identical firms (in terms of demand
and cost): the quantity-competing firm has a higher pass-through rate
than the price-competing firm has. This is in contrast with the result
that the pass-through rate under price competition is generally higher
under quantity competition.} In addition, we provide a generalization of our results to the case
of changes in both production costs and taxes (Section \ref{SectionPassThroughAndWelfareUnderProductionCostAndTaxationChanges}).

As already mentioned above, it would be possible to extend our analysis
to the case of supply chains (Peitz and Reisinger 2014). Other possible
directions include the case of two-sided platform competition (White
and Weyl 2016; and Tremblay 2018) and the case of the interactive
effects of taxation for multiple imperfectly competitive product markets.\footnote{Among many others, Ballard, Shoven, and Whalley (1985) study this
issue for perfectly competitive markets.} In addition, our methodology could be utilized to study other important
issues of pricing in general such as the welfare effects of oligopolistic
third-degree price discrimination (Adachi and Fabinger 2018). One
may also study, for example, advertising pass-through (Draganska and
Vitorino 2017).\footnote{The firm's demand can be modeled as $q_{j}=q_{j}(p_{1},...,p_{n};a_{1},...,a{}_{n})$,
where $a_{j}$ is firm $j$'s investment in advertising.} Free-riding, because of the spillover effect, may be more or less
serious, depending on the conduct index and other related indices.
Furthermore, it would be of interest to develop flexible, but analytically
solvable examples along the lines of Fabinger and Weyl (2018).

\appendix

\section{Appendix}

\subsection{Proofs and discussions for Section 2}

\subsubsection{Proof of Proposition \ref{PropositionMCt}}

\label{AppendixProofOfPropositionForMCtAndMCv}Using Equation (\ref{theta_M})
to substitute for $mc$, we first obtain a useful expression for the
markup: $p-mc=t+pv+p(1-v)\eta\theta$. Now consider an infinitesimal
change $dt$ in the unit tax that induces a change $dp$ in the equilibrium
price and a change $dq$ in the equilibrium quantity. These are related
by $dt=dp/\rho_{t}=-\eta\thinspace p\thinspace dq/\left(q\thinspace\rho_{t}\right)$.
The corresponding change in social welfare per firm is $dW=\left(p-mc\right)dq=t\thinspace dq+vp\thinspace dq+(1-v)p\eta\theta\thinspace dq$,
and the change in tax revenue per firm is $dR=\left(t+vp\right)dq+vq\thinspace dp+q\thinspace dt=\left(t+vp\right)dq-vp\eta\thinspace dq-\eta p\thinspace dq/\rho_{t}$.
Combining these relationships gives the result 
\[
MC_{t}=-\frac{dW}{dR}=-\frac{t+vp+(1-v)p\eta\theta}{t+vp-vp\eta-\frac{1}{\rho_{t}}p\eta}=\frac{(1-v)\eta\theta+\frac{t}{p}+v}{\frac{1}{\rho_{t}}\eta+v\eta-\frac{t}{p}-v}=\frac{(1-v)\theta+\epsilon\tau}{\frac{1}{\rho_{t}}+v-\epsilon\tau}.
\]

Next, consider an infinitesimal change $dv$ in the ad valorem tax
that induces a change $dp$ in the equilibrium price and a change
$dq$ in the equilibrium quantity, related by $dv=dp/(p\rho_{v})=-\eta\,dq/(q\rho_{v})$.
The change in social welfare per firm is again $dW=\left(p-mc\right)dq=t\,dq+vp\,dq+(1-v)p\eta\theta\,dq$.
The change in tax revenue per firm can be written as $\left(t+vp\right)dq+vq\,dp+pq\,dv=\left(t+vp\right)dq-vp\eta\,dq-p\eta\,dq/\rho_{v}$.
Combining these relationships leads to the result 
\[
MC_{t}=-\frac{dW}{dR}=-\frac{t+vp+(1-v)p\eta\theta}{t+vp-vp\eta-\frac{1}{\rho_{v}}p\eta}=\frac{(1-v)\eta\theta+\frac{t}{p}+v}{\frac{1}{\rho_{v}}\eta+v\eta-\frac{t}{p}-v}=\frac{(1-v)\theta+\epsilon\tau}{\frac{1}{\rho_{v}}+v-\epsilon\tau}.
\]

\subsubsection{Intuition behind Proposition \ref{PropositionMCt}}

\label{AppendixDiscussionofPropositionForMCtAndMCv}The intuition
behind Proposition \ref{PropositionMCt} for the case of unit taxation
can explained as follows. The argument for ad valorem taxation is
analogous. First, the firm's per-output profit margin is decomposed
into two parts: (1) tax payment, $t+vp=p\tau$\textbf{ }and (2) surplus
from imperfect competition, $(1-v)p\eta\theta$. Under imperfect competition,
the effects of an increase in unit tax, $dt$, on the social welfare
can be written as $dW=(p-mc)dq$, which implies that\textit{ the firm's
per-output profit margin serves as a measure for welfare change}.\footnote{The welfare change $dW=(p-mc)dq$ is further decomposed into:
\[
dW=-qdp+\{pdq+qdp-[qdt+vqdp+(t+vp)dq]-mc\cdot dq\}+[qdt+vpdp+(t+vp)dq]
\]
\[
\quad\ \ =\underset{dCS}{\underbrace{-qdp}}+\underset{dPS}{\{\underbrace{[(1-v)p-t]dq+[(1-v)dp-dt]q-mc\cdot dq}\}+}\underset{dR}{[\underbrace{qdt+vpdp+(t+vp)dq}],\quad\thinspace\thinspace}
\]
where $dPS$ can be further simplified (see below).} On the other hand, the effects of an increase in unit tax, $dt$,
on the tax revenue are:
\[
dR=\underset{(1)>0}{\underbrace{q\,dt}}+\underset{(2)>0}{\underbrace{vq\,dp}}+\underset{(3)<0}{\underbrace{(t+vp)dq}},
\]
where term (1) expresses (direct) gain, multiplied by the output $q$,
and term (2) shows (indirect) gain, due to the associated price increase,
multiplied by $vq$, whereas term (3) is the part that exhibits (indirect)
loss from the output reduction for both unit tax revenue and ad valorem
tax revenue. Now recall that $dp=\rho_{t}dt$ and $p\eta dq=-qdp$.
Thus, $qdt=qdp/\rho_{t}=-(p\eta/\rho_{t})dq$ and $vqdp=-(vqp/q)\eta dq=-(vp\eta)dq$,
which implies that
\[
dR=-(p\eta/\rho)dq-(vp\eta)dq+(t+vp)dq=[\underset{(1)\thinspace>0}{\underbrace{\left(-p\eta/\rho_{t}\right)}}+\underset{(2)\thinspace>0}{\underbrace{(-vp\eta)}}+\underset{(3)\thinspace<0}{\underbrace{(t+vp)}}].
\]
Now, in the per-price term, the denominator and the numerator in $MC_{t}$
are expressed as follows:
\[
MC_{t}=\frac{\underset{\textrm{welfare}\thinspace\textrm{loss}\thinspace\textrm{expressed}\thinspace\textrm{by}\thinspace\textrm{the}\thinspace\textrm{profit}\thinspace\textrm{margin}}{\underbrace{(1-v)\eta\theta+\tau}}}{\underset{\textrm{revenue}\thinspace\textrm{gain}}{\underbrace{\left(\frac{1}{\rho_{t}}+v\right)\eta}}+\underset{\textrm{revenue}\thinspace\textrm{loss}}{\underbrace{(-\tau)}}}.
\]

\subsubsection{Proof of Proposition \ref{PropositionConsumerAndProducerIncidence}}

\label{AppendixProofOfPropositionForItAndIv}The impact of a change
$dt$ in the tax $t$ on consumer surplus (per firm) is $dCS=-qdp=-q\rho_{t}dt$.
The impact on producer surplus is
\[
dPS=d[\left(1-v\right)pq-c\left(q\right)-tq]=-q\,dt+\left(1-v\right)p\,dq+\left(1-v\right)qdp-mc\,dq-t\,dq,
\]
\[
\Leftrightarrow dPS=-qdt+\left(1-v\right)q\rho_{t}dt+[\left(1-v\right)p-mc-t]dq.
\]
Substituting for $mc$ from Equation (\ref{theta_M}) as $mc=\left(1-v\right)\left(1-\eta\theta\right)p-t$
gives
\[
dPS=-qdt+\left(1-v\right)q\rho_{t}dt+\left(1-v\right)\eta\theta pdq=-qdt+\left(1-v\right)q\rho_{t}dt-\left(1-v\right)\theta qdp,
\]
\[
\Leftrightarrow dPS=-qdt+\left(1-v\right)q\rho_{t}dt-\left(1-v\right)\theta q\rho_{t}dt=-[1-\left(1-v\right)\left(1-\theta\right)\rho_{t}]q\,dt.
\]
The reciprocal of the incidence ratio is
\[
\frac{1}{I_{t}}=\frac{dPS}{dCS}=\frac{\left(1-v\right)\left(1-\theta\right)q\rho_{t}-q}{-q\rho_{t}}=\frac{1}{\rho_{t}}-\left(1-v\right)\left(1-\theta\right).
\]

For infinitesimal changes in ad valorem taxes, we proceed analogously.
The change in consumer surplus is $dCS=-qdp=-qp\rho_{v}dv.$ For the
change in producer surplus we have
\[
dPS=d\left(\left(1-v\right)pq-c\left(q\right)-tq\right)=-pq\,dv+\left(1-v\right)p\,dq+\left(1-v\right)qdp-mc\,dq-t\,dq.
\]
Manipulating the last four terms on the right-hand side in the same
way as before leads to
\[
dPS=-pq\,dv+\left(1-v\right)p\,dq+\left(1-v\right)qdp-mc\,dq-t\,dq,
\]
\[
dPS=-pq\,dv+\left(1-v\right)qp\rho_{v}dv-\left(1-v\right)\theta qp\rho_{v}dv=\left[\left(1-v\right)\left(1-\theta\right)\rho_{v}-1\right]qp\,dv.
\]
The reciprocal of the incidence ratio then becomes
\[
\frac{1}{I_{t}}=\frac{dPS}{dCS}=\frac{\left(1-v\right)\left(1-\theta\right)\rho_{v}q-q}{-q\rho_{v}}=\frac{1}{\rho_{v}}-\left(1-v\right)\left(1-\theta\right).
\]

\subsubsection{Intuition behind Proposition \ref{PropositionConsumerAndProducerIncidence}}

\label{AppendixDiscussionOfPropositionOnItAndIv}The intuitive reasoning
behind Proposition \ref{PropositionConsumerAndProducerIncidence}
can be provided as follows. First, the effects of an increase in unit
tax, $dt$, on the producer surplus can be decomposed into the following
five parts:
\[
dPS=\underset{(1)<0}{[\underbrace{(-q\,dt)}}+\underset{(2)<0}{\underbrace{(1-v)p\,dq}}]+\underset{(3)>0}{[\underbrace{(1-v)q\,dp}}+\underset{(4)>0}{\underbrace{(-mc\,dq)}}+\underset{(5)>0}{\underbrace{(-t\,dq)}}],
\]
where term (1) shows the (direct) loss from an increase in unit tax:
the tax increase multiplied by the output $q$, and term (2) is another
(indirect) loss from a reduction in production, multiplied by the
ad valor em tax adjusted unit price $(1-v)p$, whereas term (3) corresponds
to the (direct) gain from the associated price increase, mitigated
by $(1-v)$, due to the ad valorem tax, multiplied by the output $q$,
and finally terms (4) and (5) are (indirect) gains from cost savings
by the output reduction, $dq$, and from unit tax saving by the output
reduction, $dq$, respectively. Note here that the equation above
is rewritten as
\[
dPS=\underset{(1)<0}{[\underbrace{-q\,dt}}+\underset{(3)>0}{\underbrace{(1-v)q\,dp}}]+[(1-v)p-\underset{\textrm{marginal}\thinspace\textrm{cost}}{\underbrace{(mc+t)}}]dq.
\]
Now, in symmetric equilibrium, the marginal cost, $mc+t$, is equal
to the marginal benefit, $(1-v)p[1-\eta\theta]$, which implies
\[
dPS=\underset{(1)<0}{[\underbrace{-q\,dt}}+\underset{(3)>0}{\underbrace{(1-v)q\,dp}}]+[(1-v)p]\eta\theta\,dq.
\]
Under perfect competition, part (2) is equal to the sum of parts (4)
and (5), and thus only parts (1) and (3) survive. However, under imperfect
competition, the marginal cost is less than $(1-v)p$, thus part (2)
is greater than the sum of parts (4) and (5). The third term in the
equation above now expresses the difference between part (2) and the
sum of parts (4) and (5). Now, recall that $dp=\rho_{t}dt$ and $p\eta dq=-qdp$.
Thus, 
\[
dPS=[-q\,dt+(1-v)q\rho_{t}\,dt]-(1-v)q\theta\,dp=[-q\,dt+(1-v)q\rho_{t}\,dt]-(1-v)q\theta\rho_{t}\,dt
\]
\[
=[-1+(1-v)\rho_{t}-(1-v)\theta\rho_{t}]q\,dt=\underset{(1)<0}{[\underbrace{-1}}+\underset{(3)-\{(2)-[(4)+(5)]\}\thinspace\gtrless0}{\underbrace{(1-v)(1-\theta)\rho_{t}}}]q\,dt.\qquad
\]
On the other hand, $dCS=-\rho_{t}(qdt)$. Thus, while it is always
the case that $dCS<0$, it is possible that $dPS>0$.\footnote{One can also define $\mathit{social}$ $\mathit{incidence}$ by $SI_{t}$$\equiv dW/dPS$
and $SI_{v}$ in association with a small change in $t$ and $v$,
respectively. Hereafter, we focus on $MC_{t}$ and $MC_{v}$ as measures
of welfare burden in society, and $I_{t}$ and $I_{v}$ as measures
of loss in consumer welfare. We provide general formulas for social
incidence in the context of multi-dimensional pass-through after Section
\ref{SectionMultiDimensionalPassThroughFramework}.} 

\subsubsection{Proof of Proposition \ref{PropositionRelationshipBetweenPassThroughs}}

\label{AppendixProofOfPropositionRelatingRhotAndRhov}Let us consider
a simultaneous infinitesimal change $dt$ and $dv$ in the taxes $t$
and $v$ that leaves the equilibrium price (and quantity) unchanged,
which requires the effective marginal cost $\left(t+mc\right)/\left(1-v\right)$
in Equation (\ref{theta_M}) to remain the same. This implies the
comparative statics relationship 
\[
\frac{\partial}{\partial t}\,\left(\dfrac{t+mc}{1-v}\right)dt+\frac{\partial}{\partial v}\,\left(\dfrac{t+mc}{1-v}\right)dv=0\Rightarrow\dfrac{dt}{1-v}+\dfrac{t+mc}{\left(1-v\right)^{2}}\,dv=0\Rightarrow dt=-\dfrac{t+mc}{1-v}\,dv.
\]
Note that here we do not need to take derivatives of $mc$ even though
it depends on $q$, simply because by assumption the quantity is unchanged.
The total induced change in price, which generally would be expressed
as $dp=\rho_{t}dt+\rho_{v}p\,dv$, must equal zero in this case, implying
the result 
\[
\rho_{t}dt+\rho_{v}p\,dv\!=0\Rightarrow-\dfrac{t+mc}{1-v}\rho_{t}dv+\rho_{v}p\,dv\!=0\Rightarrow\rho_{v}\!=\left(1-\eta\theta\right)\rho_{t}\Rightarrow\rho_{v}=\frac{\epsilon-\theta}{\epsilon}\rho_{t}.
\]

\subsubsection{Intuition behind Proposition \ref{PropositionRelationshipBetweenPassThroughs}}

\label{AppendixDiscussionOfPropositionRelatingPassThroughs}To understand
this proposition (\ref{PropositionRelationshipBetweenPassThroughs})
intuitively, note that to keep prices and quantities constant, $\varDelta t$
and $\varDelta v$ must satisfy:
\[
\frac{t+\varDelta t+mc}{1-(v+\varDelta v)}=\frac{t+mc}{1-v}.
\]
Thus, the relative $\varDelta t$ that must be offset by a reduction
$-\varDelta v$ equal to $(t+mc)/(1-v)$: $\varDelta t=-(t+mc)\varDelta v/(1-v)$,
which, together with $\rho_{t}dt+\rho_{v}p\,dv=0$, leads to $(t+mc)\rho_{t}/[(1-v)p]=\rho_{v}$.
Now, recall the Lerner rule:
\[
1-\underset{\textrm{per\textrm{-}price}\thinspace\textrm{marginal}\thinspace\textrm{cost}}{\underbrace{\dfrac{t+mc}{(1-v)p}}}=\eta\theta,
\]
which implies that $(1-\eta\theta)\rho_{t}=\rho_{v}$, as Proposition
\ref{PropositionRelationshipBetweenPassThroughs} claims. Now, \textbf{$\theta/\epsilon=1-\rho_{v}/\rho_{t}$}
implies that \textbf{$\rho_{v}\leq\rho_{t}\leq\left(1-1/\epsilon\right)\rho_{v}.$} 

\subsubsection{Intuition behind Proposition \ref{Sufficient-statistics-for}}

\label{AppendixDiscussionOfPropositionOnSufficientStatistics}To gain
a perspective in Proposition \ref{Sufficient-statistics-for}, recall
from Proposition \ref{PropositionMCt} that
\[
MC_{t}=\frac{\underset{\textrm{welfare}\thinspace\textrm{loss}\thinspace\textrm{expressed}\thinspace\textrm{by}\thinspace\textrm{the}\thinspace\textrm{profit}\thinspace\textrm{margin}}{\underbrace{(1-v)\eta\theta+\tau}}}{\underset{\textrm{revenue}\thinspace\textrm{gain}}{\underbrace{\left(\frac{1}{\rho_{t}}+v\right)\eta}}+\underset{\textrm{revenue}\thinspace\textrm{loss}}{\underbrace{(-\tau)}}}.
\]
Now, Proposition \ref{Sufficient-statistics-for} states that it is
also understood as
\[
MC_{t}=\frac{\underset{\textrm{welfare}\thinspace loss\thinspace\textrm{expressed}\thinspace\textrm{by}\thinspace\textrm{the}\thinspace\textrm{profit}\thinspace\textrm{margin}}{\underbrace{(1-v)\left(1-\frac{\rho_{v}}{\rho_{t}}\right)+\tau}}}{\underset{\textrm{revenue}\thinspace\textrm{gain}}{\underbrace{\left(\frac{1}{\rho_{t}}+v\right)\eta}}+\underset{\textrm{revenue}\thinspace\textrm{loss}}{\underbrace{(-\tau)}}}.
\]

Of course, it is true that $\theta$ is expressed by the empirical
measures such as $\theta=(1-\rho_{v}/\rho_{t})\epsilon$. For example,
in the case of the assumption of Cournot competition, researchers
often may observe the number $n$ of firms and conclude that the value
of conduct index is $\theta=1/n$.\textbf{ }However, even in the case
of homogeneous products, the \textquotedblleft true\textquotedblright{}
conduct may be higher than\textbf{ }$1/n$ due to such reasons as
collusion.\footnote{See Miller and Weinberg (2017) for an empirical study of the possibility
of oligopolistic collusion in a different manner from directly estimating
the conduct parameter.} Proposition \ref{Sufficient-statistics-for} above circumvents this
difficulty in estimating $MC_{t}$ and $MC_{v}$.\footnote{Similarly, the incidence of a unit tax is expressed as
\[
\frac{1}{I_{t}}=\frac{1}{\rho_{t}}-\left(1-v\right)\left[(1-\epsilon)+\frac{\rho_{v}}{\rho_{t}}\epsilon\right],
\]
and analogously for the case of an ad valorem tax.} Conversely, one would be able to estimate $\theta$ using the proposition
above once $\epsilon$, $\rho_{t}$, and $\rho_{v}$ are estimated. 

\subsubsection{Proof of Proposition \ref{PropositionGeneralSymmetricOligopolyPassThrough}}

\label{AppendixProofOfPropositionGeneralOligopolyPassThrough}Here
we provide a proof of Proposition \ref{PropositionGeneralSymmetricOligopolyPassThrough},
as well as related intuitive arguments. Consider the comparative statics
with respect to a small change $dt$ in the per-unit tax $t$. Following
Weyl and Fabinger (2013, p.$\,$538), we define $ms\equiv-p'q$: this
is the negative of marginal consumer surplus. Then, the Learner condition
becomes:
\[
\underset{\textrm{markup}}{\underbrace{p-\frac{t+mc}{1-v}}}=\underset{\textrm{CS}}{\underbrace{\theta\thinspace ms}},
\]
where $\mathit{CS}$ is consumer surplus for the infra-marginal consumers.
Importantly, $\theta ms$ measures how much consumer surplus rises
for a small increase in output, and it is largest under monopoly.
Now consider a small change in unit tax expressed by $dt>0$. Then,
in equilibrium, 
\[
dp-\frac{dt+dmc}{1-v}=d(\theta\thinspace ms)\qquad\Leftrightarrow\qquad\underset{\textrm{change}\thinspace\textrm{in}\thinspace\textrm{marginal}\thinspace\textrm{benefit}}{\underbrace{(1-v)[\underset{>0}{\underbrace{dp}}-\underset{<0}{\underbrace{d(\theta\,ms)}}]}}=\underset{\textrm{change}\thinspace\textrm{in}\thinspace\textrm{effective}\thinspace\textrm{marginal}\thinspace\textrm{cost}}{\,\,\,\,\,\,\,\,\underbrace{\underset{>0}{\underbrace{dt}}+\underset{<0}{\underbrace{dmc}}}}
\]
Thus, using $dt=dp/\rho_{t},$ the equation is rewritten as

\[
\rho_{t}=\frac{1}{\underset{(1)>0:\thinspace\textrm{revenue}\thinspace\textrm{increase}}{\underbrace{\left(1-v\right)[dp+(-d\left(\theta\,ms\right))]}}+\underset{(2)>0:\thinspace\textrm{cost}\thinspace\textrm{savings}}{\underbrace{(-dmc)}}}dp.
\]

Now, consider term (1). Note first $d(\theta ms)=(\theta ms)'dq$
so that $d(\theta ms)=-q\epsilon(\theta ms)'dp/p$ because by definition
$dq=-q\epsilon dp/p$. Here, for a small increase $dt>0$,
\[
\underset{<0}{\underbrace{d(\theta\,ms)}}=\underset{>0}{\underbrace{-q\epsilon}}(\theta\,ms)'\underset{>0}{\underbrace{\frac{dp}{p}}}
\]
so that $(\theta\,ms)'>0$.\textbf{ }By definition, $ms\equiv-p'q=\eta p$.
Thus, $d(\theta ms)=-q\epsilon(\theta\eta p)'dp/p$. Now note that
$(\theta\eta p)'=(\theta\eta)'p+(\theta\eta)p'$. Thus,
\[
d(\theta\,ms)=-q\epsilon\left[(\theta\eta)'p+(\theta\eta)p'\right]\frac{dp}{p}\qquad\qquad\qquad\qquad\qquad\qquad
\]
\[
\Leftrightarrow d(\theta\,ms)=-q\epsilon(\theta\eta)'dp+(-q\epsilon(\theta\eta)p'dp/p)=[\theta\eta-q\epsilon(\theta\eta)']dp>0.
\]

Next, consider term (2). A change in the marginal cost, $dmc$, is
expressed in terms of $dp$\textbf{ }by $dmc=-[\left(1-v\right)\theta\eta+1-\tau]\chi\epsilon\,dp<0$.
To see this, note first that $dmc=$$\chi mc\cdot(dq/q$)$=-(\chi\epsilon\,mc)\,(dp/p$).
Then, $mc$\textbf{ }in this expression can be eliminated rewriting
$p-\theta\,ms=\left(mc+t\right)/\left(1-v\right)\Rightarrow mc=\left(1-v\right)\left(p+\theta qp'\right)-t=\left(1-v\right)\left(1-\theta\eta\right)p-t$,
which leads to $dmc=-[\left(1-v\right)\left(1+\theta\eta\right)-t/p]\chi\epsilon\,dp$.
Then, in terms of the per-unit revenue burden, $\tau\equiv v+t/p$,
that is, $dmc=-[\left(1-v\right)\left(1-\theta\eta\right)-\tau+v]\chi\epsilon\,dp=-[-\left(1-v\right)\theta\eta+1-\tau]\chi\epsilon\,dp$.
Finally, using the expressions for $dmc$\textbf{ }and $d(\theta ms)$,\textbf{
}
\[
\rho_{t}=\frac{dp}{\left(1-v\right)[dp-d\left(\theta\,ms\right)]-dmc}=\frac{1}{\underset{\textrm{revenue}\thinspace\textrm{increase}}{\underbrace{\left(1-v\right)[(1-\theta\eta)+\left(\theta\eta\right)'\epsilon q]}}+\underset{\textrm{\textrm{cost}\thinspace\textrm{savings}}}{\underbrace{\left(1-\tau\right)\epsilon\chi-\left(1-v\right)\theta\chi}}}.
\]

\[
\Leftrightarrow\rho_{t}=\frac{1}{1-v}\ \frac{1}{\underset{\textrm{revenue}\thinspace\textrm{increase}}{\underbrace{[(1-\theta\eta)+\left(\theta\eta\right)'\epsilon q]}}+\underset{\textrm{cost}\thinspace\textrm{savings}}{\underbrace{\left[-\theta+\frac{1-\tau}{1-v}\epsilon\right]\chi}}}.
\]

\subsubsection{Relationship to Weyl and Fabinger (2013)}

\label{AppendixRelationshipToWeylAndFabinger}It can be verified that
our formula for $\rho_{t}$ above is a generalization of Weyl and
Fabinger's (2013, p.$\,$548) Equation (2):
\[
\rho=\frac{1}{1+\frac{\epsilon_{D}-\theta}{\epsilon_{S}}+\frac{\theta}{\epsilon_{\theta}}+\frac{\theta}{\epsilon_{ms}}},
\]
where $\epsilon_{\theta}\equiv\theta/[q\cdot(\theta)']$, $\epsilon_{ms}\equiv ms/[ms'q]$
($ms\equiv-p'q$ is defined in the proof of Proposition \ref{PropositionGeneralSymmetricOligopolyPassThrough}
just above), and $\epsilon_{D}$ and $\epsilon_{S}$ here are our
$\epsilon$\textbf{ }and $1/\chi$, respectively. First, the denominator
in our formula is rewritten as:
\[
1-\left(\eta+\chi\right)\theta+\epsilon q\left(\theta\eta\right)'+\frac{1-\tau}{1-v}\epsilon\chi=1+\frac{\frac{1-\tau}{1-v}\epsilon_{D}-\theta}{\epsilon_{S}}+\frac{\theta}{\epsilon_{\theta}}+\theta\cdot\left(-\frac{1}{\epsilon_{D}}+\eta'\epsilon_{D}q\right)
\]
because
\[
(\theta\eta)'\epsilon q=(\theta'\eta+\theta\eta')\epsilon q=\left[\frac{\theta}{q\epsilon_{\theta}}\eta+\theta\eta'\right]\epsilon q=\frac{\theta}{\epsilon_{\theta}}+\theta\eta'\epsilon q.
\]

Next, since $\eta=-qp'/p$, it is verified that $\eta'=-\{p'p+qpp''-q[p']^{2}\}/p^{2}$,
implying that
\[
\eta'\epsilon_{D}q=\frac{p'p+qpp''-q[p']^{2}}{p^{2}}\cdot\frac{p}{p'q}\cdot q=\frac{1}{\epsilon_{D}}+\left(1+\frac{p''}{p'}q\right),
\]
where $1+p''q/p$ is replaced by $1/\epsilon_{ms}$ because $ms\equiv-p'q$
and thus $ms'=-(p''q+p')$. Then, it is readily verified that 
\[
1-\left(\eta+\chi\right)\theta+\epsilon q\left(\theta\eta\right)'+\frac{1-\tau}{1-v}\epsilon\chi=1+\frac{\frac{1-\tau}{1-v}\epsilon_{D}-\theta}{\epsilon_{S}}+\frac{\theta}{\epsilon_{\theta}}+\frac{\theta}{\epsilon_{ms}}.
\]

In summary, Weyl and Fabinger's (2013, p.$\,$548) original Equation
(2) is generalized to
\[
\rho=\frac{1}{1-v}\ \frac{1}{1+\frac{\frac{1-\tau}{1-v}\epsilon_{D}-\theta}{\epsilon_{S}}+\frac{\theta}{\epsilon_{\theta}}+\frac{\theta}{\epsilon_{ms}}}
\]
with non-zero initial ad valorem tax, which is equivalent to our formula
for $\rho_{t}$:
\[
\rho_{t}=\frac{1}{1-v}\ \frac{1}{1+\frac{1-\tau}{1-v}\epsilon\chi-\left(\eta+\chi\right)\theta+\epsilon q\left(\theta\eta\right)'}.
\]

\subsubsection{Comparison of perfect and oligopolistic competition}

\label{AppendixComparisonOfPerfectAndOligopolisticCompetition}One
can further interpret the formula for $\rho_{t}$ in comparison to
the case of perfect competition (with zero initial taxes), when the
unit tax pass-through rate is given by (see Weyl and Fabinger 2013,
p.$\,$534): $\rho_{t}=1/(1+\epsilon\chi)$. An analogous argument
can be made for $\rho_{v}$ as well.

First, through the term $-\left(\eta+\chi\right)\theta$ in the denominator
of $\rho_{t}$ in Proposition \ref{PropositionGeneralSymmetricOligopolyPassThrough},
as competitiveness becomes fiercer (i.e., a lower $\theta$) the pass-through
rate $\rho_{t}$\textit{ lowers}, that is, the pass-through becomes\textit{
smaller} as the degree of competition becomes closer to perfect competition.
This is interpreted as the negative effect of competitiveness on the
pass-through rate, via the first-order characteristics of demand and
supply, captured by $\eta$ and $\chi$, respectively.\footnote{Here, with fixed $\theta$, the denominator becomes smaller, and thus,
the pass-through rate becomes\textit{ larger} as the demand becomes\textit{
inelastic} (i.e., $\eta$ becomes larger, although $\eta$ cannot
be too large; recall the restriction, $\eta<1/\theta$) or the supply
becomes\textit{ inelastic} (i.e., $\chi$ becomes larger).} 

However, through the other term $\epsilon q\left(\theta\eta\right)'=-\epsilon q\left(-\theta\eta\right)'$,
competitiveness \textit{raises }the pass-through rate $\rho_{t}$.
To see this, suppose that $\eta$ is close to a constant. Then, $-\epsilon q\left(-\theta\eta\right)'=-q\left(-\theta\right)'$,
which implies that a larger $\left(-\theta\right)'\equiv-\partial\theta/\partial q>0$
is associated with a\textit{ higher} value of the pass-through rate.
With an abuse of notation, this situation is interpreted as the case
when $-\partial q/\partial\theta$ is small: the effect of imperfect
competition on the output reduction is small, implying\textit{ less
distortion}, an important feature if the degree of competition is
close to perfect competition. 

The argument so far is clearer if $\theta$, as often assumed, is
a constant. Then, the second term is now $-\epsilon q\left(-\theta\eta\right)'=-\epsilon q\theta\left(-\eta\right)'=\theta(\eta+1/\epsilon_{ms})$
so that
\[
\rho_{t}=\frac{1}{1-v}\ \frac{1}{\left[1+\frac{1-\tau}{1-v}\epsilon\chi\right]+\left(\frac{1}{\epsilon_{ms}}-\chi\right)\theta}.
\]
Thus, if the marginal cost is constant ($\chi=0$), $\rho_{t}$ becomes
\textit{larger} as the degree of competition becomes closer to perfect
competition. Here, with fixed $\theta$, the pass-through rate is
also\textit{ larger} as $1/\epsilon_{ms}$ becomes smaller. Recall
that $1/\epsilon_{ms}=(\varDelta ms/ms)/(\varDelta q/q)$ measures
how quick the marginal surplus lowers as a response to a decrease
in output $q$. Thus, a lower $1/\epsilon_{ms}$ is associated with
\textit{less distortion}. Overall, Weyl and Fabinger's (2013, p.\,548)
Equation (2) and our formula for $\rho_{t}$ show how it is influenced
by the industry's competitiveness which is captured by the conduct
index.

\subsubsection{Application to exchange rate changes}

\label{AppendixApplicationToExchangeRateChanges}Let us also point
out that the \textit{exchange rate pass-through} can be included naturally
in our framework of Section \ref{SectionTaxationAndWelfareInSymmetricallyDifferentiatedOligopoly}.\footnote{\label{FootnoteExchangeRatePassThroughLiterature}See, e.g., Feenstra
(1989); Feenstra, Gagnon, and Knetter (1996); Yang (1997); Campa and
Goldberg (2005); Hellerstein (2008); Gopinath, Itskhoki, and Rigobon
(2010); Goldberg and Hellerstein (2013); Auer and Schoenle (2016);
and Chen and Juvenal (2016) for empirical studies of exchange rate
pass-through.} Suppose that domestic firms in a country of interest use some imported
inputs for production. For concreteness, let us specify the profit
function of firm $j$ as $\pi_{j}=[(1-v)p_{j}-t]q_{j}-(1+a\,e)c(q_{j})$,
where the constant coefficient $a$ measures the importance imported
inputs and $e>0$ is the exchange rate. Notice that the firm's profit
is rewritten as $\pi_{j}=(1+ae)\left[\left(\!\!\begin{array}{c}
\frac{1-v}{1+ae}\end{array}\!\!p_{j}-\!\!\begin{array}{c}
\frac{t}{1+ae}\end{array}\!\!\right)q_{j}-c(q_{j})\right]$. Since the first factor on the right-hand side is constant, the firm
will behave as if its profit function was simply $\tilde{\pi}_{j}=\left[(1-\tilde{v})p_{j}-\tilde{t}\,\right]q_{j}-c(q_{j})$,
with $\tilde{v}\equiv(v+ae)/(1+ae)$ and $\tilde{t}\equiv t/(1+ae)$.
By utilizing the explicit expressions for the derivatives $\partial\tilde{v}/\partial e=(a-v)/(1+ae)^{2}$
and $\partial\tilde{t}/\partial e=-at/(1+ae)^{2}$, one can analyze
the effect of a change in the exchange rate $e$ on social welfare.
Note that this is simply interpreted as the \textit{cost pass-through}
as well (see the references in Footnote \ref{FootnoteExchangeRatePassThroughLiterature}
for empirical studies). Alternatively, one may use the results of
Section \ref{SectionPassThroughAndWelfareUnderProductionCostAndTaxationChanges}
to study the consequences of exchange rate movements.

\subsubsection{Oligopoly with multi-product firms}

\label{AppendixOligopolyWithMultiProductFirms}Here, we argue that
the results obtained in Sections \ref{SectionTaxationAndWelfareInSymmetricallyDifferentiatedOligopoly}
and \ref{SectionExpressionsForPassThrough} can be extended to the
case of multi-product firms just by a reinterpretation of the same
formulas (without modifying them).\footnote{Lapan and Hennessy (2011) study unit and ad valorem taxes in multi-product
Cournot oligopoly. Alexandrov and Bedre-Defolie (2017) also study
cost pass-through of multi-product firms in relation to the Le Chatelier\textendash Samuelson
principle.} Assume there are $n_{p}$ product categories, and the demand for
firm $j$'s $k$-th product is given by $q_{jk}=q_{jk}(\mathbf{p}_{1},\mathbf{p}_{2},..,\mathbf{p}_{n})$,
where $\mathbf{p}_{j}=(p_{j1},...,p_{jk},...,p_{jK})$ for each $j=1,2,...,n$.\footnote{See, e.g., Armstrong and Vickers (2018) and Nocke and Schutz (2018)
for recent studies of multi-product oligopoly.} The firms are symmetric, and for each firm, the product it produces
are also symmetric. The firm's profit per product is 
\[
\pi_{j}=\frac{1}{n_{p}}\sum_{k=1}^{n_{p}}\left(\left(1-v\right)p_{jk}q_{jk}-tq_{jk}-c(q_{jk})\right).
\]
We work with an equilibrium in which any firm $j$ sets a uniform
price $p_{j}$ for all of its products: $p_{jk}=p_{j}$, and consequently
sells an amount $q_{j}$ of each of them: $q_{jk}=q_{j}$.\footnote{For brevity, we do not explicitly discuss the standard conditions
for the existence and uniqueness of non-cooperative Nash equilibria
of the different underlying oligopoly games.} In this case, the profit per product equals $\pi_{j}=\left(1-v\right)p_{j}q_{j}-tq_{j}-c(q_{j}),$
which is formally the same as for single-product firms. For this reason,
we can identify the prices $p_{j}$ and quantities $q_{j}$ of Section
\ref{SectionTaxationAndWelfareInSymmetricallyDifferentiatedOligopoly}
with the prices $p_{j}$ and quantities $q_{j}$ introduced here in
this paragraph. The discussion in Section \ref{SectionTaxationAndWelfareInSymmetricallyDifferentiatedOligopoly}
was general and applies to this case of symmetric oligopoly with multi-product
firms as well. We can use the same definitions for the variables of
interest, including the industry demand elasticity $\epsilon$ and
the conduct index $\theta$.

The definitions and results for the cases of price competition and
quantity competition discussed Section \ref{SectionExpressionsForPassThrough}
are also applicable here. It may be useful to translate some of the
most important variables of that discussion into product-level variables.
For derivatives of the direct demand system, we introduce the notation:\footnote{In this notation, the first subscript counts the derivatives with
respect to the relevant price with index $k$, the second subscript
counts the derivatives with respect to the price with index $k'$
distinct from $k$, and the third subscript counts derivatives respect
to the price with index $k''$ distinct from both $k$ and $k'$.
Further, $\xi$ corresponds to derivatives with respect to prices
charged by the same firm $j$, while $\tilde{\xi}$ corresponds to
derivatives with respect to prices charged by firm $j$ and some other
firm $j'$.} 
\[
\begin{array}{cccc}
 & \xi_{1}\equiv\frac{\partial q_{jk}}{\partial p_{jk}}, & \xi_{0,1}\equiv\frac{\partial q_{jk}}{\partial p_{jk'}},\\
\xi_{2}\equiv\frac{\partial q_{jk}}{\partial p_{jk}^{2}}, & \xi_{1,1}\equiv\frac{\partial q_{jk}}{\partial p_{jk}\partial p_{jk'}}, & \xi_{0,2}\equiv\frac{\partial q_{jk}}{\partial p_{jk'}^{2}}, & \xi_{0,1,1}\equiv\frac{\partial q_{jk}}{\partial p_{jk'}\partial p_{jk''}},\\
\tilde{\xi}_{2}\equiv\frac{\partial q_{jk}}{\partial p_{jk}\partial p_{j'k}} & \tilde{\xi}_{1,1}\equiv\frac{\partial q_{jk}}{\partial p_{jk}\partial p_{j'k'}}, & \tilde{\xi}_{0,2}\equiv\frac{\partial q_{jk}}{\partial p_{jk'}\partial p_{j'k'}}, & \tilde{\xi}_{0,1,1}\equiv\frac{\partial q_{jk}}{\partial p_{jk'}\partial p_{j'k''}},
\end{array}
\]
where the derivatives are evaluated at the fully symmetric point,
where any $p_{jk}$ equals the common value $p$. For specific choices
of the demand system, these derivatives can be closely related. For
example, if the substitution pattern between two goods produced by
two different firms does not depend on the identity of the goods,
then $\tilde{\xi}_{2}=\tilde{\xi}_{0,2}=\tilde{\xi}_{1,1}=\tilde{\xi}_{0,1,1}$.
In terms of these derivatives, we can write 
\[
\begin{array}{c}
\epsilon_{F}=-\frac{p}{q}\left(\xi_{1}+\left(n_{p}-1\right)\xi_{0,1}\right),\\
\epsilon=-\frac{p}{q}\left(\xi_{1}+\left(n_{p}-1\right)\xi_{0,1}+\left(n-1\right)\tilde{\xi}_{1}+\left(n-1\right)\left(n_{p}-1\right)\tilde{\xi}_{0,1}\right),\\
\alpha_{F}=\frac{p^{2}}{q\,\epsilon_{F}}\left(\xi_{2}+\left(n_{p}-1\right)\left(\xi_{1,1}+\xi_{0,2}+\left(n_{p}-2\right)\xi_{0,1,1}\right)\right),\\
\alpha_{C}=\left(n-1\right)\frac{p^{2}}{q\,\epsilon_{F}}\left(\tilde{\xi}_{2}+\left(n_{p}-1\right)(\tilde{\xi}_{1,1}+\tilde{\xi}_{0,2}+\left(n_{p}-2\right)\tilde{\xi}_{0,1,1})\right).
\end{array}
\]
These can be substituted into the results of Proposition \ref{PropositionPassThroughAndMCUnderPriceCompetition}
to find the pass-through and the marginal cost of public funds under
price competition.

For the inverse demand system the analogous definitions are 
\[
\begin{array}{cccc}
 & \zeta_{1}\equiv\frac{\partial q_{jk}}{\partial p_{jk}}, & \zeta_{0,1}\equiv\frac{\partial q_{jk}}{\partial p_{jk'}},\\
\zeta_{2}\equiv\frac{\partial q_{jk}}{\partial p_{jk}^{2}}, & \zeta_{1,1}\equiv\frac{\partial q_{jk}}{\partial p_{jk}\partial p_{jk'}}, & \zeta_{0,2}\equiv\frac{\partial q_{jk}}{\partial p_{jk'}^{2}}, & \zeta_{0,1,1}\equiv\frac{\partial q_{jk}}{\partial p_{jk'}\partial p_{jk''}},\\
\tilde{\zeta}_{2}\equiv\frac{\partial q_{jk}}{\partial p_{jk}\partial p_{j'k}} & \tilde{\zeta}_{1,1}\equiv\frac{\partial q_{jk}}{\partial p_{jk}\partial p_{j'k'}}, & \tilde{\zeta}_{0,2}\equiv\frac{\partial q_{jk}}{\partial p_{jk'}\partial p_{j'k'}}, & \tilde{\zeta}_{0,1,1}\equiv\frac{\partial q_{jk}}{\partial p_{jk'}\partial p_{j'k''}}.
\end{array}
\]
The relations 
\[
\begin{array}{c}
\eta_{F}=-\frac{q}{p}\left(\zeta_{1}+\left(n_{p}-1\right)\zeta_{0,1}\right),\\
\eta=-\frac{q}{p}\left(\zeta_{1}+\left(n_{p}-1\right)\zeta_{0,1}+\left(n-1\right)\tilde{\zeta}_{1}+\left(n-1\right)\left(n_{p}-1\right)\tilde{\zeta}_{0,1}\right),\\
\sigma_{F}=\frac{q^{2}}{p\,\eta_{F}}\left(\zeta_{2}+\left(n_{p}-1\right)\left(\zeta_{1,1}+\zeta_{0,2}+\left(n_{p}-2\right)\zeta_{0,1,1}\right)\right),\\
\sigma_{C}=\left(n-1\right)\frac{q^{2}}{p\,\eta_{F}}\left(\tilde{\zeta}_{2}+\left(n_{p}-1\right)(\tilde{\zeta}_{1,1}+\tilde{\zeta}_{0,2}+\left(n_{p}-2\right)\tilde{\zeta}_{0,1,1})\right).
\end{array}
\]
can be substituted into the results of Proposition \ref{PropositionPassThroughAndMCUnderQuantityCompetition}
to find the pass-through and marginal cost of public funds under price
competition.

\subsection{Proofs and discussions for Section 3}

\subsubsection{Relationship between elasticities and curvatures under the direct
demand system \label{subsec: DirectElasticitiesCurvature}}

This relationship can be verified as follows. The elasticity of the
function $\epsilon_{F}(p)$ equals the sum of the elasticities of
the three factors it is composed of:
\[
\frac{1}{\epsilon_{F}(p)}p\frac{d}{dp}\epsilon_{F}(p)=\frac{1}{p}p\frac{d}{dp}p+q\left(p\right)p\frac{d}{dp}\frac{1}{q\left(p\right)}+\left(\frac{\partial q_{j}(\mathbf{p})}{\partial p_{j}}\right)^{-1}|_{\mathbf{p}=\left(p,...,p\right)}p\frac{d}{dp}\left(\frac{\partial q_{j}(\mathbf{p})}{\partial p_{j}}|_{\mathbf{p}=\left(p,...,p\right)}\right).
\]
The first elasticity on the right-hand side equals 1, the second elasticity
equals $\epsilon\left(p\right)$, and the third elasticity equals
$-\alpha_{F}\left(p\right)-\alpha_{C}$$\left(p\right)$, since 
\[
p\frac{d}{dp}\frac{\partial q_{j}(\mathbf{p})}{\partial p_{j}}|_{\mathbf{p}=\left(p,...,p\right)}=p\frac{\partial^{2}q_{j}(\mathbf{p})}{\partial p_{j}^{2}}|_{\mathbf{p}=\left(p,...,p\right)}+\left(n-1\right)p\frac{\partial^{2}q_{j}(\mathbf{p})}{\partial p_{j}\partial p_{j'}}|_{\mathbf{p}=\left(p,...,p\right)}.
\]

Note that $\alpha$ is\textit{ weakly positive} (\textit{weakly negative})
\textit{if the industry demand is convex} (\textit{concave}), and\textit{
$\alpha_{F}$ }is weakly positive (weakly negative) if the demand
as a function of firm\textbf{ }\textit{$j$}'s own price is convex
(concave). Hence, both $\alpha$ and\textit{ $\alpha_{F}$ }measure
the degree of convexity in the demand function for an industry-wide
price change and for an individual firm's price change, respectively.
Note also that $\partial(\partial q_{j}/\partial p_{j})/\partial p_{j'}$
in $\alpha_{C}$ measures the effects of firm $j$'s price change
on how many consumers rival\textit{ $j'$}\textbf{ }loses if it raises
its price. If this is negative (positive), then firm\textbf{ }\textit{$j'$}\textbf{
}loses more (less) consumers by its own price increase for a higher
value of $p_{j}$. Thus, because $\partial q_{j}/\partial p_{j'}$
is\textit{ positive} in the expression for $\alpha_{C}$, a higher
$\alpha_{C}$ also indicates more competitiveness in the industry.
It is also expected that the industry is more competitive if $\alpha$
and\textit{ $\alpha_{F}$} are higher. In effect, the equilibrium
price is characterized by $\epsilon_{F}$. However, a policy change
around equilibrium is also affected by the curvatures, which measure
\textquotedblleft second-order competitiveness\textquotedblright{}
around the equilibrium. However, Proposition \ref{PropositionPassThroughAndMCUnderPriceCompetition}
in the text shows that $\alpha$ is the only curvature that determines
the pass-through rates.

\subsubsection{Relationship between elasticities and curvatures under the inverse
demand system\label{subsec: IndirectElasticitiesCurvatures}}

In analogy with Appendix \ref{subsec: DirectElasticitiesCurvature},
the elasticity of the function $\eta_{F}\left(q\right)$ is the sum
of the elasticities of the three factors it is composed of, which
are equal to $1$, $\eta\left(q\right)$, and $-\sigma_{F}\left(q\right)-\sigma_{C}\left(q\right)$.

Now, $\sigma$ is\textit{ weakly positive} (\textit{weakly negative})
if the industry's inverse demand is\textit{ convex} (\textit{concave}),
and $\sigma_{F}$ is weakly positive (weakly negative) if the inverse
demand as a function of firm\textbf{ }$j$'s own output is convex
(concave). Here, concavity, not convexity, is related to a sharp reduction
in price in response to an increase in firm\textbf{ }$j$'s output.
Thus, $-\sigma$ and $-\sigma_{F}$ measure \textquotedblleft second-order
competitiveness\textquotedblright{} of the industry, which characterizes
the responsiveness of the equilibrium output when a policy is changed.
Note also that $\partial(\partial p_{j}/\partial q_{j})/\partial q_{j'}$
in $\sigma_{C}$ measures the effects of firm $j$'s output increase
on the extent of rival ($j'$)'s price drop if it increases its output.
If this is negative (positive), then firm $j'$\textbf{ }expects a
large (little) drop in its price by increasing its output for a higher
value of\textbf{ }$q_{j}$. Because $\partial p_{j}/\partial q_{j'}$
is negative in the expression for $\sigma_{C}$, a \textit{lower}
$\sigma_{C}$ or a higher\textbf{ }$-\sigma_{C}$\textbf{ }indicates
more competitiveness in the industry. In sum, while\textbf{ }$1/\eta_{F}$
characterizes competitiveness that determines the level of the equilibrium
quantity, $-\sigma$, $-\sigma_{F}$, and $-\sigma_{C}$ determine
competitiveness that characterizes the responsiveness of the equilibrium
output by a policy change. However, similarly to the case of price
competition, Proposition \ref{PropositionPassThroughAndMCUnderQuantityCompetition}
in the text shows that $\sigma$ is the only curvature that determines
the pass-through rates.

\subsubsection{Proof of Proposition \ref{PropositionPassThroughAndMCUnderPriceCompetition}}

\label{AppendixProofOfPassThroughFormulaUnderPriceCompetition}Since
in the case of price setting $\theta=\epsilon/\epsilon_{F}=1/(\eta\epsilon_{F})$,
we have $\left(\eta+\chi\right)\theta=\left(1+\epsilon\chi\right)/\epsilon_{F}$
and $\left(\theta\eta\right)'\epsilon q=\epsilon q\frac{d}{dq}\left(\theta\eta\right)=\epsilon q\frac{d}{dq}(\epsilon_{F}^{-1})=-\epsilon_{F}^{-2}\epsilon q\frac{d}{dq}\epsilon_{F}=\epsilon_{F}^{-2}p\frac{d}{dp}\epsilon_{F}=\left(1+\epsilon-\alpha\epsilon/\epsilon_{F}\right)/\epsilon_{F}$,
where in the last equality we utilize the expression for the elasticity
of $\epsilon_{F}\left(p\right)$ and $\alpha_{F}+\alpha_{C}=\alpha\epsilon/\epsilon_{F}$
from Subsection \ref{SubsectionElasticitiesAndCurvatures}. Substituting
these into the expression for $\rho_{t}$ in Proposition \ref{PropositionGeneralSymmetricOligopolyPassThrough}
gives 
\[
\rho_{t}=\frac{1}{1-v}\ \frac{1}{1-\frac{1}{\epsilon_{F}}(1+\epsilon\chi)+\frac{1}{\epsilon_{F}}\left(1+\epsilon-\frac{\alpha\epsilon}{\epsilon_{F}}\right)+\frac{1-\tau}{1-v}\epsilon\chi},
\]
which is equivalent to the expression for $\rho_{t}$ in the proposition.
Since for price setting $\theta=\epsilon/\epsilon_{F}$, the relationship
in Proposition \ref{PropositionRelationshipBetweenPassThroughs} implies
$\rho_{v}=\left(\epsilon-\theta\right)\rho_{t}/\epsilon=(\epsilon_{F}-1)\rho_{t}/\epsilon_{F}$,
which leads to the desired expression for $\rho_{v}$. 

\subsubsection{Intuition behind Proposition \ref{PropositionPassThroughAndMCUnderPriceCompetition}}

\label{AppendixDiscussionOfPassThroughFormulaUnderPriceCompetition}The
intuition for $\rho_{t}$ in Proposition \ref{PropositionPassThroughAndMCUnderPriceCompetition}
is as follows. First, recall from Proposition \ref{PropositionGeneralSymmetricOligopolyPassThrough}
that
\[
\rho_{t}=\frac{1}{1-v}\ \frac{1}{\underset{\textrm{revenue}\thinspace\textrm{increase}}{\underbrace{[(1\!-\!\theta\eta)\!+\!\left(\theta\eta\right)'\epsilon q]}}+\underset{\textrm{cost}\thinspace\textrm{savings}}{\underbrace{\left[\frac{1-\tau}{1-v}\epsilon-\theta\right]\chi}}}.
\]
Then, with $\theta=\epsilon/\epsilon_{F}$, $1-\theta\eta=1-1/\epsilon_{F}$,
$\left(\theta\eta\right)'\epsilon q=\left(1+\epsilon-\alpha\epsilon/\epsilon_{F}\right)/\epsilon_{F}$,
the equality above is rewritten as
\[
\rho_{t}=\frac{1}{1-v}\ \frac{1}{\underset{\textrm{revenue}\thinspace\textrm{increase}}{\underbrace{\left[\left(1-\frac{1}{\epsilon_{F}}\right)+\frac{1+\epsilon-\alpha\epsilon/\epsilon_{F}}{\epsilon_{F}}\right]}}+\underset{\textrm{cost}\thinspace\textrm{savings}}{\underbrace{\left[\frac{1-\tau}{1-v}-\frac{1}{\epsilon_{F}}\right]\epsilon\chi}}}
\]
\[
=\frac{1}{1-v}\ \frac{1}{\underset{\textrm{revenue}\thinspace\textrm{increase}}{\underbrace{\left[1+\frac{(1-\alpha/\epsilon_{F})\epsilon}{\epsilon_{F}}\right]}}+\underset{\textrm{cost}\thinspace\textrm{savings}}{\underbrace{\left[\frac{1-\tau}{1-v}-\frac{1}{\epsilon_{F}}\right]\epsilon\chi}}}.
\]

To further facilitate the understanding the connection of this result
for to Proposition \ref{PropositionGeneralSymmetricOligopolyPassThrough},
consider the case of zero initial taxes ($t=v=\tau=0$). Then, Proposition
\ref{PropositionGeneralSymmetricOligopolyPassThrough} claims that
\[
\rho_{t}=\frac{1}{1+\epsilon\chi-\theta\chi+[-\eta\theta+\epsilon q\left(\theta\eta\right)']},
\]
whereas Proposition \ref{PropositionPassThroughAndMCUnderPriceCompetition}
shows that 
\[
\rho_{t}=\frac{1}{1+\epsilon\chi-\theta\chi+[-\frac{1}{\epsilon}\cdot\frac{\epsilon}{\epsilon_{F}}+\frac{1+\left(1-\alpha/\epsilon_{F}\right)\epsilon}{\epsilon_{F}}]}=\frac{1}{1+\epsilon\chi-\theta\chi+\left(1-\frac{\alpha}{\epsilon_{F}}\right)\theta},
\]
because $\theta=\epsilon/\epsilon_{F}$. Here, the direct effect from
$-\eta\theta$ is canceled out by the part of the indirect effect
from $\epsilon q\left(\theta\eta\right)'$. The new term, which appears
as the fourth term in the denominator, shows \textit{how the industry's
curvature affects the pass-through rate}: as the demand curvature
becomes larger (i.e., as the industry's demand becomes more convex),
then the pass-through rate becomes higher, although this effect is
mitigated by the degree of competitiveness, $\theta$.

\subsubsection{Proof of Proposition \ref{PropositionPassThroughAndMCUnderQuantityCompetition}}

\label{AppendixPropositionProofOfPassThroughFormulaUnderQuantityCompetition}In
the case of quantity setting, $\theta=\eta_{F}/\eta$, so $\left(\eta+\chi\right)\theta=\left(1+\chi/\eta\right)\eta_{F}$
and $\left(\theta\eta\right)'\epsilon q=q\,(\eta_{F})'/\eta=\left(1+\eta-\sigma\eta/\eta_{F}\right)\eta_{F}/\eta$,
where in the last equality we utilize the expression for the elasticity
of $\eta_{F}\left(q\right)$ and $\sigma_{F}+\sigma_{C}=\sigma\eta/\eta_{F}$
from Subsection \ref{SubsectionElasticitiesAndCurvatures}. Substituting
these into the expression for $\rho_{t}$ in Proposition \ref{PropositionGeneralSymmetricOligopolyPassThrough}
gives 
\[
\rho_{t}=\frac{1}{1-v}\ \frac{1}{1-(1+\frac{1}{\eta}\chi)\eta_{F}+\frac{1}{\eta}\left(1+\eta-\frac{\sigma\eta}{\eta_{F}}\right)\eta_{F}+\frac{1-\tau}{1-v}\frac{1}{\eta}\chi},
\]
which is equivalent to the expression for $\rho_{t}$ in the proposition.
Since $\theta=\eta_{F}/\eta$, Proposition \ref{PropositionRelationshipBetweenPassThroughs}
implies $\rho_{v}=\left(\epsilon-\theta\right)\rho_{t}/\epsilon=\left(1/\eta-\eta_{F}/\eta\right)\rho_{t}\eta=\left(1-\eta_{F}\right)\rho_{t},$
which can be used to verify the expression for $\rho_{v}$.

\subsubsection{Intuition behind Proposition \ref{PropositionPassThroughAndMCUnderQuantityCompetition}}

\label{AppendixDiscussionOfPassThroughFormulaUnderQuantityCompetition}The
intuition for $\rho_{t}$ in Proposition \ref{PropositionPassThroughAndMCUnderPriceCompetition}
is similar to the case of price competition. Recall again that
\[
\rho_{t}=\frac{1}{1-v}\ \frac{1}{\underset{\textrm{revenue}\thinspace\textrm{increase}}{\underbrace{[(1\!-\!\theta\eta)\!+\!\left(\theta\eta\right)'\epsilon q]}}+\underset{\textrm{cost}\thinspace\textrm{savings}}{\underbrace{\left[\frac{1-\tau}{1-v}\epsilon-\theta\right]\chi}}}.
\]
Then, $\theta=\eta_{F}/\eta$ implies $\left(1/\epsilon_{S}-\eta\right)\theta=[(1/\epsilon_{S}\eta)-1]\eta_{F}$
and $\left(\theta\eta\right)'(q/\eta)=q\,(\eta_{F})'/\eta$\linebreak{}
 $=\left(1+\eta-\sigma_{F}-\sigma_{C}\right)(\eta_{F}/\eta)$. Thus,
the equality above is rewritten as
\[
\rho_{t}=\frac{1}{1-v}\ \frac{1}{\underset{\text{revenue increase}}{\underbrace{\left[\left(1-\eta{}_{F}\right)+\frac{1+\eta-\sigma\eta/\eta_{F}}{\eta}\eta_{F}\right]}}+\underset{\text{cost savings}}{\underbrace{\left[\frac{1-\tau}{1-v}\frac{1}{\epsilon_{S}\eta}-\frac{\eta{}_{F}}{\epsilon_{S}\eta}\right]}}}
\]
\[
=\frac{1}{1-v}\ \frac{1}{\underset{\text{revenue increase}}{\underbrace{\left[1+\frac{\eta_{F}-\sigma\eta}{\eta}\right]}}+\underset{\text{cost savings}}{\underbrace{\left[\frac{1-\tau}{1-v}-\eta_{F}\right]\frac{1}{\epsilon_{S}\eta}}}}.
\]

To further facilitate the understanding the connection of this result
for to Proposition \ref{PropositionGeneralSymmetricOligopolyPassThrough},
consider the case of zero initial taxes ($t=v=\tau=0$) again. Then,
Proposition \ref{PropositionPassThroughAndMCUnderQuantityCompetition}
shows that
\[
\rho_{t}=\frac{1}{1+\epsilon\chi-\theta\chi+[-\eta\cdot\frac{\eta_{F}}{\eta}+\left(1+\frac{1}{\eta}-\frac{\sigma}{\eta_{F}}\right)\eta_{F}]}=\frac{1}{1+\epsilon\chi-\theta\chi+\left(1-\frac{\sigma}{\theta}\right)\theta}
\]
because $\theta=\eta_{F}/\eta$. Here, the term $\left(1-\sigma/\theta\right)\theta$
demonstrates the \textit{effects of the industry's inverse demand
curvature,} $\sigma$\textit{, on the pass-through rate}: as the inverse
demand curvature becomes larger (i.e., as the industry's inverse demand
becomes more convex), the pass-through rate becomes higher. Interestingly,
in contrast to the case of price competition, this effect is not mitigated
by the degree of competitiveness, $\theta$. 

\subsubsection{Equilibrium prices and outputs under price and quantity competition
with the linear demand \label{subsec: Eq P and Q under P and Q Competition with L Demand}}

The equilibrium price and output under price competition are obtained
as
\[
p=\frac{1+\frac{t}{1-v}}{2-(n-1)\mu},\qquad q=\frac{1-[1-(n-1)\mu]\frac{t}{1-v}}{2-(n-1)\mu},
\]
and thus
\[
\frac{p}{q}=\frac{1}{1-[1-(n-1)\mu]\frac{t}{1-v}}\left(1+\frac{t}{1-v}\right),
\]
implying that 
\[
\text{\ensuremath{\epsilon}}=\frac{[1-(n-1)\mu]\left(1+\frac{t}{1-v}\right)}{1-[1-(n-1)\mu]\frac{t}{1-v}},\qquad\text{\ensuremath{\epsilon_{F}}}=\frac{1+\frac{t}{1-v}}{1-[1-(n-1)\mu]\frac{t}{1-v}}.
\]

Similarly, the equilibrium price and output under quantity competition
are given by
\[
p=\frac{\frac{1-(n-2)\mu}{1-(n-1)\mu}+(1+\mu)\frac{t}{1-v}}{2-(n-3)\mu},\qquad q=(1+\mu)\frac{1-[1-(n-1)\mu]\frac{t}{1-v}}{2-(n-3)\mu},
\]
and thus
\[
\frac{p}{q}=\frac{1}{1-[1-(n-1)\mu]\frac{t}{1-v}}\left(\frac{1-(n-2)\mu}{(1+\mu)[1-(n-1)\mu]}+\frac{t}{1-v}\right),
\]
implying that 
\[
\text{\ensuremath{\eta}}=\frac{1-[1-(n-1)\mu]\frac{t}{1-v}}{\frac{1-(n-2)\mu}{1+\mu}+[1-(n-1)\mu]\frac{t}{1-v}},\qquad\text{\ensuremath{\eta{}_{F}}}=\frac{1-[1-(n-1)\mu]\frac{t}{1-v}}{1+\frac{(1+\mu)[1-(n-1)\mu]}{1-(n-2)\mu}\frac{t}{1-v}}.
\]

\subsection{Proofs and discussions for Section 4}

\subsubsection{Proof of Proposition \ref{PropositionRelativeSizeOfPassThroughVectorComponents}}

\label{AppendixProofOfPropositionRelativeSizeOfPassThroughVectorComponents}Consider
an infinitesimal tax change such that the equilibrium price (and therefore
quantity) does not change: $\boldsymbol{\tilde{\rho}}\cdot d\mathbf{T}=0.$
Let us choose $d\mathbf{T}$ to have just two non-zero components:
$dT_{\ell}$ and $dT_{\ell'}$. This implies
\begin{equation}
\frac{\tilde{\rho}_{T_{\ell}}}{\tilde{\rho}_{T_{\ell'}}}=-\frac{dT_{\ell'}}{dT_{\ell}}.\label{EquationRatioOfPassThroughRatesAsRatioOfTaxChanges}
\end{equation}
Since Equation (\ref{EquationDefinitionOfGeneralizedConduct}) must
hold both before and after the tax change, it must be the case that
$1-\tau-\left(1-\nu\right)\eta\theta$ does not change, and in turn
\[
\left(-\tau_{T_{\ell}}+\nu_{T_{\ell}}\eta\theta\right)dT_{\ell}+\left(-\tau_{T_{\ell'}}+\nu_{T_{\ell'}}\eta\theta\right)dT_{\ell'}=0.
\]
Substituting for $dT_{\ell'}$ from this equation into Equation (\ref{EquationRatioOfPassThroughRatesAsRatioOfTaxChanges})
and using the definition of pass-through quasi-elasticities leads
to the desired result.

\subsubsection{Proof of Proposition \ref{PropositionCommonFactorInPassThrough}}

\label{AppendixProofOfPropositionCommonFactorInPassThrough}The same
type of reasoning as in the proof of Proposition \ref{PropositionGeneralSymmetricOligopolyPassThrough}
is useful in proving Proposition \ref{PropositionCommonFactorInPassThrough}.
In particular, comparative statics of Equation (\ref{EquationDefinitionOfGeneralizedConduct})
with respect to a tax $T_{\ell}$ leads to the desired result after
utilizing the definitions above and eliminating marginal cost using,
again, Equation (\ref{EquationDefinitionOfGeneralizedConduct}). The
calculation is a bit tedious but completely straightforward.

\subsubsection{Depreciating licenses}

\label{AppendixDepreciatingLicenses}Here we discuss the relationship
of exogenous competition to depreciating licenses in Weyl and Zhang
(2017).  In the setup of Section 2 of Weyl and Zhang (2017), there
are two agents, $S$ and $B$ (``seller'' and ``buyer''). Agent
$S$ holds an asset and declares a reservation value $p$, which influences
the tax (``license fee'') $\tilde{q}p$ the agent needs to pay to
the government. Here $\tilde{q}$ is the license tax rate (denoted
$\tau$ in the original paper). Agent $B$ may then purchase the asset
at that price $p$ from agent $S$. The value for agent $S$ is $\eta+\gamma_{S}$,
and for agent $B$ it is $\eta+\gamma_{B}$, for some common value
component $\eta$.\footnote{The original paper considers $\eta$ being determined by agent S at
the very beginning. Here we focus on the subgame after $\eta$ has
been determined. } Here $\gamma_{B}$ is a random variable with CDF $F\left(\gamma_{B}\right)$
representing heterogeneity in $B$\textquoteright s value, which is
not observed by $S$. As Weyl and Zhang (2017) show, the sale probability
$q$ (denoted the same way in the original paper) is then determined
as the solution of $P\left(q\right)=p$, where $P\left(q\right)\equiv F^{-1}(1-q)+\eta$.
Up to a constant, agent $S$'s expected profit function (utility function)
is $(P\left(q\right)-\eta-\gamma_{S})(q-\tilde{q})$ or $P\left(q\right)\left(q-\tilde{q}\right)-\left(q-\tilde{q}\right)mc$,
where we used the notation $mc\equiv\eta+\gamma_{S}$.\footnote{In the original paper, on page 4, the profit function is written as
$(M(q)-\gamma_{S})(q-\tau)+(\eta+\gamma_{S})(1-\tau)\text{-}c(\eta)$.
The last two terms are constant. The $M(q)$ in the first term corresponds
to our $P\left(q\right)-\eta$.} We recognize that this is exactly of the same form as the profit
function in the case of monopoly with constant marginal cost $mc$
subject to exogenous competition $\tilde{q}$ and inverse demand function
$P\left(q\right)$. 

\bigskip{}

\textsc{Nagoya University }

\textsc{University of Tokyo}

\end{document}